\documentclass[11pt,a4paper]{article}
\pdfoutput=1
\usepackage{jheppub}
\usepackage{latexsym}
\usepackage[latin1]{inputenc}
\usepackage{empheq}
\numberwithin{equation}{section}
\usepackage{cancel}
\usepackage{subeqnarray}
\usepackage{xcolor}

\usepackage{longtable}
\usepackage{color}
\usepackage{multirow}
\usepackage{epstopdf}
\usepackage{listings}
\usepackage[euler]{textgreek}
\usepackage{tikz}
\usepackage{cancel}
\usepackage{enumitem}
\usetikzlibrary{positioning}
\usepackage{bbold}
\usepackage{graphicx}

\usepackage{caption,graphicx,color}
\usepackage{subcaption}
\usepackage{amsmath,epsfig}
\usepackage{amssymb,amsfonts}
\usepackage{latexsym}
\usepackage[latin1]{inputenc}
\usepackage{float}
\usepackage{overpic}
\usepackage{slashed}
\usepackage{caption}
\usepackage{subcaption}
\usepackage{xcolor}

\usepackage{graphicx}
\usepackage{longtable}

\relax
\renewcommand{\theequation}{\arabic{section}.\arabic{equation}}

\def\be{\begin{equation}}
\def\ee{\end{equation}}

\newcommand{\de}{\partial}
\newcommand{\bear}{\begin{eqnarray}}
\newcommand{\bea}{\begin{eqnarray}}
\newcommand{\eear}{\end{eqnarray}}
\newcommand{\eea}{\end{eqnarray}}

\def\bsq{\begin{subequations}}
\def\esq{\end{subequations}}
\def\hri#1#2{\href{http://arxiv.org/abs/#1}{[ArXiv:#1]#2}}
\def\hre#1#2{\href{http://arxiv.org/abs/#1/#2}{[ArXiv:#1/#2]}}

\def\hrj#1#2{\href{www.doi.org/#1}{#2}}
\newbox\pippobox

\def\II{\relax{\rm I\kern-.18em I}}

\def\m{\mu}
\def\n{\nu}

\def\sp{\;\;\;,\;\;\;}

\def\f{\varphi}

\def\a{\alpha}

\title{Improved Holographic QCD on a Curved Background: an Application of Dynamical System Theory in Holography}

\author[\dagger,\sharp,*]{Elias~Kiritsis,}
\author[\dagger]{Francesco~Nitti,}
\author[\flat]{Jean-Loup~Raymond}

\affiliation[\dagger]{Universit� de Paris Cit\'e, CNRS, Astroparticule et Cosmologie, F-75013 Paris, France}
\affiliation[\sharp]{Crete Center for Theoretical Physics, Institute for Theoretical and
Computational Physics, Department of Physics
University of Crete, Heraklion, Greece}
\affiliation[*]{Arnold Sommerfeld Center for Theoretical Physics,
Ludwig-Maximilians-Universit\"at M\"unchen, 80333 M\"unchen, Germany}
\affiliation[\flat]{\'Ecole normale sup\'erieure Paris-Saclay, F-91190 Gif-sur-Yvette, France.}

\emailAdd{kiritsis@apc.in2p3.fr}
\emailAdd{nitti@apc.in2p3.fr}
\emailAdd{jean-loup.raymond@ens-paris-saclay.fr}

\preprint{\begin{flushright}
CCTP-2025-3\\
ITCP-2025/3
\end{flushright}
}

\abstract{The finite-curvature phase diagram of IHQCD, a bottom-up holographic model for large $N_c$ non-supersymmetric YM$_4$, is investigated. This holographic theory belongs to a class of Einstein-Dilaton theories that exhibit no scaling in the IR. We use advanced techniques from dynamical system theory to address this problem that is harder than other holographic setups. We classify all solutions where the dual theory is defined on a constant curvature manifold, both with positive and negative curvature. For general theories in this class a quantum phase transition occurs at finite curvature. For IHQCD in particular, we find that the phase transition occurs at zero curvature.}

\begin{document}

\maketitle

\section{Introduction}

One of the most prominent features of the AdS/CFT correspondence \cite{magoo} is the possibility of describing the strongly coupled regime of confining gauge theories, via a gravitational dual theory in higher dimensions,  \cite{witten,Malda-nunez,GPPZ}. Among  the  holographic models which display confinement, the  class of models called {\em Improved Holographic QCD} (IHQCD) plays a special role:  it is the minimal  phenomenological model which captures essentially all the strong-coupling phenomenology of four-dimensional non-supersymmetric Yang-Mills theory \cite{IHQCD}.  With the addition of extra degrees of freedom (non-abelian gauge fields and a bifundamental scalar) it constitutes the most accurate phenomenological model of  large-$N$ QCD with a (large number) $N_f$ of flavors \cite{VQCD}.

There are many bottom up models that try to reproduce aspects of large $N_c$ YM theory in four dimensions (a list can be found in \cite{Jani}).
IHQCD has been the most successful one, when it comes to comparisons  with lattice data both at zero and finite temperature, \cite{data,Panero,review}.
Moreover, although it is a bottom up model, its inception rests solidly on string theory expectations, \cite{disect}. 

All bottom-up and top-down models with confinement and a mass gap have generically  scaling IR asymptotics with  hyperscaling violation\footnote{This is due to the fact that upon reduction of the solutions to five (or four) dimensions, the leading IR asymptotics of the potential is exponential, and this translates in scaling and gapped IR asymptotics.}, \cite{GK,GK1,hy2}
The IR asymptotics of the scalar potential in IHQCD were chosen both to remove this scaling behavior and also produce linearly rising glueball trajectories\footnote{Interestingly, similar generalized Kaluza-Klein spectra have been considered in \cite{Basile}, in the context of the emergent string conjecture. We thank I. Basile and C. Montella for bringing this to our attention.} (instead of the quadratically rising ones in all other cases), \cite{IHQCD}.

In this work we shall study IHQCD when the dual quantum field theory is set  on a {\em curved} four-dimensional maximally symmetric space-time. As such, this work is a direct continuation of  \cite{AdS2,Jani}, where the behavior of more general holographic confining Einstein-Dilaton theories was  analyzed, when the dual confining QFT is set on a constant-curvature background. It is part of a more general line of work aimed at understating holographic QFTs on curved backgrounds \cite{C,F,dS,s2s2,Jani2}

In such a  model, the boundary QFT is characterized by two scales: the  scale $\Lambda_{UV}$, which parametrises the running of the coupling of the  relevant operator dual to $\varphi$, and the  boundary curvature scalar $R_{UV}$. The non-trivial dimensionless parameter of the theory can be taken as the ``dimensionless curvature'' parameter
\be \label{intro4}
{\cal R} ={ R_{UV}\over \Lambda^2_{UV}}
\ee
The goal is then to first understand the space of solutions of the gravity dual theory that are dual to the semiclassical ground-states.
Once this is done, we would like  to characterize the physical  observables  (e.g. phase diagram, spectrum) as a function of  ${\cal R}$.  For a  comprehensive introduction, motivation  and list of references, the reader is referred to \cite{Jani}.

\subsection{A very special class of holographic models}
In general $d+1$ bulk dimensions, the  theories we shall consider  consist of d-dimensional Einstein gravity minimally coupled to a dilaton field $\varphi$, schematically:
\be \label{intro1}
S =  {M_p^{d-1} }\int d^{d+1}x \sqrt{-g} \left(R - {1\over 2}(\de\varphi)^2 - V(\varphi) \right),
\ee
The subset of theories in the IHQCD class, have a potential with a single extremum (a maximum, ie. UV fixed point) and   the asymptotic large-$\varphi$ behavior takes the form:
\be \label{intro3}
\varphi \to +\infty: \quad V_{\alpha}(\varphi) \sim -V_{\infty}\, \varphi^\alpha  e^{2b_c \varphi}     \qquad   b_c = \sqrt{1\over 2(d-1)},
\ee
where $V_{\infty}> 0$  and $\alpha$ is a real parameter.  For $\alpha > 0$, the model  (like four-dimensional Yang-Mills theory) displays a mass gap and a discrete spectrum of ``glueball'' excitations. Specifically, IHQCD corresponds to  the special value $\alpha = 1/2$: this leads to to  Regge-like behavior  for the mass eigenstates, $m_n^2 \sim n$. We shall refer to the asymptotic behavior (\ref{intro3}) for general $\alpha$ as IHQCD-{\em like}.

 In \cite{Jani} {\em generic} Einstein-Dilaton holographic theories in the confining class were considered, and the phase structure of the theory was analysed as a function of ${\cal R}$. The analysis was restricted to confining  models in which  the potential behaves at large $\varphi$ as $V\sim e^{2b\varphi}$  with $b>b_c$. In such a  case,  the spectrum is  still gapped and discrete, but $m_n^2\sim n^2$ independently of the value of $\alpha$, \cite{IHQCD}. In  \cite{Jani}, a classification was given  of   the different types of solutions as a function of the boundary curvature ${\cal R}$, and the phase transitions  between different branches were analyzed. A similar analysis was carried out in \cite{AdS2} in the negative curvature case, always for $b > b_c$ strictly.

The  case $b=b_c$ is special and deserves a separate dedicated analysis for both physical and mathematical reasons:
\begin{enumerate}
\item From the physical standpoint, it is the only case which gives rise to a phenomenologically interesting model for four-dimensional YM. This means that one can actually compare the holographic results with actual physical  data like  lattice YM results, glueball spectra and  (on a positive curvature background) YM in the cosmological context.

    In contrast, although it may be interesting for model building,  it is unclear whether the general case $b>b_c$, corresponds to any physical theory.
    There are however subcases for specific discrete values of $b$, that the IR theory corresponds to higher-dimensional  holographic CFT compactified on a sphere, \cite{GK}.

\item From the mathematical standpoint, the value $b=b_c$ is on the boundary of the confining regime. As a result, the general analysis of large-$\varphi$ asymptotic solutions used in \cite{Jani} breaks down and one must resort to a more refined mathematical  method in order to find and study the ground-states of the theory.

\end{enumerate}

\subsection{The asymptotic regime of holographic RG flows as  a dynamical system}

The  IR features of the dual field theory   map  to the large-$\varphi$ asymptotic region of the potential.  To understand the IR of the dual field theory, one has  
classify {\em regular} solutions,  and identify the parameter(s) (i.e. integration constants) on which they depend. Further, one connects these integration constants to the UV parameters (which for such a theory are sumarized by the single dimensionless parameter, ${\cal R}$). Above,  by ``regular'' we mean:
 \begin{enumerate}

\item  Either strictly regular geometries which necessarily end at a finite value of $\varphi$; These were called {\em type III} solutions in \cite{Jani} and exist only when $\mathcal{R}>0$;

\item  Or solutions which extend all the way to $\varphi\to +\infty$ and  have mild singularities which are holographically acceptable (they can be resolved by turning on a small radius black hole \cite{Gubser} or by generalized uplifting to higher dimensions \cite{GK,GK1}). These solutions were named {\em type I} and {\em type II} solutions in \cite{Jani}.
\end{enumerate}
Either or both types of solutions may exist, depending on the value of ${\cal R}$. In all cases, such ``regular''  solutions are non-generic in that they constitute a set of measure zero in the full space of solutions. In particular  they have fewer integration constants than the generic ``singular'' solutions (which we call {\em type 0}). More specifically, the generic singular solutions  have two IR integration constants while,  for $b\neq b_c$,  ``regular'' solutions   have  zero or one IR integration constants, \cite{Jani}.

Since the full UV-to-IR solution is eventually found numerically,  it is  essential to know the IR analytic  behavior of the  solutions of type I and II  extending to $\varphi\to +\infty$,   and impose it as an appropriate initial condition. Otherwise the numerical integration  will invariably produce a generic, singular type 0 solution.

For a potential behaving in the IR as in (\ref{intro3}) with $b\neq b_c$,   obtaining  the analytical IR behavior of ``regular'' solutions was  achieved  in \cite{Jani}  with rather elementary methods. These fail for $b=b_c$, for reasons that shall become clear further on in this paper.

In this work, we approach the study of the asymptotic IR solutions of the model (\ref{intro1})  in the large-$\varphi$ region, by methods borrowed from dynamical systems theory. We consider a solution of the  $d+1$-dimensional bulk field equations  of the form
\be \label{intro6}
ds^2 = du^2 + e^{2A(u)} \zeta_{\mu\nu} dx^\mu dx^\nu , \qquad \varphi = \varphi(u)
\ee
where  $\zeta_{\mu\nu}$ is a $d$-dimensional constant curvature metric. Such  an ansatz is expected to contain the solutions that are  the ``ground states,''  in that they preserve the maximal symmetry of the fixed-$u$ slices, and as such they can be thought of as  vacua of a dual QFT living on a curved manifold with metric $\zeta_{\mu\nu}$.

We  study solutions in terms of a phase space, constructed using appropriate  variables  $(\tilde{W},\tilde{S})$, which may be obtained  from the  flow velocities (in the holographic direction) of the scalar field and the metric scale factor $(\dot{A}, \dot{\varphi})$:
{
\be \label{intro5}
\Tilde{W} = -\frac{2(d-1)}{\sqrt{V_\infty}} \dot{A} \exp[-b \varphi] \varphi^{-\frac{\a}{2}} , \qquad \Tilde{S} = \frac{1}{\sqrt{V_\infty}} \dot{\varphi}\exp[-b \varphi]\varphi^{-\frac{\a}{2}}
\ee
}
where $\dot{} = d/du$.
In this phase space\footnote{This is 2-dimensional for $\alpha=0$, but has to be extended  by an auxiliary third variable $Z(\varphi)$ for $\alpha\neq 0$.},  the field  equations in the large $\varphi$  regime can be put in the form of an autonomous\footnote{For a dynamical system $X'= f(X, \f)$, where the derivative is taken with respect to $\f$, this means that $f$ does not depend explicitely on $\f$. A more precise definition is given in Appendix \ref{sec:D}.} dynamical system,

\be \label{intro7}
\Tilde{W}'(\varphi)=f(\Tilde{W},\Tilde{S}) \sp \Tilde{S}'(\varphi)=  g(\Tilde{W},\Tilde{S})
\ee
where prime stands for derivative with respect to $\f$.

In this language, type I a and II ``regular'' solutions appear as {\em critical points} of this dynamical system\footnote{They are not, however, fixed points of the full holographic RG flow.}, while generic singular solutions escape to infinity.  Moreover,  the integration constant parametrizing the space of such solutions  (which can be related to UV parameters ${\cal R}$) can be understood as parametrizing small deviations away from the critical point along a stable direction. The phase space and corresponding dynamical flows are represented in figure \ref{fig:phase}.

For $b\neq b_c$, the analysis of the asymptotic solutions can be performed by elementary methods,  and it is enough to linearize  the equations around the critical points and find the stable directions. This is, in essence, what was done in \cite{AdS2,Jani}.

Things become more interesting (and non-trivial) in the critical case $b=b_c$, because of the appearence of a marginally stable direction.
To analyse the space of solutions in this case, we  use the dynamical system's concept of {\em center manifold} \cite{dyna,dynaproof}. This is, essentially, the non-linear generalization of the vector space of linear perturbations (with  zero eigenvalue)  around a critical point. Identifying the center manifold {of the critical point}, associated to ``regular'' solutions,  we are able to systematically provide a consistent (non-linear) expansion of the solutions around the critical point, which provides the desired classification. Crucially, this  includes the deformation parameters, which control the solution in the IR, and are mapped to the UV data.

The concept of center manifold is particularly useful in the presence of bifurcations: these are special values of the parameters, for which the phase portrait changes topology. For bifurcations involving only critical points, for instance when their stability changes, it follows from continuity\footnote{This will always be the case for the dynamical systems we consider here.} that one of the Liapunov exponents which control the small deformations vanishes. As a result, the problem of finding continuous deformations of the ``regular'' solutions becomes intrinsically non-linear. As we shall see, this is precisely what happens for in the case of IHQCD.

\subsection{Summary of results}

The main results of this work can be classified  in two categories:

\begin{enumerate}
\item From the formal standpoint, we develop a new method to treat holographic systems using concepts from dynamical system theory. We use this to analyze the asymptotic large-$\varphi$ regime but in its full generality there is no obstruction on using this method for  complete solutions on the whole range of $\varphi$. This is in a par with similar topological techniques used in \cite{Gukov} to study the topology of standard QFT RG flows.

\item From the physical point of view, we analyze the phase structure of holographic models in the IHQCD class,  (\ref{intro1}) and (\ref{intro3}). This extends the analysis of confining holographic theories defined on curved space-time, which was carried out in \cite{AdS2,Jani},  for exponential asymptotics  away from the critical exponent $b_c$. The main interest of this is that IHQCD with d=4 and $\a={1\over 2}$ is the closest holographic model (in the IR) to large $N_c$ YM$_4$.
\end{enumerate}

In the rest of this subsection we give a brief summary our main findings.

\paragraph{The non-critical case.} For $b\neq b_c$ we recover, using the language of dynamical systems, the known results in \cite{Jani, AdS2} about the IR asymptotics. In this formulation,  ``regular'' solutions extending to $\varphi \to \infty$ correspond to fixed points and trajectories that end on them.

One of them corresponds to the Type I solution, {and has one attractive direction for $b<b_c$ (and therefore admits a one-parameter family of small deformations corresponding to the curvature parameter). However it has  no attractive directions for $b>b_c$ (and therefore admits no small deformations).

 The second fixed point, corresponds to type II solution, has no attractive direction for $b<b_c$ and one attractive direction for $b>b_c$}.

\paragraph{Bifurcations.} Bifurcations are classified into two categories, local and global bifurcations.

{\em Local bifurcations} affect the neighborhood of a critical point of the dynamical system, typically by changing stability. In this case, one eigenvalue of the matrix governing linearized fluctuations vanishes, increasing the dimension of the center manifold.

 {\em Global bifurcations} on the other hand are bifurcations which affect a larger region of the phase space.

In our analysis we find one example of a global bifurcation occurring at  $b_G = b_c \sqrt{d}$, where the type II critical point asymptotes to infinity, destroying an invariant trajectory linking the two critical points. The value $b_G$ corresponds to Gubser's bound (which will be discussed in section 2), above which  asymptotic exponential solutions cease to exist.

More importantly, we find two occurrences of critical behavior corresponding to local bifurcations:
\begin{itemize}
\item  A first local bifurcation occurs for pure exponential asymptotics ($\alpha = 0$ in (\ref{intro3}))  at  the critical value $b=b_c$, and  this is why in this case  the  analysis performed in \cite{Jani} fails. In particular, in this case type I and II solutions merge into a single branch which we denote by type I/II, admitting one deformation parameter. This type of bifurcation, involving two critical points which collide and exchange stability, is called {\em transcritical} \cite{strogatz}.

\item For $b=b_c$ and $\alpha\neq 0$, we find a second local bifurcation at the special value $\alpha = 1/2$. Interestingly, this is exactly the value corresponding to IHQCD, indicating that this asymptotics of the potential are {\em doubly} special.
\end{itemize}
In both cases the IR solution for the dynamical variables $\tilde{W},\tilde{S}  $ defined in (\ref{intro5}) behaves,  close to the fixed point, as:
\be \label{intro8}
\tilde{W},\tilde{S} \sim \text{Constant} + \left(\text{analytic in}\, {1\over \varphi}\right) + C_\alpha \left(\text{subleading, non-analytic in} {1\over \varphi} \right)
\ee
as $ \varphi \to +\infty$,
where the leading term and the analytic subleading terms are completely fixed. The non-analytic corrections  are governed by a continuous parameter $C_\alpha$, which is the IR manifestation of the curvature parameter (\ref{intro4}).

For all  values of $\alpha$ in the interval $0<\alpha<1$, we  construct  the center manifold around the critical point\footnote{Outside the interval  $0<\alpha<1$,  one has to obtain the expression of the center manifold to higher non-linear order around the fixed point, which is more technically challenging but is doable if necessary.}, we extract the leading and subleading asymptotics,   and  we identify the non-analytic terms containing the single integration constant.

\paragraph{The phase diagram of IHQCD on a curved manifold.} Having obtained the analytic  behavior, up to the needed subleading order  of the regular solutions in the large-$\varphi$ region, we  study numerically full solutions for a typical potential with IHQCD-like asymptotics  (\ref{intro1}) and  $0<\alpha<1$, and a single UV fixed point at $\varphi=0$.

As in the case $b\neq b_c$, there are two types of ``regular'' solutions:
\begin{itemize}
\item[a)] type III, with a regular endpoint at finite $\varphi$;

\item[b)] type I/II  solutions which reach $\varphi \to +\infty$. For these, the distinction between type I and type II used in \cite{AdS2,Jani} for $b\neq b_c$ does not apply, but one can still identify a  one-parameter  family of solutions in this class, which can be continuously deformed into one another,  plus one ``isolated'' solution.
\end{itemize}
By varying the IR parameters controlling the solutions, we map out the space  of solutions as a function of the UV parameter, $\cal R$, in the whole range $-\infty < \cal R< +\infty$. This allows us to establish the phase diagram of the theory as a function of  ${\cal R}$ and the parameter $\alpha$. The resulting phase diagram can be seen in Figure \ref{fig:Ralpha} and can be summarized as follows:

\begin{itemize}

\item  $\alpha > 1/2$: the space of solutions  is split by a critical value ${\cal R}_{c+}>0$ of the dimensionless curvature parameter:  for any ${\cal R} >{\cal R}_{c+}$, solutions exist which end at a finite $\varphi_0$ (type III), while for any curvature in the range  $-\infty < {\cal R} <{\cal R}_{c+}$  we find only type I/II  solutions extending to $\varphi \to +\infty$. For type III solutions,  the limit  ${\cal R} \to +\infty$ correspond to $\varphi_0\to 0$ (the UV fixed point), while as  ${\cal R} \to {\cal R}_{c+}$ from above,  $\varphi_0\to +\infty$.  At the critical curvature ${\cal R}_{c+}$ the type III solution merges with the type I/II branch.
\item $\alpha < 1/2$: In this case  type III solutions  exist for all positive curvature values (but not negative curvature), while  type I/II solutions reaching $\varphi \to +\infty$  only exist for  negative curvature. There is still a critical value ${\cal R}_{c-}<0$, but in this case it  separates two different branches of type I/II solutions.
\item $\alpha = 1/2$: the critical curvature ${\cal R}_{c+}\to 0$ as $\alpha \to 1/2^+$. At exactly this point  the non-analytic asymptotic behavior of the type I/II solutions  changes from power-law to logarithmic. A detailed calculation  shows that the curvature parameter ${\cal R}$ of  solutions reaching $\varphi \to \infty$ is {\em strictly negative}. This shows that for IHQCD proper, only type III regular solutions exist at  positive curvature.
\end{itemize}

\paragraph{The free energy and quantum phase transitions.} At the boundaries of parameter space which separate  type I/II and type III solutions, one can, in principle, find  phase transitions as  a function of $\mathcal{R}$. Such transitions have been found and studied for $b>b_c$  in \cite{Jani}. In all cases studied in \cite{Jani} the transition occurs  at a strictly positive curvature $\mathcal{R}$. It  was found to be first order for $b$ above a certain value $b_E$  (as we shall discuss in more detail  in section 5), and  higher order for lower values of $b$, $b_c<b<b_E$.  For $b=b_c$, we expect the transition, if it  exists, to be also continuous and of higher order.

This is confirmed, as a result of explicit  computations. To analyse the transition, we compute numerically the (renormalized) free energy and we study its behavior near the boundaries between the type I/II and type III solutions, for different values of $\alpha$ in the interval $0<\alpha < 1$.
\begin{itemize}
\item
For $\alpha >1/2$  the value  ${\cal R}_{c+} >0 $ corresponds to a  phase transition of second or higher order.
\item For $\alpha <1/2$ there is no  phase transition at finite positive curvature. A transition occurs at $\mathcal{R} = 0$ (which corresponds to the value $\mathcal{R}_{c-}$ defined above), in which the free energy density and its derivative are continuous.

\item The value $\alpha = 1/2$  is the endpoint of the positive curvature phase transition. In  this case, all non-negative curvature solutions are type III. In particular, {\em there is no curvature-driven phase transition in IHQCD at positive curvature.}  The transition to type I/II solutions occurs at $\mathcal{R} =0 $  but this time the numerical analysis is not accurate enough to indicates that the free energy density has some  discontinuous derivative with respect to the curvature at $\mathcal{R} = 0$.
    As we know that the theory is gapped, one might be tempted to compare to gapped QFTs. Indeed theories with a finite number of massive particles have generically singular derivatives near $\mathcal{R}=0$ due to the conformal anomaly. However, it is not clear that this persists for theories with an infinite number of massive fields, like is the case here.

\end{itemize}

This paper is organized as follows.
Section 2 contains a comprehensive review of confining holographic Einstein-dilaton theories in flat at zero and finite curvature.  We give an overview of the various regimes as a function of the large-$\f$ asymptotics of the potential, and describe the space of solutions and phase transitions which were found for $b\neq b_c$ and were discussed in \cite{AdS2,Jani}.

In Section 3 we introduce the tools from dynamical system theory in the case of a pure exponential IR asymptotics ($\alpha=0$). We show how to treat the equations as a 2-dimensional autonomous system; we recover the form of the asymptotic solutions in the non-critical case $b\neq b_c$; we introduce the concept of center manifold  and use it to solve the non-linear problem associated to the IR asymptotic  solution  in the  in the critical case $b=b_c$.

In Section 4 we treat the most interesting, IHQCD-like case of a power-law corrected critical exponential potential ($b=b_c, \alpha \neq 0$). We show how to translate the   IR problem into a three-dimensional autonomous system, which we analyse again using the center manifold (which in this case is two-dimensional). We derive leading and subleading asymptotics of the solutions at large $\varphi$ for $0<\alpha < 1$.

In Section 5 we consider a theory with a simple scalar potential and study full UV-IR RG flows at finite curvature. We compute the solutions numerically and  use the analytic expressions obtained in the previous sections to control the IR asymptotics of the numerical computation. We map out the space of solutions, compute the free energy, and analyze the phase transitions between solutions with different geometric properties.

Several technical details, as well as concepts from dynamical systems theory, are collected in the Appendix.

\section{RG flows for holographic QFTs on curved spacetimes}

This section provides a self-consistent introduction to the holographic approach to confining theories based on Einstein-Dilaton gravity, and of the philosphy underlying Improved Holographic QCD. We also review  the holographic dictionary and  previous results about holographic duals to   QFTs on zero and finite curvature manifold. We provide the classification of the solutions and the phase diagram  for non-critical values of the parameter controlling the IR potential asymptotics, ($b\neq b_c$), for later comparison with the results obtained from section 3 on in the critical case and in IHQCD.  The interested reader is referred to \cite{IHQCD,disect,AdS2,Jani} for details.

\subsection{Setup and equations of motion}

We study the holographic flows for a $(d+1)$-dimensional Einstein-dilaton theory on asymptotically Anti de Sitter (AdS) space-time.  They are parametrized by the dependence on a  radial (holographic)  direction and  co-dimension-one slices,  whose geometry are that of $d$-dimensional de Sitter (dS) space-time (or, in the Euclidean version, a $d$-dimensional sphere $S^d$). The metric in the asymptotically AdS region is such that, via the holographic duality, these geometries are dual to ground states in a $d$-dimensional holographic QFT on de Sitter space (or a Euclidean field theory on $S^d$).

The theory, on the gravity side, is described by the following two-derivative action:
\begin{equation}
    S[g, \varphi] = M_p^{d-1}\int_{\mathcal{M}} du d^{d} x \sqrt{-g} \left( R - \frac{1}{2}\partial_a\varphi \partial^a\varphi - V(\varphi)\right),
    \label{action}
\end{equation}
where, as usual, the metric is dual to the energy-momentum tensor and the dilaton to the most relevant scalar operator of the dual theory.

Since we want to describe the ground state of the dual QFT on a constant positive curvature manifold, the natural ansatz for the metric is (in domain-wall coordinates)
\begin{equation}
    ds^2 = du^2 + e^{2A(u)} \zeta_{\mu \nu} dx^\mu dx^\nu,
    \label{metric}
\end{equation}
where $\zeta_{\mu\nu}$ is a fiducial $d$-dimensional positive-constant-curvature metric.
Although the equations we are solving are the same for any positive constant curvature metric $\zeta_{\m\n}$, we shall be mostly interested in this metric describing $S^d$ in the Euclidean case and dS$_d$ in the Minkowski signature case.

The associated ansatz for the dilaton is
\begin{equation}
    \varphi = \varphi(u).
    \label{ansatzphi}
\end{equation}

The equations of motions obtained upon varying the action (\ref{action})  are, for the ansatz (\ref{metric}-\ref{ansatzphi}):
\begin{align}
    2(d-1) \Ddot{A} + \dot{\varphi}^2 + \frac{2}{d}e^{-2A(u)}R^{(\zeta)} &=0, \label{rgflow1}\\
    d(d-1) \dot{A}^2 - \frac{1}{2} \dot{\varphi}^2 + V(\varphi) - e^{-2A(u)}R^{(\zeta)} &=0, \label{rgflow2} \\
    \Ddot{\varphi} + d\dot{A}\dot{\varphi} - V' &= 0, \label{rgflow3}
\end{align}
where we have denoted
$$\dot{f} \equiv \frac{df}{du}\sp f' \equiv \frac{df}{d\varphi}\;.
$$
The Ricci scalar of the fiducial  metric $\zeta_{\mu\nu}$ is denoted $R^{(\zeta)}$. Note that this is not the physical dS/S curvature as seen by the dual QFT, which instead must be extracted by a near-boundary Fefferman-Graham expansion, as described in  \cite{C,F} (and subsection \ref{ssec:uv}).

A solution of  equations (\ref{rgflow1}-\ref{rgflow3}) is uniquely associated to a holographic RG flow and therefore to the ground state of a unique dual QFT. Considering the Euclidean theory or the theory on another  Einstein  manifold (non-maximally symmetric) does not affect the equations of motion (\ref{rgflow1}-\ref{rgflow3}). Therefore, the results of this article are valid for generic positive constant curvature  metrics of general  signature.

\subsection{ The scalar potential and the QFT UV parameters}
\label{ssec:uv}
Before discussing how to solve equations (\ref{rgflow1}-\ref{rgflow3}), in this subsection we identify the parameters that characterize the dual quantum field theory in the far UV (hence referred to as UV-parameters) and which are in one-to-one correspondence with the boundary conditions to be imposed at the asymptotic AdS boundary  on the gravity side.

Here, we are considering  a $d$-dimensional QFT which is asymptotically conformal in the UV, and we are  taking  the simplified assumption that there is only one relevant operator, of conformal  dimension $\Delta$,  which is important for the dynamics and is responsible for the theory flowing to the IR.  This assumption is reflected on the gravity side by the fact that there is a single scalar field  $\varphi$ coupled to the metric in the bulk\footnote{This is equivalent to the assumption that the ``running" does not turn-on other scalar fields that are apart of the gravitational theory, so that we can effectively set them all to zero during the flow.}.

When a  QFT such as the one described above, is put on a constant curvature manifold, conformal invariance is explicitly broken by two relevant parameters:
\begin{enumerate}
\item  The dimensionful coupling $j$, of dimension $d-\Delta$,  multiplying  the relevant operator driving  the flow;

\item  The scalar curvature of the space-time where the QFT is set to be defined, which we denote by $R^{UV}$.
\end{enumerate}

Out of these two dimensionful parameters, we can construct a single dimensionless combination,
\be \label{UV0}
{\cal R}  = {R^{UV} \over \Lambda_{UV}^2}
\ee
where $\Lambda_{UV} \equiv  |j|^{1/(d-\Delta)}$ is the mass scale of the relevant coupling turned on in the UV.

This {\em dimensionless curvature} (\ref{UV0}) is the single  parameter which controls the theory and measures the curvature effects on the RG-flow. It also characterizes the solutions on the dual gravity side \cite{C}.

The UV fixed point and the dimension of the relevant operator are encoded in the potential $V(\varphi)$ in the UV region (to be identified below).
Since the purpose of the present article is to describe theories whose infrared dynamics are close to four-dimensional Yang-Mills (YM$_4$), a discussion of possible UV asymptotics are in order.
YM$_4$ is asymptotically free in the UV, and one can reproduce a similar asymptotically-free running of the coupling by using a dilaton potential, $V(\f)$, with leading  modified logarithmic-AdS asymptotic as has been described in \cite{IHQCD}.
 This mimics the RG-flow driven by the marginally relevant operator $Tr F^2$ with $\Delta = 4$.

Of course, an asymptotically free theory is not expected to have a holographic description in the UV, \cite{disect}. However, as argued in \cite{IHQCD,disect}, one may think of such a description as providing appropriate near-boundary conditions, that prepare the theory properly in the IR, where a holographic description may expected to be a better approximation\footnote{YM$_4$ at large $N_c$ is not expected to be a holographic theory in the IR. It was argued in \cite{disect} that the success of the SVZ approach indicated that it could come close to a holographic theory.}.

A different strategy  has been to ignore matching the UV  region and use a truly relevant perturbation in the UV, \cite{GN}. This is not crucial for the classification of the solutions and the IR dynamics, although it quantitatively affects the dynamics. Several studies over the years have indicated that the most important ingredient for the IR dynamics is the asymptotics of the dilaton potential $V(\f)$ as $\f\to \infty$, \cite{data}. We shall therefore use a simplified potential such that the operator driving the flow is a relevant one, and we shall take its conformal weight to be (in a general  QFT dimension)  $\Delta< d$.

In such a holographic theory, the UV is  realized around a point in field space (which we take without loss of generality to be $\varphi=0$), where the potential  has a local maximum, and an expansion of the form:
\be \label{UV1}
V(\varphi) = -{d(d-1) \over \ell^2} - {1\over 2 \ell^2 }\Delta(d-\Delta) \varphi^2 + {\cal O}(\varphi^3) \qquad \varphi \to 0
\ee
In this region, the metric  is asymptotically AdS  and $\varphi(u)$ has  scaling behavior to leading order:
\be \label{UV2}
ds^2 = du^2 + e^{-2u/\ell + 2c }\left[1 + {\cal O}\left(e^{2u/\ell}\right)\right] \zeta_{\mu\nu}dx^\mu dx^\nu , \quad \varphi(u) = \varphi_- \ell^{\Delta_-} e^{\Delta_- u} +  \ldots
\ee
where $c$ and $\varphi_-$ are, as it turns out, integration constants of Einstein's equations (the former dimensionless, the latter with mass dimension $(d-\Delta)$).  We have introduced here the standard AdS/CFT notation $\Delta_- \equiv d-\Delta$. The dots denote subleading orders whose size depends on $\Delta$ and which will be specified more carefully later on.

According to the  standard holographic dictionary, the physical UV parameters discussed earlier in this subsection are identified as:
\be \label{UV3}
R^{UV} \equiv e^{-2c} R^{(\zeta)} , \qquad j \equiv \varphi_-
\ee
where $ R^{(\zeta)}$ is the curvature scalar of the  (fiducial)  metric $\zeta_{\mu\nu}$.  The UV parameters are therefore  encoded in   the integration constants as they  appear in the  expansion of the solution in  the asymptotically  AdS region (\ref{UV2}).

The system of Einstein's equations  (\ref{rgflow1}-\ref{rgflow3}) is third order in derivatives.  It has therefore, three integration constants. Regularity of the solution fixes one combination of them, and we are left with the two parameters discussed above. Furthermore, only one combination of them corresponds to the truly physical dimensionless parameter ${\cal R}$ defined in (\ref{UV0}), which characterizes the solutions, and that we can determine in terms of the boundary data as:
\be \label{UV4}
{\cal R} = R^{UV} |\varphi_- |^{-2/\Delta_-}.
\ee

The first order formalism we introduce in the next section, is such that it allows us to keep track precisely of the dimensionless curvature parameter ${\cal R}$ and nothing else.

\subsection{First-order formalism}

The differential system of equations (\ref{rgflow1}-\ref{rgflow3}) is relatively complicated and to  study it, we first recast it into a simpler form. We exclude solutions that are fixed points of the RG flow, for which $\dot{\varphi} = 0$. Assuming this, the chain rule then gives:
\begin{equation}
    \dot{f} = f' \dot{\varphi}
    \label{chain}
\end{equation}
This allows us to transform (\ref{rgflow1}-\ref{rgflow3}) into a first order system by introducing a new set of variables:
\begin{align}
    W(\varphi) &\equiv -2(d-1)\dot{A}, \label{defw}\\
    S(\varphi) &\equiv \dot{\varphi},\label{defs}\\
    T(\varphi) &\equiv e^{-2A(u)}R^{(\zeta)}\label{defT},
\end{align}
where in the above equations it is assumed that the functions on the right hand side are considered as functions of $\varphi$, obtained by inverting $\varphi(u) \to u(\varphi)$. Therefore, the slice curvature $T$ and the ``superpotentials'' $W$, and $S$ are functions of $\varphi$ only. This procedure can be applied piecewise in any region where $\varphi(u)$ is monotonic.

The functions $W,S$ and $T$ have very interesting properties, that have been extensively studied in the general curvature case in \cite{C}. In particular, $T$ has the sign of the curvature of the boundary, and this sign is constant along flows.

Using the definitions (\ref{defw}-\ref{defT}) the system (\ref{rgflow1}-\ref{rgflow3}) becomes:
\begin{align}
    &-W' S + S^2 + \frac{2}{d} T = 0\label{eqn:solveT},\\
    & \frac{d}{4(d-1)}W^2 -\frac{1}{2} S^2 + V - T = 0 \label{eqn:algT},\\
    &S'S - \frac{d}{2(d-1)}W S - V' = 0,\label{eqn:WfunS}
\end{align}
Once these equations are solved, one can then find the functions $A(u),\varphi(u)$ by a direct  integration of equations (\ref{defw}-\ref{defs}). Notice that $T$ appears only algebraically in this system, so it is possible to eliminate one of the equations by solving for it in (\ref{eqn:solveT}):
\begin{equation}
    T(\varphi) = \frac{d}{2} S\left( W' - S\right) \label{eqn:T}.
\end{equation}
Re-injecting the value of $T$ in the previous system, we  obtain a two-dimensional dynamical system:
\begin{align}
    W' &= \frac{1}{2(d-1)} \frac{W^2}{S} + \frac{d-1}{d} S + \frac{2}{d}\frac{V}{S},\label{eqn:w} \\
    S' &= \frac{d}{2(d-1)} W + \frac{V'}{S}.\label{eqn:s}
\end{align}
This system is strongly nonlinear, and the right hand side depends explicitly on $\varphi$ through $V(\varphi)$. It is therefore non-autonomous. Its study for a generic potential is made challenging by both $V$ and $V'$ appearing as forcing for the system.

The advantage of using equations (\ref{eqn:w}-\ref{eqn:s}) over the original Einstein's equations is that, once regularity is imposed, the only integration constant entering the solution is the dimensionless curvature parameter (\ref{UV4}), as  it was shown in \cite{C}. This can be understood heuristically  from the fact that the  integration constant $\varphi_-$ in (\ref{UV3}) is  really only an integration constant of the flow equation $\dot{\varphi} =S$, therefore it cannot enter in the first order system (\ref{eqn:w}-\ref{eqn:s}). The first order system  (\ref{eqn:w}-\ref{eqn:s}) has two  integration constants. Fixing one by regularity, leaves a single independent integration constant, that can be identified with ${\cal R}$.

We expect, therefore, that solutions of the first order system come in {\em continuous one-parameter families,} parametrized by ${\cal R}$. This system has been studied extensively in literature in the past, both in the flat sliced case $T=0$, \cite{IHQCD,thermo,multirg}, the positively curved case, \cite{C,F,dS,Jani2}, and the negative-curved case, \cite{C,AdS1}, in the interior of the $\f$-space.

From now on,  we assume that in the limit $\varphi \to +\infty$  the potential takes the asymptotic form:
 \begin{equation}
     V \sim -V_{\infty}e^{2 b \varphi} \varphi^{\alpha}, \qquad {\mathbf \varphi \to +\infty}
     \label{potentialasymp}
 \end{equation}
with $b \geq 0 $ and $\alpha$ real. For this kind of potentials, the system  (\ref{eqn:w}-\ref{eqn:s})   needs to be studied near the boundaries of the scalar space, $\f=\pm\infty$. In the flat case, $T=0$, this has been done in \cite{IHQCD,thermo,multirg}. In the positively curved and negatively curved cases, for asymptotically exponential potentials\footnote{ie. as in (\ref{potentialasymp}) with $\a=0$.}, it has been studied in \cite{Jani,AdS2}. However, the techniques used in the past, were not powerful enough to address the most interesting case, with asymptotics (\ref{potentialasymp}) with $b=b_c$ (defined in (\ref{intro3})  and $\a\not=0$. The interest in this cases stems from the fact that such asymptotics are necessary
in order for the theory to not scale in the IR, \cite{IHQCD,GK}. The case $b=b_c, \a={1\over 2}$ has been argued in \cite{IHQCD} to approximately describe the low-energy dynamics of YM$_4$.

We are particularly interested in the question of whether a given asymptotic solution admits or not a continuous deformation, which according to the discussion above corresponds to moving between  solutions with different values of ${\cal R}$.  As we shall see, this deformation  appears at subleading order with respect to the leading asymptotics.

Before we turn to the analysis for general $\alpha$ and $b$, which we shall do  in the next two sections, we give a short  review  of what we know about the IR behavior (with  or without space-time curvature) of holographic theories with potential asymptotics given by (\ref{potentialasymp}) .

\subsection{ Confining IR asymptotics in flat space} \label{subsec:flat}

In  this subsection  we briefly review the properties of the flat-space holographic QFT model, i.e. when we set $\zeta_{\mu\nu} = \eta_{\mu\nu}$ (the Minkowski metric) in (\ref{metric}) and  $R^{(\zeta)} = 0$. In this case, the first order system  of the previous section  reduces to the single {\em superpotential} equation,
\be \label{flat1}
 {1\over 2} W^{'2} - {d\over 4(d-1)} W^2 = V
\ee
and we have $T=0$ and  $S = W'$. The solution (up to a trivial integration of the flow equations) is completely characterized by the function $W(\varphi)$, which is controlled by a single integration constant of the first order equation  (\ref{flat1}).

We shall now review the properties of the solutions in the large-$\varphi$ region, which maps to  the far infra-red regime of the theory,  and how these properties  determine the physics of the dual QFT in the IR  as a function of  the  parameters $\alpha, b$ controlling the asymptotic form of the potential in  equation (\ref{potentialasymp}).

There exist two critical values for the $b$ parameter, across which the behavior changes significantly:
\be \label{flat2}
b_G \equiv \sqrt{d\over 2(d-1)}, \qquad b_c \equiv \sqrt{1\over 2(d-1)}
\ee

All qualitative features are essentially determined by where $b$ is situated with respect to these critical values, and for all values of $b$ {\em except $b=b_c$} the value of $\alpha$ is largely irrelevant, except for small subleading corrections. In  the case  $b=b_c$ however the value of $\alpha$ plays a major role. We shall discuss this case last.

We first discuss the role of $b_G$, which we call (somewhat inappropriately\footnote{The name stems from the fact that only for $b<b_G$ solutions exist which satisfy Gubser's {\em criterion} concerning good IR singularities, \cite{IHQCD,thermo}.})  {\em Gubser's bound}.

\paragraph{Below Gubser's bound $0\leq  b < b_G$.}
 In  this case there exist two kinds of solutions, which are characterized by their behavior as $\varphi \to \infty$:
\begin{enumerate}
\item {\bf Generic} solutions come in a continuous family. The leading behavior of the  superpotential, in the large-$\varphi$ region, is
\be \label{flat3}
W(\varphi) \simeq C e^{b_G \varphi} \qquad \varphi \to +\infty
\ee
where $C$ is an integration constant. The leading exponential  is independent of the potential $V(\f)$.

\item A unique {\bf Special} solution (with no deformation parameters), whose leading asymptotics is:
\be \label{flat4}
W(\varphi) \simeq W_s e^{b \varphi} \varphi^{\alpha/2} \quad W_s \equiv \sqrt{2 V_{\infty}\over b_G^2 -b^2}, \qquad \varphi \to \infty .
\ee
Unlike the generic solution, in the special solution,  the leading coefficient is fixed and one can show that this solution has no free parameters at any subleading order.

Note that we are in the regime $b<b_G$, so the special solution's exponential behavior is less steep than any of the generic solutions.
\end{enumerate}
Both solutions reach a curvature singularity as $\varphi \to +\infty$. However, one can argue that the singularity in the special solution is an acceptable one in holography, whereas the one in the generic solution is not.
In fact, the generic solutions fail to satisfy some reasonable physical requirements. For one thing, the generic solutions fail Gubser's test \cite{Gubser}, i.e. they cannot be turned into regular small-mass black holes by a small deformation, as it was shown in \cite{thermo}. This requirement is satisfied by the special solution.
Another hint that the generic solutions are unphysical is that only the special solution can be obtained from a regular geometry by a process of (generalized) dimensional reduction on a sphere or a torus \cite{GK,GK1}. This indicates that the singularity can resolved by uplifting to higher dimensions.

From now on, we discard the generic solutions as ``bad'' holographic backgrounds and we only consider as acceptable solutions those with  the special asymptotics in (\ref{flat4}).

\paragraph{Above Gubser's bound $ b > b_G$.}

The first thing to notice in this regime is that the special solution  (\ref{flat4}) ceases to exist, because the superpotential becomes complex.

Moreover, one can show, \cite{thermo}, that for $b>b_G$  it is impossible for a solution (with flat slicing)  to reach an asymptotic regime $\varphi \to +\infty$: the solution will always ``bounce'' \cite{multirg} at a finite value of $\varphi$ and decrease again towards the origin of field space, and  beyond\footnote{what will happen eventually depends on the details of the potential as $\varphi \to -\infty$. In general the behavior is expected to be BKL-like, \cite{BKL}. A similar system was analyzed in the cosmological case in \cite{BKL2}.}.

Independent of the fate of the generic solution, the range $ b > b_G$ does not contain any asymptotic solution satisfying Gubser's criterion (the way it was stated in \cite{Gubser}) because only the special asymptotics  (\ref{flat4}) can be slightly deformed into a regular solution (a small black hole or, as we shall see later, putting the theory on a  sphere with a very small curvature). \\

From now on we  shall assume $b<b_G$. In this regime, one has to further differentiate between holographic theories which display confinement , and those who do not. The boundary between these two cases is set by $b=b_c$, defined in (\ref{flat2}):
\begin{itemize}
\item {\bf Confining regime $b_c < b < b_G$.}\\
In this regime the flat space theory has a gaped and discrete spectrum of  glueball-like excitations, whose masses $m_n$   scale at large excitation number  as:
\be \label{flat5}
m_n \simeq \Lambda\, n\sp n\gg 1
\ee
where $\Lambda$ is a characteristic IR scale of the theory, which  is linearly related to  the  scale $\Lambda_{UV}$ discussed in section 2.2 and defined in terms of the UV coupling  equation below equation (\ref{UV0}).

 The theory displays a first order confining/deconfining phase  transition at a finite temperature $T_c \sim \Lambda$, whose gravity dual manifestation is a Hawking-Page-like transition between black hole solutions and ``thermal gas'' solutions, \cite{thermo}.

{ With the extra assumption that $\varphi$ is a non-critical string theory dilaton, } the Wilson loop, $$\langle Tr[P~e^{i\oint_{\gamma} A}]\rangle=e^{W(\gamma)}$$ in the low-temperature phase satisfies an area law at large separation,
$$W(\gamma)~~~ \sim  ~~~ \Lambda^2 \times Area(\gamma)~~~\gg~~~ 1\;.$$
and a Coulomb-like law in the high-temperature phase.

 All these solutions at $T=0$  are scaling in the IR with hyperscaling violation, \cite{GK,hy2}.
They cannot therefore describe the IR physics of a theory like 4d Yang-Mills, \cite{disect}.

\item {\bf Non-confining regime $0\leq b < b_c$.}

In this regime, the flat space  theory has a continuous spectrum of  excitations  (dual to a continuous spectrum of  normalizable modes in the gravity dual) starting at $m=0$ \cite{IHQCD}. The Wilson loop, computed holographically, behaves as an inverse power-law at large separation; the theory does not display a Hawking-Page-like phase transition at finite temperature \cite{thermo}. All these features lead us to  call this range of values {\em non-confining regime}.

Note that this regime includes the case $b=0$, in which the potential goes to a constant as $\varphi=0$ and has the same qualitative behavior of theories with an  infrared conformal fixed point  (in which the potential has a minimum at a finite value of $\varphi$).

\item {\bf Critical case: $b =  b_c$}

In this case, whether the theory confines or not  (in the sense described above, i.e. discrete spectrum and Wilson line area law) depends on the value of $\alpha$  in (\ref{potentialasymp}):
\begin{itemize}
\item $\alpha >0 $: The potential belongs to the confining class, with all the features described above, except for a different behavior of the mass spectrum:
\be \label{flat6}
m_n = \left\{ \begin{array}{ll}\Lambda\, n^{\alpha} & \quad 0< \alpha< 1 \\
\Lambda \, n  & \quad \alpha \geq 1 \end{array} \right. ;
\ee
\item $\alpha < 0$: the potential belongs to the non-confining class (continuous spectrum, no Wilson loop area law) ;
\item  $\alpha = 0$: this is a ``hybrid'' situation, in which there is still an area law for the Wilson loop, but  the spectrum is continuous with a finite mass gap above zero, \cite{Umut}.
\end{itemize}
\end{itemize}
In all the cases above there is no scaling in the IR, and therefore such asymptotics are more suitable in describing YM in the IR, \cite{disect}.

The case of {\em Improved Holographic QCD} (IHQCD)  \cite{IHQCD}, which will be  the main focus of this work, falls in the critical, confining case for $d=4$:
\be \label{flat7}
\textrm{IHQCD:} \qquad b = b_c = {1\over \sqrt{6}} , \qquad \alpha = {1\over 2}
\ee
According to equation (\ref{flat6}), it is characterized by a Regge-like asymptotic spectrum of glueballs,
$$m_n \sim \Lambda \sqrt{n}\;.$$
  This is the phenomenologically interesting case, since one can construct a full potential $V(\varphi)$ with these asymptotics that can reproduce at a quantitative level many properties of four-dimensional Yang-Mills theory, \cite{data,Panero}. It constitutes  one of the building blocks of phenomenologically realistic holographic theories which include flavor, such as V-QCD \cite{VQCD}.
In particular, it is in this case that the finite temperature behavior near the deconfinement phase transition contains the well known $T^2$ term, that has been observed in lattice calculation of YM$_4$, \cite{Umut}.

\subsection{IR structure of curved-space RG-flows for non-critical exponential asymptotics} \label{subsec:noncrit}

When a non-zero curvature is turned on, there is one more continuous parameter on which the solutions depend: it is the dimensionless curvature (\ref{UV4}). This is a  parameter of the UV theory, but it must also appear encoded in the IR behavior of the solution. Accordingly, at finite curvature, the asymptotic solutions should admit a continuous integration constant which  encodes the dimensionless curvature.

We are particularly interested in identifying this continuous deformation parameter in the asymptotic regime. A complete classification of the IR solutions for positive curvature was obtained in \cite{Jani}  in the non-critical case  $b\neq b_c$,  and $\alpha =0$. An  analogous classification for negative curvature was discussed in \cite{AdS2}. In the rest of this section,  we review these results, while in the rest of the paper we focus on the critical case $b=b_c$.

\subsubsection{Solutions extending to infinity: Type I and II singular  endpoints}

We first  discuss the case of solutions which extend to $\varphi \to \infty$. In the presence of curvature, for $b<b_G$ (which we henceforth always assume), we always find the {\em generic} singular solution with asymptotics (\ref{flat3}). Following \cite{Jani}, we call this class  {\bf Type 0} solutions. As before, such solutions are holographically unacceptable, for the same reasons as in the flat case. We shall not discuss them further.

There are also solutions with special asymptotics, which extend the special asymptotics (\ref{flat4}) to finite curvature. These are singular, in the same way as the asymptotics (\ref{flat4}), but unlike the Type 0 asymptotics,  these singularities are holographically acceptable. To leading order,  they behave as
\be \label{exp1}
S \simeq S_0 e^{b\varphi}, \quad W \simeq W_0 e^{b\varphi}, \quad T \simeq T_0 e^{2b\varphi}\sp \varphi \to \infty\;,
\ee
where $S_0, W_0$ and $T_0$ are constants which can take two possible  {\em fixed} values: with the same nomenclature introduced  in \cite{AdS2}, these are:
\begin{itemize}
\item {\bf Type I} asymptotics
\be \label{TypeI}
S_0 = \sqrt{2 V_{\infty}\over d-1}, \quad W_0 = b\sqrt{8(d-1)V_{\infty}}, \quad
T_0 = 2d(b^2 - b_c^2) V_{\infty}
\ee
\item {\bf Type II} asymptotics:
\be \label{TypeII}
S_0 = b \sqrt{2V_{\infty}\over b_G^2 - b^2}, \quad W_0 =\sqrt{2V_{\infty}\over b_G^2 - b^2}, \quad
 T_0 = 0
\ee
\end{itemize}
Notice that at $b=b_c$ the two sets of coefficients coincide, as one can  observe  using (\ref{flat2}). This will be important later.

In the case of type I asymptotic solutions, the sign of the curvature (which is the same as the sign of  $T_0$, by equation (\ref{defT})) is determined by $b$: type I  solutions have  positive curvature for $b > b_c$ and negative curvature for $b<b_c$.  As was shown in \cite{AdS2}, these solutions admit a single continuous integration constant {\em only} in the case of negative curvature. This integration constant  appears at subleading order in the expansion around $\varphi = +\infty$, which is of the form:
\be \label{exp2}
S \simeq  S_0 e^{b\varphi} + S_1 e^{\lambda_I \varphi}+\cdots
\ee
where $S_1$ is an arbitrary constant  which controls the value of the curvature parameter ${\cal R}$,  and  similar expressions hold for $W$ and $T$, which are completely determined in terms of $S_1$. The exponent of the second term in (\ref{exp2}) is given by
\be \label{exp3}
\lambda_I = {1\over 2} \left[b(d+1) - \sqrt{b^2(d-9)(d-1) + 4}\right].
\ee
The expression (\ref{exp2}) is an asymptotic solution only if the second term is small compared to the first one.  This is the case,  if and only if, $b < b_c$. In the opposite case, $b>b_c$, one must instead set $S_1=0$.   One therefore concludes that, for $b<b_c$ (i.e. in the non-confining regime)  type I solutions exist for generic negative curvatures, but in the confining regime $b>b_c$ type I solutions exist for a single specific value of the curvature parameter ${\cal R} = {\cal R}_c$ \cite{Jani}.

For type II asymptotics, the situation is opposite. First, note that the sign of the curvature is not specified to leading order as  $T=0$. If we calculate to the next order, we find a similar behavior as in  (\ref{exp2}),
\be \label{exp4}
S \simeq  S_0 e^{b\varphi} + S_1 e^{\lambda_{II} \varphi}+\cdots
\ee
with  $S_0$  given in (\ref{TypeII}) , where $S_1$ is again an arbitrary constant and the subleading exponent is now:
\be \label{exp5}
\lambda_{II} = b + 2{b_c^2 - b^2\over b}
\ee
One can show that the sign of $T$ is the same as the sign of $S_1$. As $S_1$ is an arbitrary real number,   these solutions exist for both signs of the curvature. However, the term controlled  by $S_1$  in (\ref{exp4}) is actually subleading  only   for $b > b_c$, i.e. in the confining regime. In this regime,  it captures  the curvature deformation  of the special asymptotics (\ref{flat4}) (in the case $\alpha=0$),  as it appears  in the IR.

As shown in \cite{Jani}, this classification of the solution, and how their existence is related   to the sign of the boundary curvature, hold no matter what is the subleading behavior of the potential beyond the leading exponential (\ref{potentialasymp}).  The discussion above provides the physical interpretation of the transcritical bifurcation that will be discussed in section \ref{lanal}.

In the non-confining regime on the other hand, we have to set $S_1=0$ not to spoil the leading asymptotics in (\ref{exp4}), and this corresponds to setting the slice curvature to zero. So the only type II solutions in the non-confining case, are the ones with flat slices presented in subsection \ref{subsec:flat}.

\subsubsection{Solutions ending at finite $\varphi$:  Type III regular endpoints}

For {\em positive} curvature slices, and for any value of $b$ and $\alpha$,   there also exist completely regular  solutions with an endpoint at a coordinate $u_0$  where $\varphi$ stops at a finite value, $\f_0$, and the slice manifold shrinks to zero regularly\footnote{More precisely, these are regular endpoints in the Euclidean signature. In the Lorentzian signature the endpoint corresponds to a horizon.}:
\be \label{exp6}
\varphi (u) \to \varphi_0,  \qquad ds^2  \to du^2 + (u-u_0)^2 d\Omega_d^2
\ee
We call these solutions {\bf Type III}.  The free IR parameter is now given by $\varphi_0$, which controls the curvature parameter ${\cal R}$. In particular,  $\varphi_0$ approaching a maximum of the potential corresponds to large UV curvature, as one can show using the near-boundary asymptotic expansion of the solution \cite{C}. What happens in the limit $\varphi_0 \to +\infty$ on the other hand, depends on the value of $b$.  In the  non-confining regime, $b<b_c$, in the limit $\varphi_0\to \infty$, we obtain ${\cal R} \to 0$ and the solution reduces to the flat  solution. In the  confining regime, $b>b_c$,  instead, in the limit {$\varphi_0\to +\infty$}, ${\cal R}$ asymptotes to the special value ${\cal R}_c>0$ corresponding to the unique  Type I solution.

In contrast, for  {\em negative curvature} one cannot have a type-III endpoint \cite{AdS2}: the solution either reaches infinity as type I or type II,  or it displays a  {\em bounces} in the scale factor and flows to another AdS boundary\footnote{There are also solutions connecting to IR-like regions of the type I or II, but it is unclear whether these have a holographic interpretation \cite{AdS2}.} (which can be the same extremum of the potential where the flow started, or a different one).

\subsubsection{Positive curvature holographic RG-flow zoology}

The structure of the space of solutions for positive curvature can be summarized as follows (the interested reader is referred to   \cite{Jani} for details):
\begin{itemize}
\item In the {\bf non-confining regime,  $b<b_c$}  the only solutions which exist for ${\cal R} >0$  are the regular ones, Type III. The IR parameter controlling the curvature, is the position $\varphi_0$ of the endpoint in field space. {$\f_0$ near a maximum of the potential  and large $\varphi_0$ correspond to the large and small curvature limit, respectively}.
\item In the {\bf confining regime, $b>b_c$}, there exist two continuous families of solutions for ${\cal R}>0$: Type II (which reach infinity in field space)  and Type III (which stop at a finite value of $\varphi$, $\f_0$). These solutions merge, and both approach the Type I solution at the critical curvature ${\cal R}_c$. This limit is obtained in taking  the integration constant of each solution to infinity ($\varphi_0$ in the case of Type III, $S_1$ in the case of Type II).

As shown in \cite{Jani}, in the confining regime there is always a phase transition between type III solutions (high curvature) and type II solutions (low curvature). Depending on the value of $b$, the trasition may be first order or higher. In the latter case the transition occurs exactly at ${\cal R} = {\cal R}_c$ where ${\cal R}_c$ was defined in \cite{Jani}.

The result  that the Type II solution approaches the Type I in the limit  $S_1\to \infty$ can be understood in terms of generalized dimensional reduction of a higher-dimensional  theory on an internal sphere \cite{Jani}. Here we  will re-derive it  in terms of the critical points of a dynamical system.
\end{itemize}

\subsubsection{The case of Improved Holographic Yang-Mills}

The  classification above does not apply to  the most phenomenologically interesting case of $b=b_c$ and in particular for the IHQCD value $\alpha = 1/2$. The reason is that precisely at $b=b_c$ the two types of asymptotics  I and II become degenerate, as one can see by comparing the coefficients of the leading terms in (\ref{TypeI}) and (\ref{TypeII}).

 Another indication  that  $b=b_c$ is special is that for positive curvature, it is the boundary  between existence and non-existence of solutions with special asymptotics  reaching $\varphi \to \infty$. Therefore, what happens at this point requires further analysis.

As it turns out, translating the problem in the language of dynamical systems, the special point $b=b_c$ corresponds to a bifurcation. At this point, the equations governing the deformation of the system become non-linear.  Furthermore, we shall see that a second  bifurcation appears {\em precisely} at the IHQCD value $\alpha =  1/2$.

Describing   the asymptotic solutions in the critical case $b=b_c$ (whether they exist or not, and how their IR integration constant corresponding to the curvature arises) will be the goal of the rest of this work.

\section{IR Exponential potentials and Dynamical Systems\label{ssub}}
As a warm-up  to illustrate how we can apply dynamical system theory in the present setting, we first consider the  case of a potential which matches  the asymptotic form (\ref{potentialasymp}) for   $\alpha=0$,
\begin{equation}
    V(\varphi) = -V_{\infty} e^{2 b\varphi},\label{asympVexp}
\end{equation}
We shall analyze asymptotic solutions extending to $\varphi \to +\infty$, rederive some of the results reviewed in section \ref{subsec:noncrit} for $b\neq b_c$, and derive the asymptotics for the critical case $b=b_c$. The results we find are still valid in the case of a full potential where (\ref{asympVexp}) is only valid asymptotically, with any additive (exponentially smaller) terms or any power of $\varphi$ multiplying the leading exponential in the case $b \neq b_c$.

In practice, all the equalities in the following subsections will turn into approximate equalities which give the leading IR behavior of a potential that deviates from (\ref{asympVexp}) at finite and small $\varphi$.


\subsection{Reduction to an autonomous system}

The main observation to connect to dynamical system theory is that we can express the equations of motion in terms of an autonomous dynamical system. In the asymptotic region where  (\ref{asympVexp}) holds, there is a proportionality relation  between $V$ and its derivative:
\begin{equation}
    V' = 2b V,
    \label{asympder}
\end{equation}
so the equations (\ref{eqn:w}-\ref{eqn:s}) can be rewritten in terms of $V$ only. The  following  change of variables then eliminates $V$ from the equations,  
 \begin{align}
     W &= \sqrt{V_{\infty}}\Tilde{W}e^{b\varphi} & S &= \sqrt{V_{\infty}}\Tilde{S}e^{b\varphi}\label{eqn:cov}
 \end{align}
and equations (\ref{eqn:w}-\ref{eqn:s}) take the form of an autonomous dynamical system:
\begin{align}
    \Tilde{W}' &= -b\Tilde{W} + \frac{1}{2(d-1)} \frac{\Tilde{W}^2}{\Tilde{S}} + \frac{d-1}{d} \Tilde{S} - \frac{2}{d\Tilde{S}},\label{eqn:Wtilde} \\
    \Tilde{S}' &= -b\Tilde{S} + \frac{d}{2(d-1)} \Tilde{W} - \frac{2b}{\Tilde{S}}\label{eqn:Stilde}.
\end{align}
This modified system appears to be more complicated than (\ref{eqn:w}-\ref{eqn:s}), but we can now rely on the powerful tools of autonomous dynamical system theory to study it. Solutions in this formalism are still equivalent to solutions to the equations of motion in the initial theory. Indeed, given solutions $\Tilde{W}$ and $\Tilde{S}$, one can invert the change of variables (\ref{eqn:cov}) to obtain the values of $W$ and $S$, and from there one recovers the expression for $T$ using (\ref{eqn:T}). Using the definitions of $W$, $S$ and $T$ one can finally obtain $A$ and $\varphi$ from equations (\ref{defw}-\ref{defs}).

\subsection{Null isoclines analysis}

To analyse this autonomous system, we look for null isoclines, $\Tilde{W}'=0$ and $\Tilde{S}'=0$, given by the two algebraic equations:
\begin{align}
    \frac{1}{2(d-1)} \Tilde{W}^2 -b \Tilde{S} \Tilde{W}  + \frac{d-1}{d} \Tilde{S}^2 - \frac{2}{d} &= 0,\label{nullclineW}\\
    -b\Tilde{S}^2 + \frac{d}{2(d-1)} \Tilde{W} \Tilde{S} - 2b &= 0.\label{nullclineS}
\end{align}
The intersections of these two curves are   critical points of the dynamical system. We  obtain, generically, two such points, associated to Type I and II solutions. The critical  points are given by\footnote{There is a trivial  sign ambiguiity in these solutions that correspond to reversing the direction of the flow $u\to -u$. We fix it so that $S>0$.} :
\begin{align}
    \text{Type I:} & & \Tilde{S}_I &= \sqrt{\frac{2}{d-1}} & &\Tilde{W}_I = 2b\sqrt{2(d-1)}
         \label{1}
         \\
    \text{Type II:} & & \Tilde{S}_{II}
     &= b \sqrt{\frac{2}{b_G^2 -b^2}} & &\Tilde{W}_{II} = \sqrt{\frac{2}{b_G^2 -b^2}}
     \label{2}
\end{align}
The  critical points correspond to the asymptotic solutions found previously for (\ref{TypeI}) and (\ref{TypeII}). Notice that there are two critical points for $b<b_G$, but only one for $b>b_G$, as the functions $S$ and $W$ must be real-valued.  We emphasize that the fixed points of the dynamical systems are {\em not} fixed points of the RG flows, but instead they are IR asymptotic behaviors of the RG flow. In this formalism, the RG fixed points are on the line $S = 0$.

\smallbreak

 To probe {the range of curvature for which each of} these solutions exists, we calculate $T$ (using (\ref{eqn:T})) at the respective critical points:
\begin{align}
    &\text{Type I:} &T_{I} &= 2d V_{\infty}e^{2b\varphi}(b^2 - b_c^2), \label{T1}\\
    &\text{Type II:} &T_{II} &= 0.\label{T2}
\end{align}
When $T$ is zero at the critical point, subleading corrections can a priori give it any sign. This implies that amongst the considered solutions,  type I only exists in the positive curvature case if $b_c \leqslant b < b_G$, while type II is a priori acceptable for $0 < b < b_G$.
\smallbreak
\begin{figure}[htp]
\centering
\begin{subfigure}[t]{0.45\textwidth}
\includegraphics[width=\textwidth]{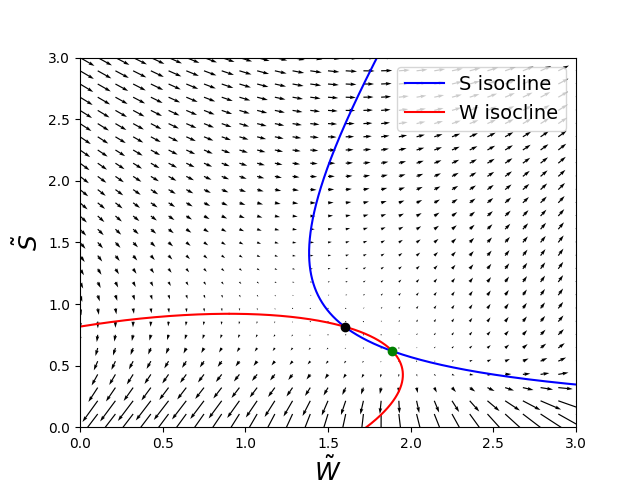}
\caption{$b<b_c$, close to $b_c$}
\end{subfigure}

\begin{subfigure}[t]{0.45\textwidth}
    \includegraphics[width=\textwidth]{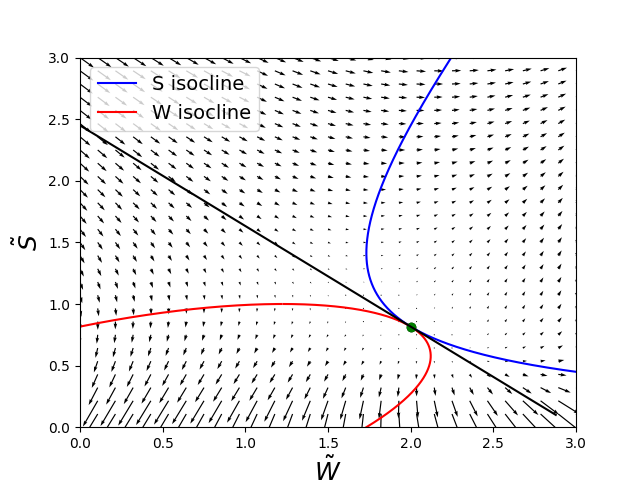}
    \caption{$b=b_c$}
\end{subfigure}
\begin{subfigure}[t]{0.45\textwidth}
\includegraphics[width=\textwidth]{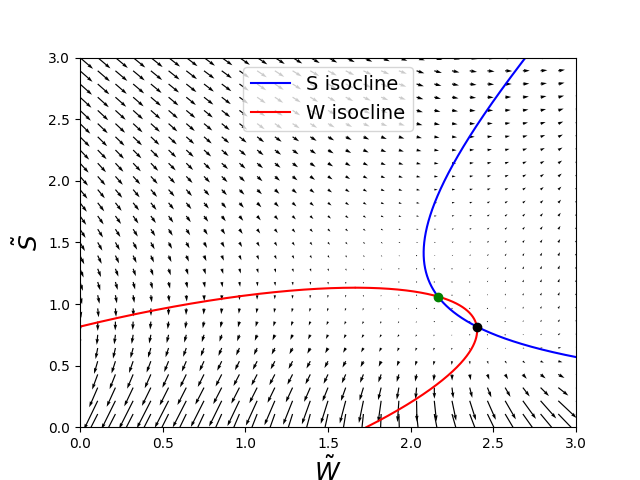}
\caption{$b>b_c$, close to $b_c$}
\end{subfigure}

\caption{Phase plane of the system in dimension $d=4$, with unit $V_{\infty}$, for $b<b_c$ (a), $b=b_c$ (b) and $b>b_c$ (c). The Type II critical point is indicated in green, and the Type I critical point is indicated in black. The blue curves correspond to $\Tilde{S}'=0$. The red curves correspond to $\Tilde{W}'=0$. The black line in the case $b = b_c$ is the common tangent to the two null isoclines.}
\label{fig:phase}
\end{figure}

The phase plane $(\tilde{W},\tilde{S})$ of the dynamical system is shown in figure \ref{fig:phase} for different values of $b$. The blue lines are the isoclines correesponding to $\tilde S'=0$. The red lines are the isoclines corresponding to $\tilde W'=0$. They intersect at the critical points. Notice that the critical points become degenerate for $b=b_c$.

The next step is to determine the stability of these critical points. In the stable case, there are orbits that end on the fixed point. Any initial condition on these orbits corresponds to a trajectory that ends on the critical point. These trajectories are holographic RG flows that share first order asymptotic with the critical solution.
Therefore, these solutions come in continuous families.

In the unstable case, the trajectories move away from the critical point, and the critical solution is isolated. Therefore, it is necessary to study the stability of the critical points, in order  to understand which holographic RG flow solutions come with a deformation parameter. The physical meaning of the continuous deformation from the point of view of the dual field theory is that different solutions in the continuous family correspond to different values of the {\em physical} curvature, $\mathcal{R}$, of the boundary theory.

The null isoclines $\Tilde{S}'=0$ and $\Tilde{W}'=0$ delimit regions of the phase plane in which the sign of $\Tilde{W}'$ and $\Tilde{S}'$ is constant. Therefore, examining the signs in each region of the phase plane suffices to determine the stability of critical points, assuming eigenvalues of the system all have non-zero real part at the critical point. This type of critical point is called hyperbolic.

The coordinates of the critical points in (\ref{1}), (\ref{2}) show that Type I and Type II solutions merge into one at $b=b_c$. The signs of $\Tilde{S}'$ and $\Tilde{W}'$ and the different regions  delimited by the red and blue curves in figure \ref{fig:phase} suggest an inversion of stability of the critical points at $b = b_c$. This bifurcation would imply that at $b = b_c$, the critical point is non-hyperbolic, i.e. one of its eigenvalues has zero real part. Consequently, we need to perform direct computations to conclude on the stability of the critical points.

\subsection{Linear analysis\label{lanal}}

To study the stability of fixed points, we must linearize
 the system of equations around the fixed-point solutions by setting
\be
\Tilde{W} =\Tilde{W}_{I,II}+\delta \Tilde{W}_{I,II}
\sp
\Tilde{S}=\Tilde{S}_{I,II}+\delta \Tilde{S}_{I,II}
\label{defdelta}
\ee
and expand the dynamical system equations in powers of $\delta \Tilde{W}$ and $\delta \Tilde{S}$:
\begin{equation}
    \begin{pmatrix}
        \delta \Tilde{W}_I'\\
        \delta \Tilde{S}_I'
    \end{pmatrix} = \begin{pmatrix}
        b & - \frac{b^2}{b_c^2} + \frac{2(d-1)}{d}\\
        \frac{d}{2(d-1)} & (d-2)b
    \end{pmatrix}     \begin{pmatrix}
        \delta \Tilde{W}_I\\
        \delta \Tilde{S}_I
    \end{pmatrix} + h.o.t..
    \label{firstorderI}
\end{equation}

\begin{equation}
    \begin{pmatrix}
        \delta \Tilde{W}_{II}'\\
        \delta \Tilde{S}_{II}'
    \end{pmatrix} = \begin{pmatrix}
        - b + \frac{1}{b(d-1)}& \frac{d-2}{d} \\
        \frac{d}{2(d-1)} & \left(d\left(\frac{b_c^2}{b^2}\right)-2\right)b
    \end{pmatrix}     \begin{pmatrix}
        \delta \Tilde{W}_{II}\\
        \delta \Tilde{S}_{II}
    \end{pmatrix} + h.o.t.,
    \label{firstorderII}
\end{equation}
where $h.o.t$ contains every higher order contributions in $\delta \Tilde{S}$ and $\delta \Tilde{W}$. We diagonalize these matrices and obtain their eigenvalues $\lambda_{x, I/II}$ and $\lambda_{y, I/II}$ at the associated critical point:
\begin{align}
    \text{Type I:} & & \lambda_{x, I} &= \frac{1}{2} \left[b (d-1)-\sqrt{4 + b^2 (d-1)(d-9)} \right]\nonumber\\
    & &  \lambda_{y, I} &= \frac{1}{2} \left[b (d-1)+\sqrt{4 + b^2 (d-1)(d-9)} \right]\label{eigenBI}\\
    \text{Type II:} & & \lambda_{x, II} &= - 2b + \frac{1}{b(d-1)} \nonumber \\
    & &  \lambda_{y, II} &= - b + \frac{d}{2(d-1) b} \label{eigenBII}
\end{align}
We recover the Efimov bound, \cite{Aharony2, s2s2}, above which eigenvalues of type I solutions become complex:
\begin{equation}
    b_E = \sqrt{\frac{4}{(d-1)(9-d)}}.
    \label{befimov}
\end{equation}

We also obtain that for $b \neq b_c$, both critical points are hyperbolic fixed points. For $b=b_c$ instead, the single  critical point is non-hyperbolic as expected: one of the eigenvalue vanishes (namely $\lambda_{x,II}$ in (\ref{eigenBII})), the other is positive ($\lambda_{y,II}$).

For $b\neq b_c$, the Hartman-Grobman theorem applies \cite{dyna,dynaproof}. It states that in a neighborhood of a critical point, the flow of the dynamical system is homeomorphic  to the flow of the linearized dynamical system. This means that the solution will behave locally like the solution of the linearized system. Moreover, in the linear case all the solutions are completely classified. For positive eigenvalues, this implies instability, and for negative eigenvalues this implies stability. As a consequence, continuous families of RG flows that end on a hyperbolic critical point in the IR, exist if and only if the critical point has at least one eigenvalue with a strictly negative real part.

\smallbreak

From the eigenvalues, we conclude that I and II are respectively a saddle point (I) and fully unstable (II) for $b<b_c$, merge at $b=b_c$, then separate and switch stability into a fully unstable point (I) and a saddle point (II) for $b>b_c$. Consequently, the system undergoes a transcritical bifurcation for $b=b_c$.

By the previous discussion, this also implies that, for $b>b_c$, type II solutions come in a one-parameter  continuous family, since it is possible to turn on the deformation in the stable direction without spoiling the leading large-$\varphi$ asymptotics. On the other hand, type I solutions are isolated and do not admit a continuous deformation. We obtain directly from (\ref{eigenBII}) that in the stable direction:

\begin{equation}
    \delta \Tilde{S}_{II}' = \left( -2b + \frac{1}{b(d-1)}\right) \delta \Tilde{S}_{II} + \mathcal{O}(\delta \Tilde{S}_{II}^2),
    \label{odeSII}
\end{equation}
which yields the following continuous family of asymptotic solutions for $S$, parametrized by an integration constant $C$:
\begin{equation}
    S_{II}(\varphi) = \sqrt{\frac{2\left(\frac{b}{b_G}\right)^2}{1-\left(\frac{b}{b_G}\right)^2}} e^{b\varphi}  + Ce^{(-b + \frac{1}{(d-1)b})\varphi} + o(e^{(-b + \frac{1}{(d-1)b})\varphi})
    \label{SII}
\end{equation}
Using equation (\ref{eqn:T}), we obtain at second order for $T_{II}$ in the limit $\varphi \to \infty$:
\begin{equation}
    T_{II} (\varphi) \sim - \left(1 + (d-2)\frac{b_c^2}{b^2} \right) \sqrt{\frac{2\left(\frac{b}{b_G}\right)^2}{1-\left(\frac{b}{b_G}\right)^2}} ~C~ e^{\frac{1}{(d-1)b}\varphi}.
    \label{TII}
\end{equation}

Consequently, Type II solutions do not only exist in flat space as equation (\ref{T1}) seemed to suggest.  As the sign of $T$ depends on $C$, this implies that type II solutions can exist for both negative and positive curvature for $b>b_c$.

Finally, for $b> b_G$,  there is no continuous family of RG flows that end on a critical point in the IR,  as the only critical point remaining, namely type I, is fully unstable in that range. Hence the type I solution is a special solution for $b>b_G$.

\subsection{The critical case $b=b_c$}

In this subsection we shall study the critical case, $b=b_c$.
As already mentioned, in this case there is a single, degenerate critical point.

In the following, we denote by an index $*$ the values at this unique critical point.  The critical point is given by $(\Tilde{W}_*, \Tilde{S}_*) = (2, \sqrt{\frac{2}{d-1}})$. In addition, the two isoclines are tangent at the critical point, and the tangent is given by:
\begin{equation}
    \frac{d \Tilde{W}}{d\Tilde{S}}_* = \frac{-(d-2)\sqrt{2(d-1)}}{d}.
    \label{eqn:tan}
\end{equation}

\subsubsection{Linear analysis and center manifold theory\label{center}}

The perturbation equations (\ref{firstorderI}-\ref{firstorderII}) both reduce,  at the critical value $b=b_c$,  to:
\begin{equation}
    \begin{pmatrix}
        \delta \Tilde{W}'\\
        \delta \Tilde{S}'
    \end{pmatrix} = \begin{pmatrix}
        \frac{1}{\sqrt{2(d-1)}} & \frac{d-2}{d}\\
        \frac{d}{2(d-1)} & \frac{d-2}{\sqrt{2(d-1)}}
    \end{pmatrix}     \begin{pmatrix}
        \delta \Tilde{W}\\
        \delta \Tilde{S}
    \end{pmatrix} + h.o.t..
    \label{firstordercrit}
\end{equation}

The eigenvalues of this matrix are $\lambda_x = 0$ and $\lambda_y = \sqrt{\frac{d-1}{2}}$. Therefore, this critical point is indeed non-hyperbolic. To analyse non-hyperbolic fixed points, we shall use center manifold theory \cite{dyna}.

We denote by  $E_{c}$, $E_u$ and $E_s$ the vector spaces corresponding respectively to solutions having zero, positive and negative real part eigenvalues at the critical point.

The center manifold theorem implies that there exist manifolds $W_c$, $W_u$ and $W_s$ that are tangent to the corresponding vector spaces and invariant under the flow of the dynamical system\footnote{For generic systems  $W_u$ and $W_s$ are unique but $W_c$ may  fail to be unique. In our case however, $E_s$ and $W_s$ are trivial, and this implies that $W_c$ is unique.}.
We collect in Appendix \ref{sec:D} more details.
More rigorous statements and proofs of the theorem used, can be found in Chapter 3.2 of \cite{dyna} and \cite{dynaproof}.

Intuitively, $W_c$ corresponds to the set of trajectories in phase space that do not decay or grow exponentially away from the critical point. Our dynamical system has only positive or null eigenvalues at the critical point, therefore its stable manifold $W_s$ is trivial. Hence, the continuous families of solutions we are looking for are necessarily included in $W_c$. We can then project the dynamical system on $W_c$, in order to reduce the degree of the differential system and make it simpler to solve.

Since the center manifold $W_c$ in the critical case is one-dimensional, it corresponds exactly to a curve in phase space that contains the critical point. Calculating the eigenvector corresponding to the null eigenvalue, we recover that $W_c$ is tangent to null isoclines.

\subsubsection{Flow lines of the vector field in phase space}

In view of the previous discussion, the objective is now to obtain the behavior of $S$ on the trajectories that approach the critical point in the null direction. To achieve this goal, we need a higher order expansion of the space $W_c$, as the linear approximation is not enough.

The autonomous system considered is two-dimensional, therefore the expansion can be obtained\footnote{We are assuming sufficient regularity of the equations, which is the case here.} using chain rule and our equations (\ref{eqn:Wtilde}) and (\ref{eqn:Stilde}):
\begin{equation}
    \frac{d \Tilde{W}}{d\Tilde{S}} = \frac{\frac{d \Tilde{W}}{d\varphi}}{\frac{d \Tilde{S}}{d\varphi}}
\end{equation}
The full trajectory equation obtained is :
\begin{equation}
    \frac{d \Tilde{W}}{d\Tilde{S}}\left( -b_c \Tilde{S} + \frac{d}{2(d-1)} \Tilde{W} - \frac{2b_c}{\Tilde{S}} \right) =  -b_c \Tilde{W} + \frac{1}{2(d-1)} \frac{\Tilde{W}^2}{\Tilde{S}} + \frac{d-1}{d} \Tilde{S} - \frac{2}{d\Tilde{S}} \label{eqn:traj}
\end{equation}
The solution to this equation gives the  curve $W_c$,  once we impose that it passes through the critical point and it is tangent to the null isocline there. To find an approximate solution close to the critical point,
we  expand  equation (\ref{eqn:traj})  in powers of $\delta \Tilde{S}$, and we taylor-expand $\Tilde{W}$ and its first derivative, around the fixed point:
\begin{align}
    \Tilde{W}(\delta \Tilde{S}) &= \Tilde{W}_* + \left(\frac{d\Tilde{W}}{d\Tilde{S}}\right)_* \delta \Tilde{S} + \frac{1}{2} \left(\frac{d^2\Tilde{W}}{d\Tilde{S}^2}\right)_* \delta \Tilde{S}^2 + \frac{1}{6}\left(\frac{d^3\Tilde{W}}{d\Tilde{S}^3}\right)_* \delta \Tilde{S}^3 + O(\delta \Tilde{S}^4),\label{CM1D}\\
    \frac{d\Tilde{W}}{d\Tilde{S}}(\delta \Tilde{S}) &= \left(\frac{d\Tilde{W}}{d\Tilde{S}}\right)_* + \left(\frac{d^2\Tilde{W}}{d\Tilde{S}^2}\right)_* \delta \Tilde{S} + \frac{1}{2}\left(\frac{d^3\Tilde{W}}{d\Tilde{S}^3}\right)_* \delta \Tilde{S}^2 + O(\delta \Tilde{S}^3).\label{CM1Dder}
\end{align}
We substitute the expressions (\ref{CM1D}) and (\ref{CM1Dder}) in equation (\ref{eqn:traj}) and regroup terms order by order in $\delta \Tilde{S}$. We then solve the equation order by order for the values of the derivatives of $\Tilde{W}(\tilde S)$.

There is no contribution to order zero in $\delta \Tilde{S}$ as we sit at the critical point. For the first order terms to cancel, we obtain two solutions for $ \left(\frac{dW}{dS}\right)_* $, one of them being given by equation (\ref{eqn:tan}). The other one is unstable and corresponds to the flat case. It is described in appendix \ref{sec:A}. We discard it from our analysis  as we already know that the tangent direction is the only one that can produce stable solutions. We therefore use the stable solution. Cancellations of the second and third order terms in $\delta \Tilde{S}$ require,
\begin{align}
    \left(\frac{d^2\Tilde{W}}{d\Tilde{S}^2}\right)_* &= \frac{2(d-1)^2(d-4)}{d^2}, & \left(\frac{d^3\Tilde{W}}{d\Tilde{S}^3}\right)_*&= -3\sqrt{2} \frac{(d-1)^{\frac{5}{2}} (d^2 - 16) }{d^3}. \label{eqn:derWS}
\end{align}
In the special  case $d=4$, all the subsequent derivatives vanish, and the center manifold conicides with the vector space $E_c$:  for d=4, the center manifold is a straight line in phase space. In this case  an exact solution for the trajectory can be found, as shown  in appendix \ref{sec:C}. From  now on we discuss the case of general $d$.

\subsubsection[Asymptotics of S on the trajectory]{Asymptotics of $S$ on the trajectory}
\label{sec343}

To calculate the asymptotics of $S$ on this trajectory, we substitute (\ref{CM1D}) into equation (\ref{eqn:Stilde}) using  (\ref{eqn:tan}) and (\ref{eqn:derWS}). We obtain a differential equation for perturbations up to third order\footnote{The leading term in this equation is insufficient for our purpose as shown in Appendix \ref{sec:B}, therefore we also include the third order term.}:
\begin{equation}
    \delta \Tilde{S}' = -\frac{2(d-1)}{d} \delta \Tilde{S}^2 + \frac{4(d-1) \sqrt{2(d-1)}}{d^2} \delta \Tilde{S}^3 + \mathcal{O}(\delta \Tilde{S}^4).\label{eqn:deltasp}
\end{equation}
Projecting the differential equation on the null trajectory has reduced the degree of the differential system by one. The first order term in $\delta \Tilde{S}$ vanishes as a consequence of projecting on the null eigenvalue direction. Equation (\ref{eqn:deltasp}) is an autonomous differential equation of first order  and is therefore separable. We integrate it to obtain:
\begin{equation}
        \sqrt{\frac{2}{d-1}} \log\left(\left| -2\frac{d-1}{d \delta \Tilde{S}(\varphi)} + \frac{4 (d-1) \sqrt{2(d-1)}}{d^2}\right|\right) + \frac{d}{2(d-1) \delta \Tilde{S}(\varphi)} = \varphi -\f_{\infty}- C_0
        \label{eqn:exactexplog}
\end{equation}
where we have neglected the higher order terms.
$\f_{\infty}$ is the starting point of the evolution of $\delta \Tilde{S}$ which evolves in the interval, $\f\in [\f_{\infty},+\infty)$. Although it is redudant with the integration constant $C_0$, we keep it as a placeholder and as a reminder that the solution must be intended as valid in the large-$\varphi$ limit.

The sign of the integration constant\footnote{We shall consistenly denote the IR integration constants of our type I/II solutions by  $C_{\a}$ where $\a$ denotes the subleading exponent in the asymptotics of the scalar potential $V(\f)$ in (\ref{potentialasymp}). In our case here $\a=0$.  }, $C_0$,
determines if {the flow is stable or unstable. In the case $C_0<0$, it is stable,} $\f$ grows to $+\infty$ and $\delta \Tilde{S}$ goes to $0$. {However, if $C_0>0$ the flow is unstable and $\delta \Tilde{S}$ goes to $+\infty$ for a finite value of $\f$, and does not reach $\f \to \infty$.}

As we have seen, there is stability if and only if $C_0<0$. This type of null eigenvalue direction is {called} semi-stable, {in the sense that the behavior of the flow depends on the initial condition (encoded in $C_0$) in this direction. This phenomenon can be seen} from the phase plane in figure \ref{fig:phase}. Indeed, in the null eigenvalue direction, the arrows point in the same direction on both sides of the critical point.

The left hand side of (\ref{eqn:exactexplog}) can be inverted explicitly, which is done in Appendix \ref{sec:B}. We obtain the asymptotics of $\delta \Tilde{S}$, which can be done either directly from (\ref{eqn:exactexplog}) or expanding the analytical expression of Appendix \ref{sec:B}:
\begin{equation}
    \delta \Tilde{S}(\varphi) = \frac{d}{2(d-1)} \frac{1}{\varphi} + \frac{d}{(d-1)\sqrt{2(d-1)}} \frac{\log\varphi}{\varphi^2} +\left( \frac{d}{2(d-1)}\right)^2 \frac{C_0+\f_{\infty}}{\varphi^2} + o(\varphi^{-2}). \label{asympexpS}
\end{equation}
The integration constant $C_0$ appears at subleading order, allowing us to identify explicitly the deformation of the solution coming from semi-stability. The reason why the higher-order corrections $\sim (\log\f)/\f^2$ are leading compared to the integration constant term is the vanishing of the first order term in $\delta \Tilde{S}$. This is a characteristic of bifurcations with real eigenvalues.
\smallbreak
We now obtain $\delta \Tilde{W}$ by replacing $\delta \Tilde{S}$ in its Taylor expansion in (\ref{CM1D}),
\begin{equation}
     \delta \Tilde{W} = \frac{-(d-2)}{\sqrt{2(d-1)} \varphi} - \frac{d-2}{d-1} \frac{\log \varphi}{\varphi^2} - \frac{d(d-2)}{2(d-1)\sqrt{2(d-1)}} \frac{C_0+\f_{\infty}}{\varphi^2} + \frac{d-4}{4\varphi^2}  + o(\varphi^{-2}).
     \label{asympexpW}
\end{equation}
Finally, we compute $T$ using equation (\ref{eqn:T}):
\begin{align}
    T = &\frac{d}{\sqrt{2(d-1)}} e^{\sqrt{\frac{2}{d-1}} \varphi} \left( -\frac{1}{\varphi} - \sqrt{\frac{2}{d-1}} \frac{\log \varphi}{\varphi^2} \right. \\
    &- \left. \frac{1}{\varphi^2} \left(\frac{d (C_0 +\f_{\infty})}{(d-1)\sqrt{2(d-1)}} -\frac{3(d-4)}{4(d-1)} \right) + o(\varphi^{-2})\right)
    \label{asympexpT}
\end{align}
We find that $T$ has a negative leading-order contribution, regardless of the sign $C_0$. This implies that solutions of type I/II do not exist in the critical case $b = b_c$ with positive slice curvature. As a result, only non-singular type III solutions exist in this case for positively curved slices.

\section{Solving IHQCD in the IR} \label{sec:ihqcd}

We now want to refine the  results of the previous section and search for a similar characterisation of the possible solutions with a potential relevant for IHQCD, which includes a power-law multiplying the exponential asymptotics.

We therefore move on to consider a potential which matches the general IR asymptotics (\ref{potentialasymp}):
\begin{equation}
    V(\varphi) = -V_{\infty} e^{2 b_c \varphi} \varphi^\alpha. \label{asympV}
\end{equation}
The  leading exponential is chosen to be critical\footnote{In the non-critical  case $b \neq b_c$, the extra power law does not affect any of the qualitative properties  of the solutions we found in the previous section.}.

As in the previous sections, all the equalities below will turn into approximate equalities which give the leading IR behavior of a potential that deviates from (\ref{asympV}) at finite and small $\varphi$.

\subsection{Reduction to an autonomous system and new null isocline}

For the potential (\ref{asympV}) we have
\begin{equation}
    V' = \left(2b_c + \frac{\alpha}{\varphi}\right)V.
    \label{potentialasympIHQCD}
\end{equation}
This potential poses a new difficulty, as the equations do not depend only on $V(\varphi)$, but also on $\varphi$ explicitly. Similarly to equation (\ref{eqn:cov}) we  change of variables to:
\begin{align}
    W &= \sqrt{V_{\infty}}\Tilde{W}e^{b_c\varphi}\varphi^{\alpha/2} & S &= \sqrt{V_{\infty}}\Tilde{S}e^{b_c\varphi}\varphi^{\alpha/2}\label{eqn:cov3d}
\end{align}
We obtain that we can again remove $V(\varphi)$ from the equations (\ref{eqn:w}-\ref{eqn:s}), but we are now left with explicit $\varphi$ dependence.

To obtain an autonomous system, we introduce a new auxiliary dynamical variable in the system:
\begin{equation}
    Z(\varphi)\equiv-\frac{\alpha}{\varphi}.
    \label{defZ}
\end{equation}
We consider the asymptotic $\varphi \to +\infty$ behavior of the dynamical system. This implies that the sign of $Z$ is fixed and that the phase space has a boundary $Z=0$.

Introducing the auxiliary variable $Z(\varphi)$ restores the autonomous nature of the system at the cost of having to handle an additional dimension in phase space. However $Z$  is not a free dynamical variable because it is constrained to satisfy  (\ref{defZ}). As a result, the manifold on which the dynamics takes place is still two-dimensional.

Reexpressing the system (\ref{eqn:w}) and (\ref{eqn:s}) in terms of $Z$ and with this potential, we obtain:
\begin{align}
    \Tilde{W}' &= -b_c \Tilde{W} + \frac{1}{2(d-1)} \frac{\Tilde{W}^2}{\Tilde{S}} + \frac{d-1}{d} \Tilde{S} - \frac{2}{d\Tilde{S}} + \frac{Z\Tilde{W}}{2},\label{eqn:Wtilde2} \\
    \Tilde{S}' &= -b_c \Tilde{S} + \frac{d}{2(d-1)} \Tilde{W} - \frac{2b_c}{\Tilde{S}} + Z\left(\frac{\Tilde{S}}{2} + \frac{1}{\Tilde{S}}\right),\label{eqn:Stilde2}\\
    Z' &=  \frac{1}{\alpha} Z^2. \label{eqn:z2}
\end{align}

The extra integration constant associated to $Z$  is fixed by imposing equation (\ref{defZ}), which acts as a constraint.   The new null isocline ($Z'=0$)  is the curve $Z = 0$, allowing us to identify the previous exponential potential case as the plane at the boundary of the phase space. In particular, critical points of the new system have $Z_* = 0$. The two equations above (\ref{eqn:Wtilde2}) and (\ref{eqn:Stilde2}), when we set $Z=0$ become our old equations (\ref{eqn:Wtilde}) and (\ref{eqn:Stilde}).
Therefore the new critical points are the same as the ones covered in the previous section\footnote{This argument holds also for $b\neq b_c$}, and trajectories that reach them in the IR end on the plane as $\varphi \to +\infty$.

{More generally, we can make the system autonomous for any potential, and without restricting to the asymptotic region, by adding an auxiliary variable $Z(\varphi)$ and making it a three dimensional dynamical system.}

\subsection{Linear analysis}

To begin with, we expand the right hand side of the system (\ref{eqn:Wtilde2})- (\ref{eqn:z2}) and calculate the perturbation equation at the unique critical point:
\begin{equation}
    \begin{pmatrix}
        \delta \Tilde{W}'\\
        \delta \Tilde{S}'\\
        \delta Z'
    \end{pmatrix} = \begin{pmatrix}
        \frac{1}{\sqrt{2(d-1)}} & \frac{d-2}{d} & 1\\
        \frac{d}{2(d-1)} & \frac{d-2}{\sqrt{2(d-1)}} & \frac{d}{\sqrt{2(d-1)}}\\
        0 & 0 & 0
    \end{pmatrix}     \begin{pmatrix}
        \delta \Tilde{W}\\
        \delta \Tilde{S}\\
        \delta Z
    \end{pmatrix} + h.o.t.,
    \label{firstorderIHQCD}
\end{equation}
where we have set $\delta Z \equiv Z$ due to $Z_* = 0$. The main feature of this new system is that the critical point now has a degenerate zero eigenvalue. The corresponding two-dimensional eigenspace  is found by diagonalizing the matrix appearing in (\ref{firstorderIHQCD}). We define two new coordinates, using eigenvectors that span this eigenspace:
\begin{align}
    X_1 &\equiv -\frac{d}{\sqrt{2(d-1)^3}} \delta \Tilde{W} + \frac{\delta \Tilde{S} - d \delta Z}{d-1}\label{defx1}\\
    X_2 &\equiv \delta Z, \label{defx2}
\end{align}
There is of course a freedom in the definition of $X_1$ and $X_2$. We keep $X_2 = \delta Z$ for simplicity. The coefficient of $\delta Z$  in the definition of $X_1$ can be modified without changing the solution for $\delta \Tilde{S}$. A third coordinate, associated with the positive eigenvalue, is defined by:
\begin{equation}
    Y \equiv \frac{d}{\sqrt{2(d-1)^3}} \delta \Tilde{W} + \frac{d-2}{d-1} \delta \Tilde{S} + \frac{d}{d-1} \delta \Tilde{Z} \label{defy}
\end{equation}
One can then decompose the full system written in terms of these variables isolating the linear contribution:
\begin{equation}
    \begin{pmatrix}
        X_1'\\
        X_2'\\
        Y'
    \end{pmatrix} = \begin{pmatrix}
        0 & 0 & 0\\
        0 & 0 & 0\\
        0 & 0 & \sqrt{\frac{d-1}{2}}
    \end{pmatrix}     \begin{pmatrix}
        X_1\\
        X_2\\
        Y
    \end{pmatrix} + \begin{pmatrix}
        f_1(X_1, X_2, Y)\\
        f_2(X_1, X_2, Y)\\
        g(X_1, X_2, Y)
    \end{pmatrix}
    \label{strucxyz}
\end{equation}
where $f_1$, $f_2$ and $g$ do not contribute to linear order. These functions are obtained by taking the linear combinations of equations (\ref{eqn:Wtilde2}-\ref{eqn:z2}) indicated in (\ref{defx1}-\ref{defy}) and subtracting the linear contribution. To second order in $X_1$, $X_2$ and $Y$, $f_1$, $f_2$ and $g$ are given by:
\be
    f_1(X_1, X_2, Y) = -\frac{d-1}{2d} (2 X_1 + Y)^2  - \frac{1}{2}(4X_1 + 3Y)X_2 - \frac{d X_2^2}{\alpha (d-1)} + \mathcal{O}(X_1^3,X_2^3,Y^3)
    \ee
    \be
    f_2(X_1, X_2, Y) = - \frac{d-1}{2d} ((d-4)X_1^2 + 2(d-2)X_1 Y + (d-1)Y^2) -
    \ee
     $$
     - \frac{1}{2}((d-6)X_1 + (d-5)Y)X_2 + \frac{2d X_2^2}{\alpha (d-1)}  +\mathcal{O}(X_1^3,X_2^3,Y^3)
     $$
     \be
    g(X_1, X_2, Y) = \frac{X_2^2}{\alpha}
    \label{f1f2g}
\ee

\subsection{The center manifold}
\label{sec:4.3}

Like in the purely exponential case, the first order is not sufficient to obtain the right asymptotics of $\delta \Tilde{S}$, so we use again a higher order expansion of the center manifold. Due to the presence of the third variable, $Z$, the approach we used in section 3.4.2 needs to be modified. The strategy is the following.
\begin{enumerate}
    \item We  define the vector spaces $E_{c/u/s}$ of the linearized solutions (center, unstable and stable solutions respectively), and their non-linear counterparts $W_{c/u/s}$, as  they were defined in section \ref{center}.

    \item We derive the second order approximation $W_c^{(2)}$ to the center manifold by expanding to second order the function describing $W_c$.

    \item Starting with the non-linear definition of $W_c$, we pullback the dynamical system on $W_c^{(2)}$ by expanding the solution to second order around the fixed point.

    \item We explicitly solve the approximate ODE obtained.
\end{enumerate}
Step 1. to 3. will allow us to reduce the non-linear system by systematically excluding the unstable direction, which reduces the number of equations of the problem.
Step 4. then provides a separable differential equation after making use of (\ref{defZ}) and (\ref{defx2}). This is the higher-dimensional analogue of what was done in the previous section, where we reduced the problem from degree two to one by first finding the curve $W_c$ and calculating $\delta \Tilde{S}$ on it.

{{\bf Step 1} was already achieved partially in the previous subsection.} Given that our system has a one-dimensional $E_u$ (with coordinate $Y$) and a two-dimensional $E_c$ (with coordinates $(X_1, X_2)$), $W_c$ is now a surface of dimension two. Moreover, at linear order, it is tangent to $E_c$, ($E_s$ and $W_s$ are as before trivial), justifying that one-parameter solutions can, as before, only exist on $W_c$.

We now proceed with {\bf step 2}. By center manifold theory \cite{dyna, dynaproof}, there exists a smooth function $h(X_1, X_2)$ such that $W_c$ is described by the surface $Y = h(X_1, X_2)$. More details about center manifold theory are collected in appendix \ref{sec:D}. {We shall now derive a partial differential equation whose solution is  $h$.}

On one hand, using the chain rule, we obtain
\begin{equation}
    Y' = \frac{\partial h}{\partial X_1} X_1' + \frac{\partial h}{\partial X_2} X_2' \label{chainruleCM}
\end{equation}
On the other hand, the differential equations (\ref{strucxyz}), reduced on the center manifold $Y = h(X_1, X_2)$ take the form,
\begin{align}
    \begin{pmatrix}
        X_1'\\
        X_2'
    \end{pmatrix} &= \begin{pmatrix}
        f_1(X_1, X_2, h(X_1, X_2))\\
        f_2(X_1, X_2, h(X_1, X_2))
    \end{pmatrix}\label{eqn:CMx}\\
        Y' &= \sqrt{\frac{d-1}{2}} h(X_1, X_2) + g(X_1, X_2, h(X_1, X_2)).\label{eqn:CMy}
\end{align}
Combining these expressions, we obtain a non-linear partial differential equation of first order for $h(X_1, X_2)$:
\begin{equation}
   \frac{\partial h}{\partial X_1} f_1(X_1, X_2, h) + \frac{\partial h}{\partial X_2} f_2(X_1, X_2, h) = \sqrt{\frac{d-1}{2}} h + g(X_1, X_2, h),
    \label{eqn:manifold}
\end{equation}
along with the boundary conditions at the fixed point  $(X_1,X_2)=(0,0)$:
\begin{align}
    h(0, 0) &= 0 & \frac{\partial h}{\partial X_1}(0,0) &= 0 & \frac{\partial h}{\partial X_2}(0,0) &= 0. \label{CMboundary}
\end{align}
These  conditions impose that $W_c$ contains the fixed point and it is tangent, at the fixed point,  to the null-eigenspace $E_c$ spanned by $X_1$ and $X_2$.

With the  conditions in (\ref{CMboundary}), $h(X_1,X_2)$ has a general expansion around the fixed point which starts at quadratic order:
\begin{equation}
    h(X_1, X_2) = \frac{1}{2}\frac{\partial^2 h}{\partial X_1^2} X_1^2 + \frac{\partial^2 h}{\partial X_1 \partial X_2} X_1 X_2 + \frac{1}{2}\frac{\partial^2 h}{\partial X_2^2} X_2^2 + \mathcal{O}\Big(\textrm{cubic in } X_1,X_2 \Big)
    \label{CMexpansion}
\end{equation}
This ansatz is general under the assumption that $X_1$ and $X_2$ are of the same order in $\varphi$, which will be verified a posteriori. Expanding (\ref{eqn:manifold}) to second order in $X_1$ and $X_2$ fixes the value of the second derivatives in $h$:
\begin{align}
    \frac{\partial^2 h}{\partial X_1^2} &= \frac{d-4}{d} \sqrt{2(d-1)}\label{d2hdx12}\\
    \frac{\partial^2 h}{\partial X_1 \partial X_2} &= \frac{d-6}{d\sqrt{2(d-1)}}\label{d2hdx1dx2}\\
    \frac{\partial^2 h}{\partial X_2^2} &= \frac{-d}{\alpha} \left(\frac{2}{d-1}\right)^{\frac{3}{2}}.
    \label{d2hdx22}
\end{align}

{This is the conclusion of step 2, and the second order (non-linear) approximation of the space $W_c^{(2)}$ is generated by  the second partial derivatives of $ h(X_1, X_2)$.}

To perform {\bf step 3} and pullback the dynamical system on $W_c^{(2)}$, we replace $h(X_1, X_2)$ in equation (\ref{eqn:CMx}) by its quadratic  approximation, (\ref{CMexpansion}) and use the expressions for $f$ and $g$ from equations (\ref{f1f2g}). This results in  two non-linear ordinary differential equations,
\begin{align}
    X_1' &= \frac{-2(d-1)}{d} X_1^2 - 2 X_1 X_2 - \frac{d }{\alpha(d-1)} X_2^2 + \mathcal{O}(\varphi^{-3})\label{eqnx1p}\\
    X_2' &= \frac{X_2^2}{\alpha}\label{eqnx2p}
\end{align}

Moreover, the solution for $X_2$ is already known from (\ref{defZ})
, therefore we need to impose that
\be
X_2 = \delta Z =  -\frac{\alpha}{\varphi}\;,
\label{xx2}\ee reducing the system to a single non-linear ordinary differential equation,
\begin{equation}
    X_1' = \frac{-2(d-1)}{d} X_1^2 + \frac{2 \alpha }{\varphi} X_1 - \frac{d \alpha}{(d-1)\varphi^2} + \mathcal{O}(\varphi^{-3})
    \label{eqnx1}
\end{equation}
We must solve this equation to second order in $\frac{1}{\varphi}$, in order to construct the center manifold and obtain the information we are  seeking. We do so in the next subsection.

\subsection{Asymptotic solution on the center manifold}

So far we have completed the first three steps announced at the beginning of the previous subsection.
We are now proceeding with {\bf step 4} which involves solving equation (\ref{eqnx1}) which is a
 Riccati equation.  While Riccati  equations are generically not solvable by quadrature, this one can be reduced to a separable form by first setting,
\begin{equation}
    X_1(\varphi) = \frac{d}{2(d-1) \varphi} U(\varphi),
    \label{ansatzx1}
\end{equation}
which gives the following equation for $U$:
\begin{equation}
    \frac{U'}{U^2 - (2\alpha + 1)U + 2\alpha} = -\frac{1}{\varphi} + \mathcal{O}(\varphi^{-2}) \label{mainU}.
\end{equation}
This equation can be integrated with elementary methods (see Appendix \ref{sec:F}), giving the general solution:
\be \label{solU-f}
\log\left(\left| {U-U_1 \over U-U_2}\right|\right) = (1-2\alpha)\log \varphi + c
\ee
where $c$ is an integration constant and $U_{1,2}$ are the two roots of the polynomial $U^2 - (2\alpha + 1)U + 2\alpha$, namely:
\begin{align}
    U_1 &= 2\alpha, & U_2 = 1.
    \label{solu1u2}
\end{align}
These two roots correspond to two special constant solutions $U(\varphi) = U1, U(\varphi) = U2$ which, via equation (\ref{ansatzx1}), give two particular solutions of equation (\ref{eqnx1}):
\be \label{solx1x2}
X_1 = \frac{\alpha d}{(d-1)\varphi}, \qquad X_1 = \frac{d}{2(d-1)\varphi}.
\ee

 Notice that the two roots (\ref{solu1u2})  become degenerate for the IHQCD value $\alpha = \frac{1}{2}$. In this case the expression (\ref{solU-f})  ceases to be valid and the solution  takes a different form. This case will be dealt with separately in section \ref{subsubsec:ihqcdbif}.

\subsubsection{General power-law correction $\alpha$}

\begin{figure}[H]
    \centering
    \begin{subfigure}[t]{0.9\textwidth}
        \begin{tikzpicture}
        \node (img) {\includegraphics[width=\textwidth]{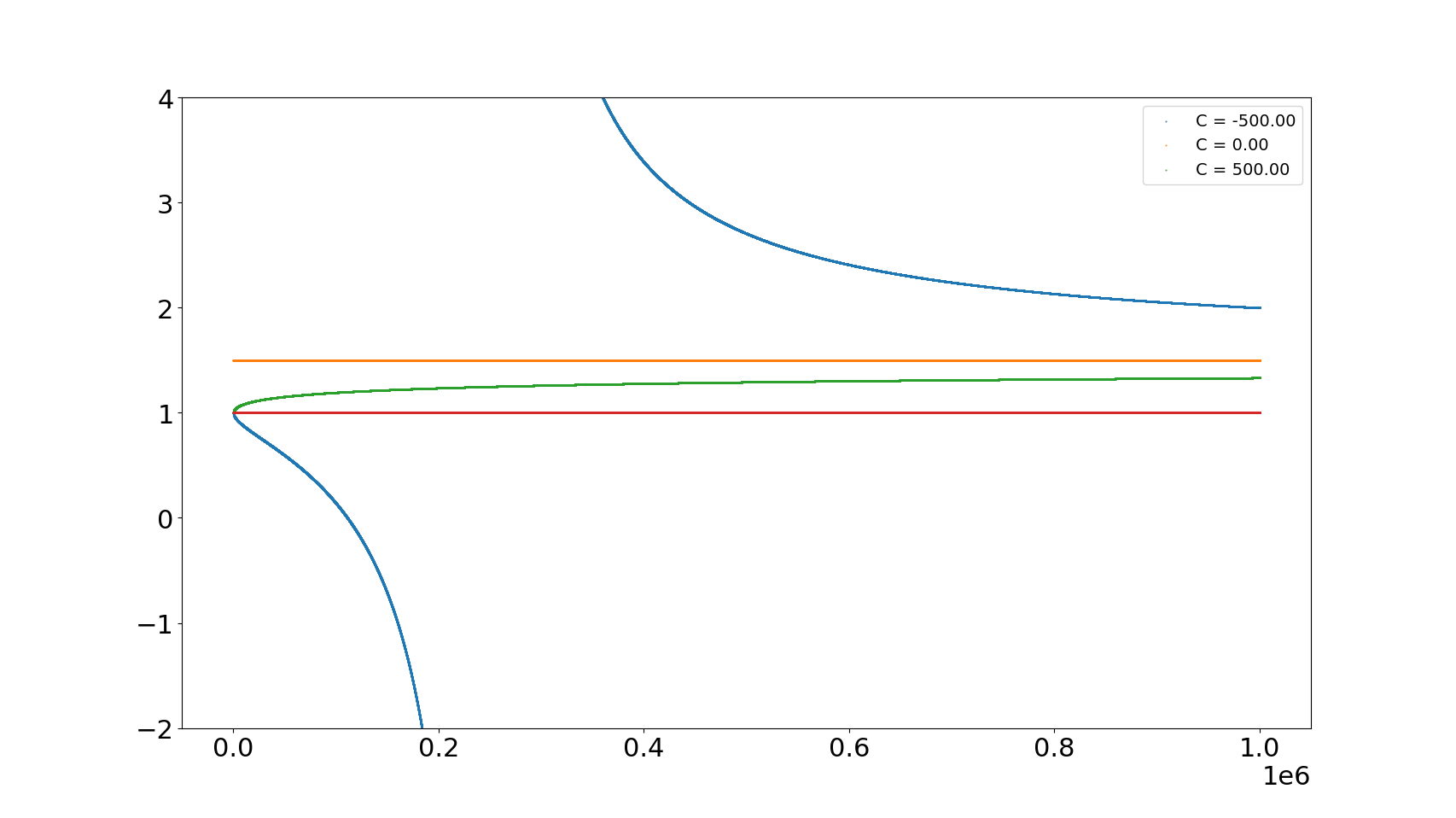}};
        \node[below=of img, node distance=0cm, xshift=3cm, yshift=1.2cm]{$\varphi$};
        \node[left=of img, node distance=0cm, rotate=0, anchor=center,xshift=1.5cm, yshift=2.5cm] {$U$};
        \end{tikzpicture}
    \end{subfigure}
    \caption{$U$ as a function of $\varphi$ at $\alpha=\frac{3}{4}$, for $C_{3\over 4} = -500$ (blue), $C_{3\over 4}=0$ (extremal solution corresponding to $U_1$, yellow), and $C_{3\over 4}=500$ (green). The extremal solution, $C_{3\over 4}\to \pm\infty$, corresponding to $U_2$ is plotted in red.}
    \label{fig:Uvarphi34}
\end{figure}

We first consider a generic asymptotic power-law correction with exponent $\alpha\neq 1/2$. In this case the roots $U_1$ and $U_2$ are distinct, and after some algebra reported in appendix \ref{sec:F}, we obtain:
\begin{equation}
    U(\varphi) = \frac{2\alpha + C_{\a} \varphi^{1-2\alpha}}{1 + C_{\a} \varphi^{1-2\alpha}} \times (1+ \mathcal{O}(\varphi^{-1}))
    \label{soluconst},
\end{equation}
where $C_{\a}$ is a real integration constant.
$U_1$ and $U_2$ are ``extremal points" for this solution, in the sense that taking the limit $C_{\a}\to \pm \infty$ gives $U_2$, and taking the limit $C_{\a} \to 0$ gives $U_1$. The graph of $U$ as a function of $\varphi$ for several values of $C_{\a}$ is shown in figure \ref{fig:Uvarphi14} for $\alpha = \frac{1}{4}$ and in figure \ref{fig:Uvarphi34} for $\alpha = \frac{3}{4}$. As $\alpha$ goes through $\frac{1}{2}$ the extremal solutions are interchanged:

\begin{figure}[H]
\centering
\begin{subfigure}[t]{0.9\textwidth}
\begin{tikzpicture}
\node (img) {\includegraphics[width=\textwidth]{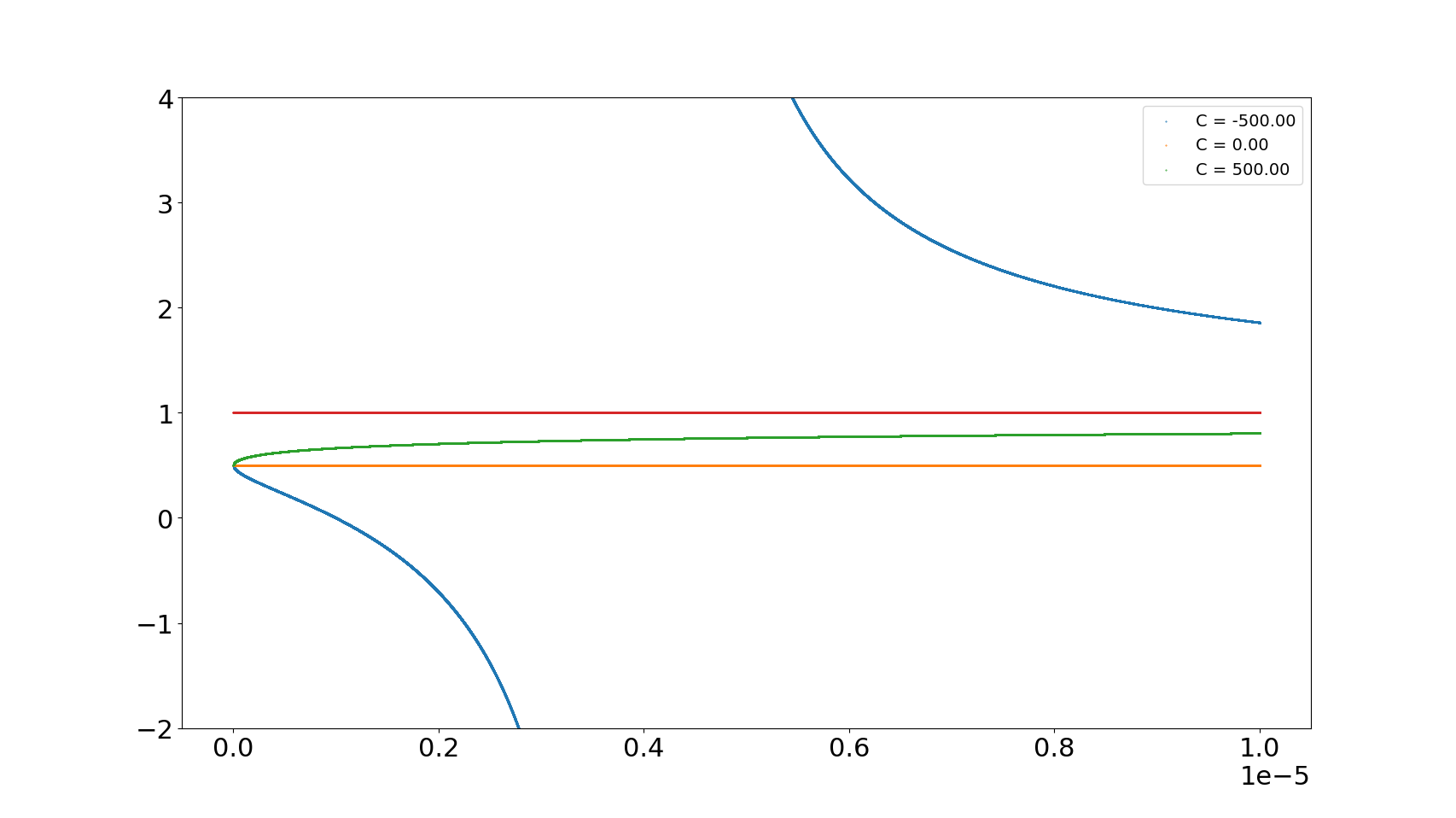}};
\node[below=of img, node distance=0cm, xshift=3cm, yshift=1.2cm]{$\varphi$};
\node[left=of img, node distance=0cm, rotate=0, anchor=center,xshift=1.5cm, yshift=2.5cm] {$U$};
\end{tikzpicture}
\end{subfigure}
\caption{$U$ as a function of $\varphi$ at $\alpha=\frac{1}{4}$, for $C_{1\over 4} = -500$ (blue), $C_{1\over 4}=0$ (extremal solution corresponding to $U_1$, yellow), and $C_{1\over 4}=500$ (green). The extremal solution, $C_{1\over 4}\to \pm\infty$, corresponding to $U_2$ is plotted in red.}
\label{fig:Uvarphi14}
\end{figure}

From $U$ in (\ref{soluconst}) we construct $X_1$ from (\ref{ansatzx1}) and then together with $X_2$ in  (\ref{xx2}) and the expansion of $h(X_1,X_2)$ in (\ref{CMexpansion}) we reconstruct $\delta \Tilde{S}, \delta \Tilde{W}, \delta Z$, and finally obtain $S$, $W$ and $T$:
\begin{equation}
    S(\varphi) =e^{\frac{\f}{\sqrt{2(d-1)}}} \f^{\frac{\alpha}{2}} \left( \sqrt{\frac{2}{d-1}} + \frac{d}{2(d-1) \varphi} \frac{2\alpha +C_{\a} \varphi^{1-2\alpha}}{1 + C_{\a}\varphi^{1-2\alpha}} + \mathcal{O}(\varphi^{-2}) \right), \label{exactX}
\end{equation}
\begin{equation}
    W(\varphi) = e^{\frac{\f}{\sqrt{2(d-1)}}}\f^{\frac{\alpha}{2}} \left(  2 + \frac{1}{\sqrt{2(d-1)} \varphi} \frac{2\alpha +C_{\a} \varphi^{1-2\alpha} (2\alpha (d-1) - (d-2))}{1 + C_{\a}\varphi^{1-2\alpha}} + \mathcal{O}(\varphi^{-2}) \right), \label{exactW}
\end{equation}
and from (\ref{eqn:T}) we obtain:
\begin{equation}
    T(\f) = \frac{d}{\sqrt{2(d-1)}} e^{\sqrt{\frac{2}{d-1}} \varphi} \varphi^{\alpha - 1}\left(\frac{C_{\a} (2\alpha - 1) \f^{1-2\alpha}}{1+ C_{\a}\varphi^{1-2\alpha}} + \mathcal{O}(\varphi^{-1}) \right). \label{asT}
\end{equation}

Note, that we obtain these solutions by integrating the system for large $\varphi$, so the expressions (\ref{exactX}) and (\ref{asT}) are to be interpreted as valid when expanded in powers of $\varphi^{-1}$. In particular, we obtain two different branches with different asymptotics if $\alpha$ is smaller or larger than $\frac{1}{2}$, depending on which of the constant term or the $\varphi^{1-2\alpha}$ term is dominant in the fraction.

\vskip 0.5cm

\underline{Branch  $\alpha < \frac{1}{2}$}:

\vskip 0.5cm

After expanding $S$, $W$ and $T$, we obtain:

$$
    S_-(\varphi) =e^{\frac{\f}{\sqrt{2(d-1)}}} \f^{\frac{\alpha}{2}} \left( \sqrt{\frac{2}{d-1}} + \frac{d}{2(d-1) \varphi} \left( 1 +  \frac{2\alpha-1}{C_{\a}}\varphi^{2\alpha - 1} +\right.\right.
$$
\be
\left.\left.+\mathcal{O}\left(\varphi^{4\alpha - 2}\right) \phantom{\frac{2\alpha-1}{C_{\a}}}\hskip -1.3cm\right)  + \phantom{\sqrt{\frac{2}{d-1}}}\hskip -1.5cm \mathcal{O}\left(\varphi^{-2}\right) \right),
 \label{asympsdown}
\ee
$$
    W_-(\varphi) = e^{\frac{\f}{\sqrt{2(d-1)}}}\f^{\frac{\alpha}{2}} \left(2 + \frac{1}{\sqrt{2(d-1)} \varphi} \left( 2\alpha (d-1) - (d-2)  + \right. \right.
$$
\be
 \left. \left. +\frac{(d-2)(1-2\alpha)\f^{2\alpha- 1}}{C_{\a}}  + \mathcal{O}\left(\varphi^{4\alpha - 2}\right)  \right)  + \mathcal{O}\left(\varphi^{-2}\right) \right), \label{Wdown}
\ee
\begin{equation}
    T_-(\f) = \frac{d}{\sqrt{2(d-1)}} e^{\sqrt{\frac{2}{d-1}} \varphi} \varphi^{\alpha - 1} \left(2\alpha - 1 + \frac{1-2\alpha}{C_{\a}} \varphi^{2\alpha-1} + \mathcal{O}(\varphi^{4\alpha - 2}) + \mathcal{O}(\varphi^{-1}) \right).\label{TCminus}
\end{equation}
The sign of $T$ is negative, regardless of the value of $C_{\a}$.

We observe that the expansion is valid for $\alpha > 0$. For $\alpha \leqslant 0$, the order of the term $\varphi^{2\alpha - 1}$ is higher than the $\mathcal{O}(\varphi^{-1})$, so the expansion would need to be pushed to higher order to draw any conclusion
\footnote{More precisely, the order of the expansion that we need to   reach, depends on the value of $\alpha$. For $ -\frac{n}{2} < \alpha <\frac{-n+1}{2}$, an expansion to order $\mathcal{O}(\frac{1}{\varphi^{n+1}})$ of $\delta \Tilde{S}$ is needed, for which we need to take into account all terms $\frac{X_1^{m+2 - k}}{\varphi^k}$ in the differential equation, for $k$ integer running from $0$ to $m+2$, and every $m \leqslant n$. From the exponential potential results, we expect logs to appear when $2\alpha - 1$ is exactly a negative integer as it was the case for $\alpha$ = 0.}


\vskip 0.5cm

\underline{Branch  $\alpha > \frac{1}{2}$}:

\vskip 0.5cm

For such values of $\alpha$, rearranging the terms, the expansion becomes:

$$
    S_+(\varphi) =e^{\frac{\f}{\sqrt{2(d-1)}}} \f^{\frac{\alpha}{2}} \left( \sqrt{\frac{2}{d-1}} + \frac{d \alpha}{(d-1) \varphi} \left( 1 + \frac{(1-2\alpha)}{2\alpha}C_{\a} \varphi^{1-2\alpha}  + \right.\right.
$$
\be
+ \left.\left. \mathcal{O}\left(\varphi^{2-4\alpha}\right)\phantom{\frac{(1-2\alpha)}{2\alpha} }\hskip -1.7cm \right) +\phantom{ \sqrt{\frac{2}{d-1}}}\hskip -1.3cm \mathcal{O}\left(\varphi^{-2}\right) \right),\label{asympsup}
\ee
$$
    W_+(\varphi) = e^{\frac{\f}{\sqrt{2(d-1)}}}\f^{\frac{\alpha}{2}} \left(  2 +  \frac{1}{\sqrt{2(d-1)}\varphi} \left( 2\alpha + C_{\a} (2\alpha-1)(d-2)\varphi^{1-2\alpha}  +
    \right.\right.
$$
\be
\left.\left.+
 \mathcal{O}\left(\varphi^{2-4\alpha}\right) \right)  +\phantom{ \frac{1}{\sqrt{2(d-1)}\varphi}}\hskip -2cm
 \mathcal{O}\left(\varphi^{-2}\right) \right),\label{Wup}
\ee
\begin{equation}
    T_+ = \frac{d}{\sqrt{2(d-1)}} e^{\sqrt{\frac{2}{d-1}} \varphi} \varphi^{\alpha- 1} \left( (2\alpha - 1) C_{\a} \f^{1-2\alpha}  + \mathcal{O}(\varphi^{2-4\alpha}) + \mathcal{O}(\varphi^{-1})\right) \label{TC}
\end{equation}
This time, the sign of $T$ is given by the sign of $C_{\a}$. This expansion is reliable as long as $\alpha < 1$, after which higher orders have to be considered\footnote{To access values of $\frac{n}{2}<\alpha< \frac{n+1}{2}$, we would a priori need to take into account terms of order $\frac{X_1^{m+1 - k}}{\varphi^k}$ in the differential equation, with again $m\leqslant n$. This time, we expect logs to appear for positive half integers starting from $\alpha = 1$.}

\subsubsection{The IHQCD bifurcation} \label{subsubsec:ihqcdbif}

IHQCD corresponds to the critical value $\a={1\over 2}$. In this case, the polynomial in the denominator of the left hand side of equation (\ref{mainU}) has only one root with multiplicity two. After some algebra, presented in appendix \ref{sec:F}, we obtain the following asymptotics for $S$, $W$:
\begin{equation}
    S(\f) = e^{\frac{\f}{\sqrt{2(d-1)}}} \f^{\frac{1}{4}} \left( \sqrt{\frac{2}{d-1}}+ \frac{d}{2(d-1)\f} \left(1 + \frac{1}{\log(\f) - C_{1\over 2}} \right)  +  \mathcal{O}(\varphi^{-2}) \right),
    \label{IHQCDdeltas}
\end{equation}
\begin{equation}
    W(\f) = e^{\frac{\f}{\sqrt{2(d-1)}}} \f^{\frac{1}{4}} \left( 2 + \frac{1}{\sqrt{2(d-1)}\f} \left(1 -  \frac{d-2}{\log(\f) - C_{1\over 2}} \right)  +  \mathcal{O}(\varphi^{-2}) \right),
    \label{IHQCDW}
\end{equation}
where $C_{1\over 2}$ is a real integration constant. We obtain the following expression for $T$ in this case:
\begin{equation}
    T = \frac{d}{\sqrt{2(d-1)}} e^{\sqrt{\frac{2}{d-1}}\varphi} \f^{-\frac{1}{2}} \left( -\frac{1}{\log(\varphi) - C_{1\over 2}}  + \mathcal{O}(\varphi^{-1}) \right).
    \label{IHQCDsolT}
\end{equation}
This gives a negative sign for $T$, regardless of the value of the integration constant, $C_{1\over 2}$.

Interestingly, the positive curvature  solutions which extend all the way to $\varphi \to +\infty$,   do not immediately appear at $\alpha>0$ (which is the critical value for confinement/deconfinement in flat space)  but for $\alpha > \frac{1}{2}$, which is exactly the physical value used in IHQCD. In particular, {\em this result excludes acceptable singular solutions for positively curved spaces in IHQCD}. The change of behavior between the two particular solutions found at $\alpha = \frac{1}{2}$ is a sign of another bifurcation in this point.
\smallbreak

The method used here might miss unstable solutions as they can be outside of the center manifold on which we are projecting. However, such solutions  grow  exponentially away from the fixed point, because there is no stable manifold outside the center manifold. For $b=b_c$, there is just one specific solution, therefore as unstable solutions diverge away from the critical point, they either bounce\footnote{This case was covered in \cite{C}} or become generic type $0$ solutions, which are not acceptable. For $b\neq b_c$, one has to study the topology of the dynamical system to draw a conclusion. Calculating a suitable index, we can show that there is a unique solution connecting the two critical points. More details and references for this method are given in appendix \ref{sec:D}. This solution is included in the stable manifold of the saddle critical point, and in the unstable manifold of the other critical point. For all other solutions, the same conclusion as for $b=b_c$ holds, and again the flows either bounce or become generic type $0$ solutions.

Having obtained a complete classification of the asymptotic solutions, we shall proceed to connect the IR asymptotics we found with UV data in order to obtain complete RG-flow solutions. To do so, we have to consider the full potential for the whole range of $\varphi$, and abandon the simple asymptotic form in equation (\ref{asympV}) that we used so far. As it is not possible to solve the full equations analytically, we shall resort to numerical integration.

\section{Full RG flow solutions}
To capture the full flow from the UV to the IR, we have to give the full potential.
We shall chose a relatively simple potential with a single maximum (UV fixed point) and  the large $\f$ asymptotics discussed in this paper.
A convenient choice is:
\begin{align}
    V(\varphi) &= -\frac{d(d-1)}{\ell^2} + \frac{m^2}{2}\varphi^2 - 4 {V_{\infty}\over \ell^2}  \left(\sinh\left(\frac{\varphi}{\sqrt{2(d-1)}}\right)\right)^2  |\varphi|^{\alpha} \nonumber \\
    &+  \frac{2 V_{\infty} |\varphi|^{2+\alpha}}{(d-1)\ell^2} + \frac{ V_{\infty} |\varphi|^{4+\alpha}}{3(d-1)^2\ell^2},
    \label{numpotential}
\end{align}

\begin{figure}[H]
\centering
\begin{tikzpicture}
\node (img) {\includegraphics[width=0.3\textwidth]{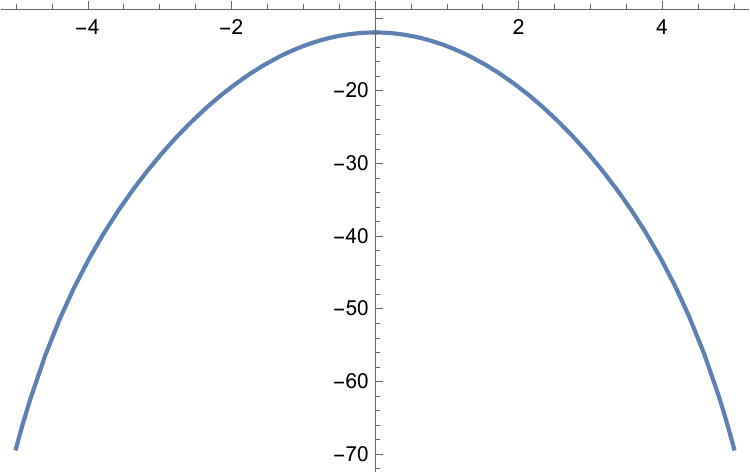}};
\draw (0.5, -2) node[below]{$V$};
\draw (2, 1) node[below]{$\varphi$};
\end{tikzpicture}
\caption{Potential $V(\varphi)$}
\label{fig:potential}
\end{figure}

This potential has a  UV fixed point located at $\varphi = 0$, which we use as a proxy for the asymptotically-free UV regime of YM$_4$. In \ref{numpotential} {we have included powers of $\ell$ such that $V_\infty$ is dimensionless.} In choosing the form of the potential we have subtracted the first term in the $\sinh^2$ expansion around zero so that the parameter $m^2$ represents the  mass of the scalar field fluctuations around the maximum at $\varphi=0$. We also removed the next term in the expansion so that we slow further the flow in the UV.

In this simplified model, the marginally relevant Yang-Mills coupling is replaced in the UV  by a relevant coupling, whose dimension $\Delta_->0$ is related to $m^2$ by the usual AdS/CFT relation $\Delta(\Delta-4) = m^2{\ell^2}$. This greatly simplifies the  numerical integration in the UV without changing the qualitative features of the solutions.

\subsection{Method}

We want to connect the IR continuous parameter {$C_{\a}$, defined in (\ref{asympexpS}), (\ref{soluconst}) and (\ref{IHQCDdeltas}) depending on the potential considered,} to the UV data, namely the curvature and vacuum expectation value (vev) of the relevant operator driving the flow. For convenience, we rewrite the system of equations into a single one by taking a derivative of (\ref{eqn:s}) and substituting (\ref{eqn:w}) into it. We obtain the following second order differential equation for $S$
\begin{equation}
    d S^3 S''- \frac{d}{2} S^4 - S'^2 S^2  -
 \frac{d}{d-1} S^2 V + (d + 2) S S' V' - d S^2 V'' - V'^2 = 0.
 \label{odesecorderS}
\end{equation}
From any given solution, we want to extract the normalized vev $\mathcal{C}$ and curvature parameter $\mathcal{R}$, defined in terms of the leading and subleading terms $\varphi_-$ and $\varphi_+$ terms in the expansion of $\varphi$ in the UV, as $u\to -\infty$:
\begin{align}
   & \varphi(u) = \varphi_- {\ell^{\Delta_-}} e^{\Delta_- \frac{u}{\ell}} +... +\varphi_+ {\ell^{\Delta_+}} e^{\Delta_+ \frac{u}{\ell}}  +... &  &\\
   &\mathcal{R} = R^{UV} |\varphi_-|^{-\frac{2}{\Delta_-}} & \mathcal{C} &= \frac{(2\Delta_+ - d)\Delta_-}{d} \varphi_+ |\varphi_-|^{-\frac{\Delta_+}{\Delta_-}}
   \label{UVexp}
\end{align}
where $R^{UV}$ is the Ricci curvature of the manifold on which the UV QFT lives, and the conformal dimension is given by:
\begin{equation}
    \Delta_{\pm} = \frac{1}{2}\left( d \pm \sqrt{d^2 + 4 {\ell}^2 m^2}\right).
    \label{conformaldim}
\end{equation}
We use standard quantisation, in which the dimension of the operator is given by $\Delta_+$ and the dimension of the coupling is given by $\Delta_-$. We obtain $\mathcal{R}$ and $\mathcal{C}$ through the following expansion of $S$ and $W$ \cite{C}:
\begin{align}
    S(\varphi) &= \left(\frac{\Delta_-}{\ell} \varphi + \dots\right) + \left( \frac{d}{\Delta_-} \frac{\mathcal{C}}{\ell} \varphi^{-1 + \frac{d}{\Delta_-}} + \dots\right) + \dots \label{UVS} \\
    W(\varphi) &= \left(\frac{2(d-1)}{\ell}+ \frac{\Delta_-}{2\ell}  \varphi^2+ \dots\right) + \left( \frac{\mathcal{R}}{d\ell} \varphi^{\frac{2}{\Delta_-}} + \dots\right) + \left(\frac{\mathcal{C}}{\ell}\varphi^{\frac{d}{\Delta_-}} + \dots \right), \label{UVW}
\end{align}
where the dots indicate subleading terms in $\varphi$ around $\f = 0$.

To obtain $S$ and $W$, we use numerical integration. The details and settings of the procedure can be found in appendix \ref{sec:E}. The boundary conditions are specified at a large but finite $\varphi_{ini}$ using the asymptotics that we calculated analytically {in equations (\ref{asympexpS}), (\ref{exactX}) and (\ref{IHQCDdeltas}) }. We shall measure all dimensionfull parameters of the theory in units of the AdS length $\ell$. This is equivalent to setting $\ell=1$.
 The rest of the parameters of the theory, $d$, $V_{\infty}$, $m^2$ and $\Delta_\pm$, are fixed to
\be
 d= 4\sp
    V_{\infty} = 1 \sp  m^2 = -\frac{15}{4}~~\to~~ \Delta_- = \frac{3}{2}\sp \Delta_+ = \frac{5}{2}
    \label{Values}
\ee

\subsection{Numerical Results}

{We shall now present our results from the numerical study of the equations with different values of $\a$.}

\subsubsection{Curvature and scalar vev\label{5.2}}

In all cases, we have two classes of solutions:

\vskip 0.5cm

$\bullet$ {\bf Type I/II solutions}, that start at the UV fixed point of the potential at $\f=0$, corresponding to the AdS$_{5}$ boundary and which go all the way to $\f\to+\infty$. Their single integration constant is denoted by $C_{\a}$ and is eventually translated to the value of the dimensionless curvature $\mathcal{R}$. $C_{\a}$ can take all real values. These are the solutions whose asymptotics  have been analysed in detail in the previous sections of this paper.

For each value of $C_{\a}$, we have a unique ``regular" solution\footnote{As discussed, type I/II solutions have mild naked singularity, however they are acceptable holographic solutions according to the Gubser's criterion.}.
The slice curvature of such solutions can be generically positive, zero or negative.

In the special case of $\a={1\over 2}$, the integration constant $C_{1\over 2}$ can only take {negative} values as indicated by the numerics. The reason is that for $C_{1\over 2}{>}0$, the flow misses the UV fixed point of the potentials at $\f=0$ and ends generically at a bad singularity at $\f\to -\infty$.
This phenomenon was found in confining asymptotically exponential potentials with $b_c<b<b_{G}$ in \cite{AdS2}.
In particular, it was found in \cite{AdS2} that by further tuning the value of the integration constant, the solution can be made to end up at a ``regular" asymptotics at $\f\to -\infty$. A discrete infinit set  of such ``regular" solutions was found, that have no AdS$_{5}$ boundary. There, they  were interpreted as holographic interfaces between topological QFTs.

Numerically, these solutions are constructed by starting the numerical evolution from near $\f\to\infty$ using the type I/II  analytical asymptotics, and solving until the solution stops at the maximum of the potential at $\f=0$. From the Fefferman-Graham expansion of the solutions around $\f=0$, we can extract the dimensionless curvature, $\mathcal{R}$,  and the dimensionless scalar vev, $\mathcal{C}$ as functions of $C_{\a}$.

\vskip 0.5 cm

$\bullet$ {\bf Type III (regular) solutions}, which start at the UV fixed point at $\f=0$ and go up to a value $\f_0$ of $\f$. This value $\f_0$ is the (positive) real constant that classifies the solutions in this class. The detailed asymptotics at the solution near $\f_0$ are described in appendix \ref{III}.

For each value of $\f_0$, we have a unique regular solution.
$\f_0$ can take any real value. However, as the potential is symmetric under $\f\to -\f$, we need only consider values $\f_0\geq 0$, which is what we do in the sequel.
The curvature for type III solutions can be only positive or zero\footnote{Negative curvature slices cannot shrink regularly to zero, \cite{AdS1}.}.
When $\f_0\to 0$, then the solutions has $\mathcal{R}\to +\infty$, \cite{C}.

Numerically, these solutions are constructed by starting the numerical evolution from near $\f\to\f_0$ using the type III  analytical asymptotics of appendix \ref{III}, and solving until the solution stops at the maximum of the potential at $\f =0$. From the Fefferman-Graham expansion of the solutions around $\f=0$, we can extract the dimensionless curvature, $\mathcal{R}$,  and the dimensionless scalar vev, $\mathcal{C}$ as functions of $\f_0$.

Before moving further we would like to compare the solutions we shall be presenting here, with the solutions found in previous works, in the confining case, both for negative curvature, \cite{AdS2}, and for positive curvature, \cite{Jani}.

For $\mathcal{R}<0$ and exponential potentials with $b_c<b<b_{G}$, in \cite{AdS2} the following types of solutions were found:
\begin{itemize}
\item  Solutions that interpolate between the UV maximum and  $\f\to \pm \infty$. They are like the type I/II solutions here. Such solutions had sometimes non-monotonic scale factors associated to A-bounces (places where $\dot A=0$) and $\f$-bounces (places where $\dot \f=0$).
Such solutions also exist here, but are harder to find. We show examples of such solutions on figure \ref{bounc} for $\a={1\over 4}$. We shall not   study them further, but we expect a similar behavior as in the cases studied in \cite{AdS2}.
\item Solutions that start and end at the UV maximum at $\f=0$. They were interpreted as holographic interfaces. We also have such solutions here, but we shall not study them and they will not be mentioned below.
\item Solutions with no AdS$_{d+1}$ boundary. These are solutions that start ``regularly" at $\f=+\infty$ do not stop at the top of the potential, and end ``regularly" at $\f=\pm\infty$ after
having an integer number of $\f$-bounces in-between. Such solutions exist also in this case, but we shall not study them  in this paper.
\end{itemize}

For $\mathcal{R}>0$ and $b_c<b<b_{G}$  the solution space was analyzed in \cite{Jani}. In this case   the scale factor is monotonic  and  all ``regular''  solutions start at the UV  and end either  at finite $\varphi_0$ (type III)  or in one of the asymptotic solutions which extend to $\varphi \to  \pm \infty$ (types I and II),  possibly going through one or more  $\varphi$-bounces.

We now proceed to present the results of our numerical studies.
In the subsequent plots, the  data for type I/II solutions are plotted in blue and the ones for type III solutions are shown for comparison in red.

Firstly, we plot $\mathcal{R}$ as a function of the (IR) integration constants, $C_{\a}$ or $\f_0$. For each of these constants there is a unique solution. The constants  $C_{\a}$ are defined in (\ref{soluconst}) and (\ref{IHQCDdeltas}) depending on the case considered for the critical solution. Secondly, we plot the vev, $\mathcal{C}$, as a function of $\mathcal{R}$.

We need also a few definitions:
\begin{itemize}

\item For a fixed $\a$, we define $\mathcal{R}_{c+}$ as the upper bound of $\mathcal{R}$ over the set of type I/II solutions with $C_{\a}\geq 0$.\label{R+}

\item Similarly we define $\mathcal{R}_{c-}$ as the upper bound of $\mathcal{R}$ over the set of type I/II solutions with $C_{\a}\leq 0$.

\end{itemize}

\begin{figure}[H]
\centering
\begin{subfigure}[t]{0.70\textwidth}
\begin{tikzpicture}
\node (img) {\includegraphics[width=\textwidth]{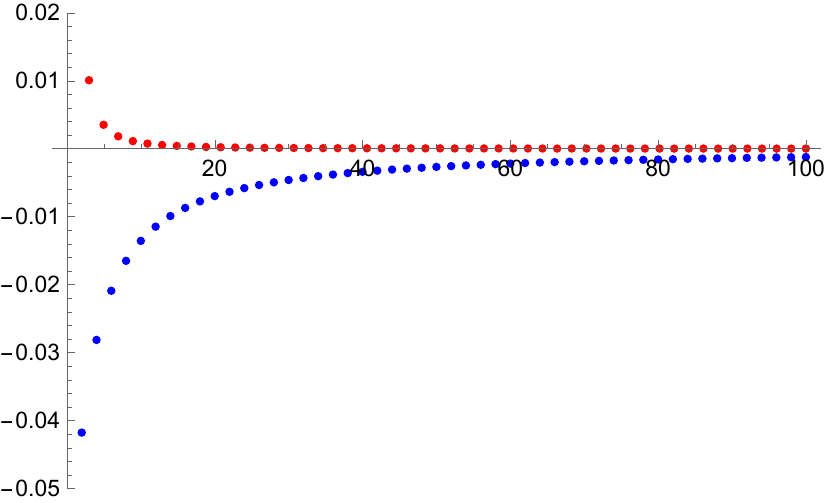}};
\node[below=of img, node distance=0cm, xshift=4cm, yshift=5.2cm]{${\color{blue}-C_0}, {\color{red}\varphi_0}$};
\node(leg)[below=of img, node distance=0cm, xshift=4cm, yshift=4.2cm]{${\color{blue}\bullet}$ Type I/II};
\node[below=of img, node distance=0cm, xshift=3.9cm, yshift=3.6cm]{${\color{red}\bullet}$ Type III};
\node[left=of img, node distance=0cm, rotate=0, anchor=center,xshift=2.5cm, yshift=3cm] {$\mathcal{R}$};
\end{tikzpicture}
\end{subfigure}
\caption{Dimensionless curvature $\mathcal{R}$ vs IR parameter ($C_0$ or $\f_0$) in the critical exponential case, $b=b_c$, $\alpha = 0$.}
\label{fig:Rconstexp}
\end{figure}
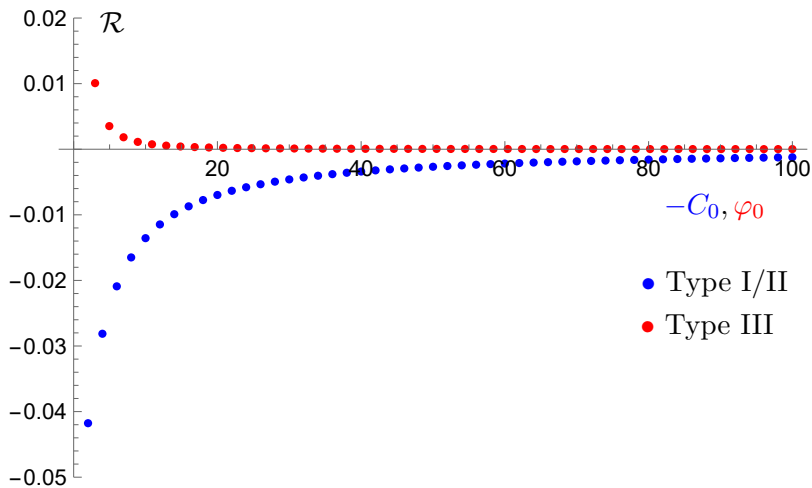

\begin{figure}[H]
    \centering
    \begin{subfigure}[t]{0.70\textwidth}
    \begin{tikzpicture}
    \node (img) {\includegraphics[width=\textwidth]{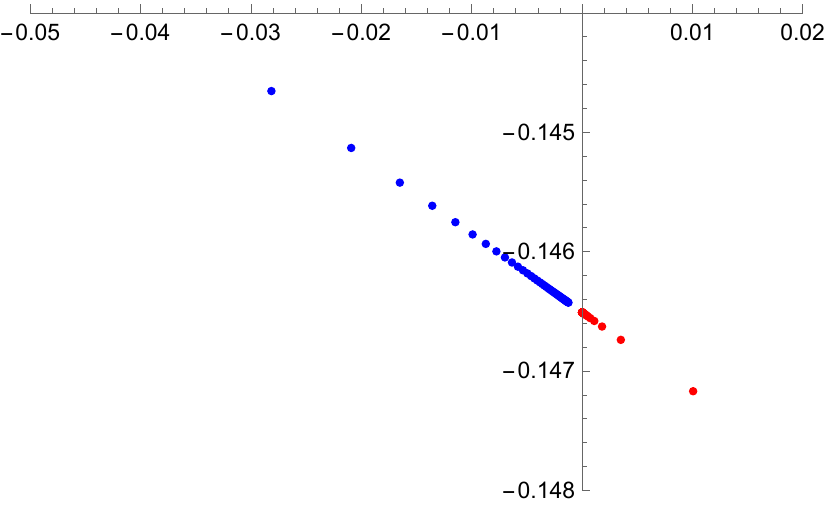}};
    \node[below=of img, node distance=0cm, xshift=2.5cm, yshift=1.5cm]{$\mathcal{C}$};
    \node[left=of img, node distance=0cm, rotate=0, anchor=center,xshift=1.5cm, yshift=2.2cm] {$\mathcal{R}$};
    \node(leg)[below=of img, node distance=0cm, xshift=4cm, yshift=6.2cm]{${\color{blue}\bullet}$ Type I/II};
    \node[below=of img, node distance=0cm, xshift=3.9cm, yshift=5.6cm]{${\color{red}\bullet}$ Type III};
    \end{tikzpicture}
    \end{subfigure}
    \caption{Dimensionless vev, $\mathcal{C}$, vs dimensionless curvature, $\mathcal{R}$ in the critical exponential case, $b=b_c$, $\alpha = 0$.}
    \label{fig:RCexp}
\end{figure}
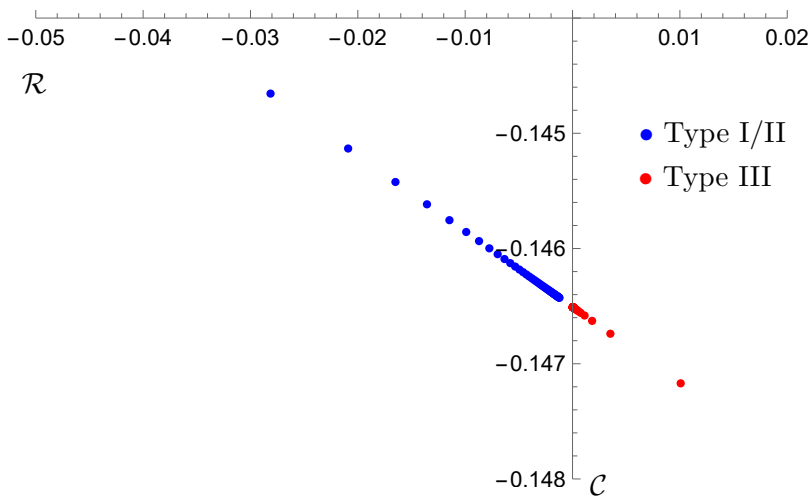

The exponential critical case, ($b=b_c$, $\a=0$), is shown in figure \ref{fig:Rconstexp} and \ref{fig:RCexp}. {From now on, the parameter $\f_\infty$ introduced in \ref{eqn:exactexplog} is set to zero to simplify the notation.} We find that for type I/II solutions, $\mathcal{R}$ is always negative, with $\mathcal{R} \to 0$ as ${C_0} \to -\infty$. We obtain $\mathcal{R}_{c-} = 0$ in this case.

Semi-stability, discussed previously, removes a certain range of $C_0>0$, which are not shown in the plot, for which the solution misses the UV fixed point. For large $\mathcal{R}$, the numerical error also becomes important, but we observe qualitatively that the RG flow misses the UV fixed point. It first goes through $S = 0$ for a negative value of $\varphi$, which indicates a bouncing RG flow as discussed in \cite{multirg, C}. Further increasing $C_0$ gives solutions where $S$ diverges like $S \sim e^{b_G \varphi}$ which we already discarded as unacceptably singular. The results obtained for the vev plot show a linear slope coherent with the results from the literature for $b\neq b_c$.

We shall now compare these results that refer to the pure exponential asymptotics, to the case where the exponential asymptotics are corrected by a power law. Recall that the definition of the different integration constants $C_{\a}$ can be found, depending of the potential considered, in (\ref{asympexpS}), (\ref{soluconst}) or (\ref{IHQCDdeltas}).

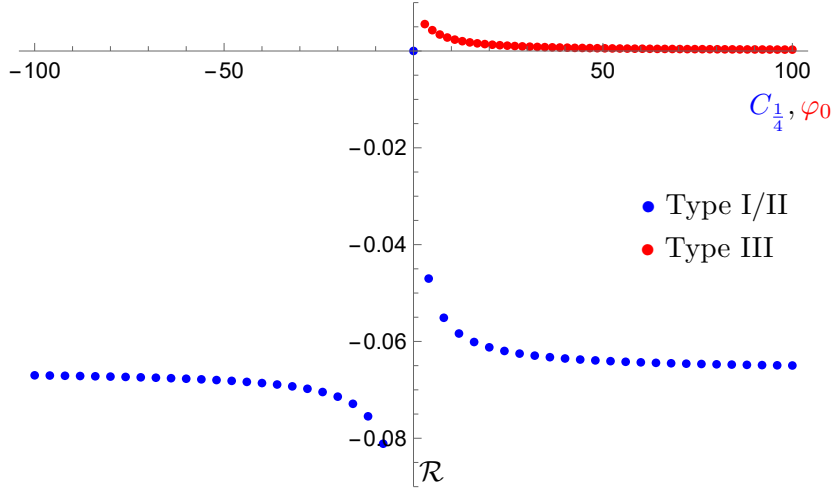
\begin{figure}[H]
    \centering
    \begin{subfigure}[t]{0.70\textwidth}
    \begin{tikzpicture}
    \node (img) {\includegraphics[width=\textwidth]{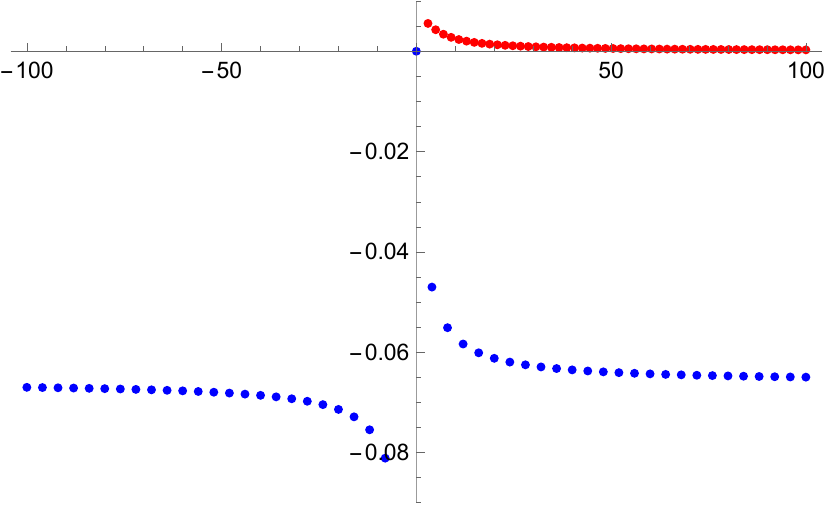}};
    \node[below=of img, node distance=0cm, xshift=5cm, yshift=6.5cm]{${\color{blue}C_\frac{1}{4}}, {\color{red}\varphi_0}$};
    \node[left=of img, node distance=0cm, rotate=0, anchor=center,xshift=6.7cm, yshift=-3cm] {$\mathcal{R}$};
    \node(leg)[below=of img, node distance=0cm, xshift=4cm, yshift=5.2cm]{${\color{blue}\bullet}$ Type I/II};
    \node[below=of img, node distance=0cm, xshift=3.9cm, yshift=4.6cm]{${\color{red}\bullet}$ Type III};
    \end{tikzpicture}
    \end{subfigure}
    \caption{Dimensionless curvature $\mathcal{R}$ vs IR parameter ($C_{1\over 4}$ or $\f_0$) in the exponential-Power law case, $b=b_c$, $\alpha = \frac{1}{4}$}
    \label{fig:Rconstalphainf12}
\end{figure}

\begin{figure}[H]
    \centering
    \begin{subfigure}[t]{0.70\textwidth}
    \begin{tikzpicture}
    \node (img) {\includegraphics[width=\textwidth]{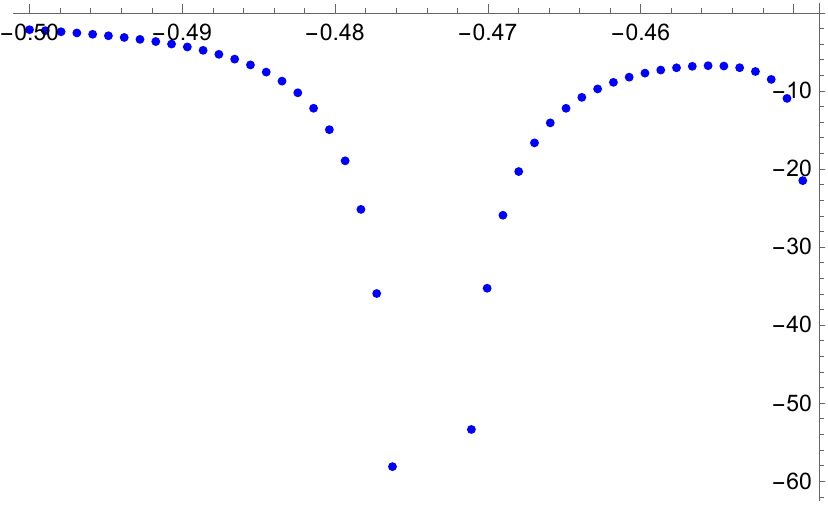}};
    \node[below=of img, node distance=0cm, xshift=-5cm, yshift=8.2cm]{${\color{blue}C_\frac{1}{4}}, {\color{red}\varphi_0}$};
    \node[left=of img, node distance=0cm, rotate=0, anchor=center,xshift=12cm, yshift=-3cm] {$\mathcal{R}$};
    \node(leg)[below=of img, node distance=0cm, xshift=-3cm, yshift=4.2cm]{${\color{blue}\bullet}$ Type I/II};
    \end{tikzpicture}
    \end{subfigure}
    \caption{A blow-up of figure \ref{fig:Rconstalphainf12} near $\mathcal{R}=0$ that shows again the dimensionless curvature $\mathcal{R}$ vs IR parameter ($C_{1\over 4}$) in the exponential-Power law case, $b=b_c$, $\alpha = \frac{1}{4}$. The non-monotonicity of the curve is associated to the presence of bounces (solutions with non-monotonic $\varphi$). The point at which $\mathcal{R}$ diverges corresponds to a pure vev solution $\varphi_- = 0$ see e.g. \cite{C}.}
    \label{bounc}
\end{figure}
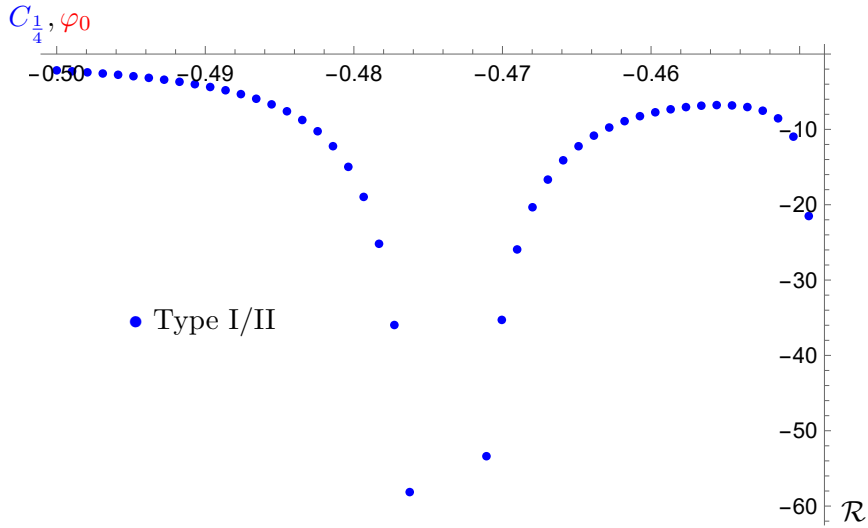

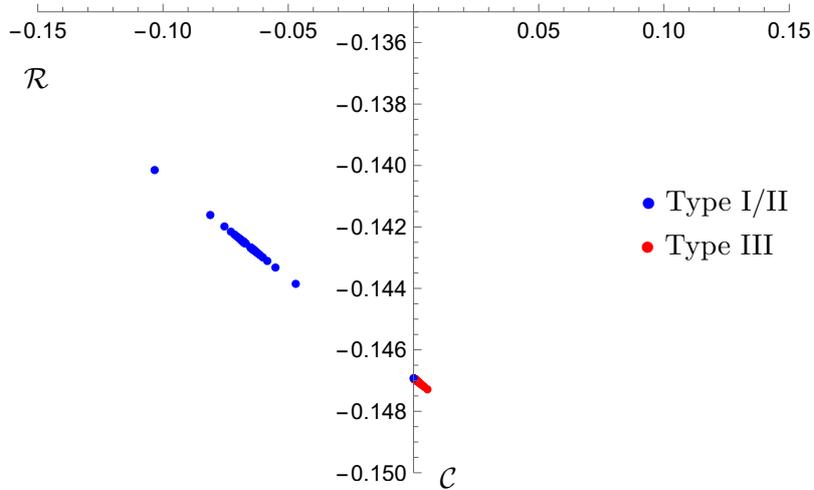
\begin{figure}[H]
    \centering
    \begin{subfigure}[t]{0.70\textwidth}
    \begin{tikzpicture}
            \node (img) {\includegraphics[width=\textwidth]{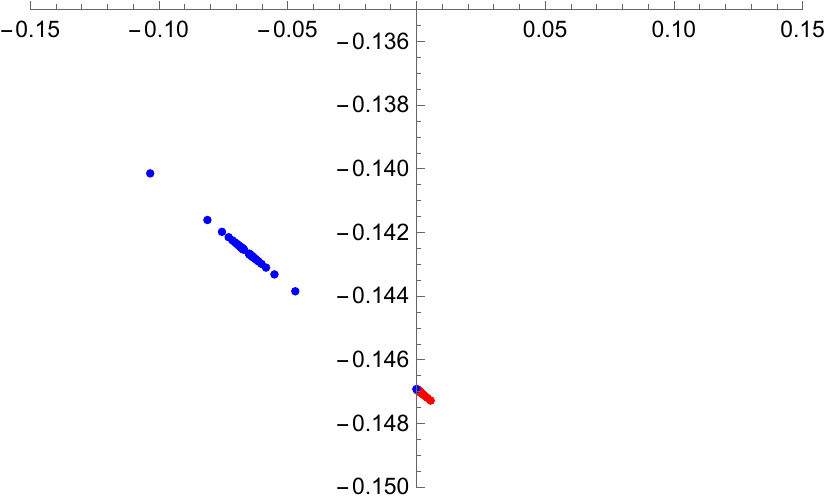}};
            \node[below=of img, node distance=0cm, xshift=0.5cm, yshift=1.5cm]{$\mathcal{C}$};
            \node[left=of img, node distance=0cm, rotate=0, anchor=center,xshift=1.5cm, yshift=2.2cm] {$\mathcal{R}$};
            \node(leg)[below=of img, node distance=0cm, xshift=4cm, yshift=5.2cm]{${\color{blue}\bullet}$ Type I/II};
            \node[below=of img, node distance=0cm, xshift=3.9cm, yshift=4.6cm]{${\color{red}\bullet}$ Type III};
    \end{tikzpicture}
    \end{subfigure}
    \caption{Dimensionless vev, $\mathcal{C}$,  vs dimensionless curvature, $\mathcal{R}$ in the exponential-Power law case, $b=b_c$, $\alpha = \frac{1}{4}$}
    \label{fig:RCalphainf12}
\end{figure}

In the power law case, (shown for $\alpha = \frac{1}{4}$ in figures \ref{fig:Rconstalphainf12} and \ref{fig:RCalphainf12}), {for $\a<\frac{1}{2}$ we find again} that in this case $\mathcal{R}_{c-}$ is finite and negative while $\mathcal{R}_{c+} = 0$. We observe a similar linear behavior in the vev plot. The points in the vev plot become denser when one approaches  $\mathcal{R}_{c-}$.

\begin{figure}[H]
    \centering
    \begin{subfigure}[t]{0.70\textwidth}
        \begin{tikzpicture}
            \node (img) {\includegraphics[width=\textwidth]{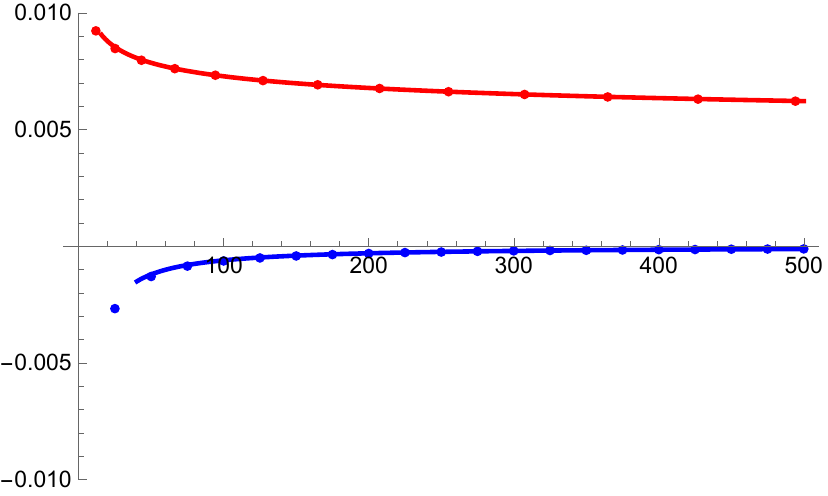}};
            \node[below=of img, node distance=0cm, xshift=4.5cm, yshift=4cm]{${\color{blue}-C_\frac{1}{2}}, {\color{red}\varphi_0}$};
            \node[left=of img, node distance=0cm, rotate=0, anchor=center,xshift=3cm, yshift=3cm] {$\mathcal{R}$};
            \node(leg)[below=of img, node distance=0cm, xshift=4cm, yshift=2.7cm]{${\color{blue}\bullet}$ Type I/II};
            \node[below=of img, node distance=0cm, xshift=3.9cm, yshift=2.1cm]{${\color{red}\bullet}$ Type III};
            \end{tikzpicture}
    \end{subfigure}
    \caption{Dimensionless curvature $\mathcal{R}$ vs IR parameter ($C_{1\over 2}$ or $\f_0$) in the critical case, $b=b_c$, $\a = \frac{1}{2}$. The dots correspond to the numerical data. The continuous red line corresponds to $\mathcal{R}=0.060/(4.8 + \log(\f_0))$. The continuous blue line corresponds to $\mathcal{R}={0.069\over C_{1\over 2}}$.}
    \label{fig:Rconstalpha12}
\end{figure}
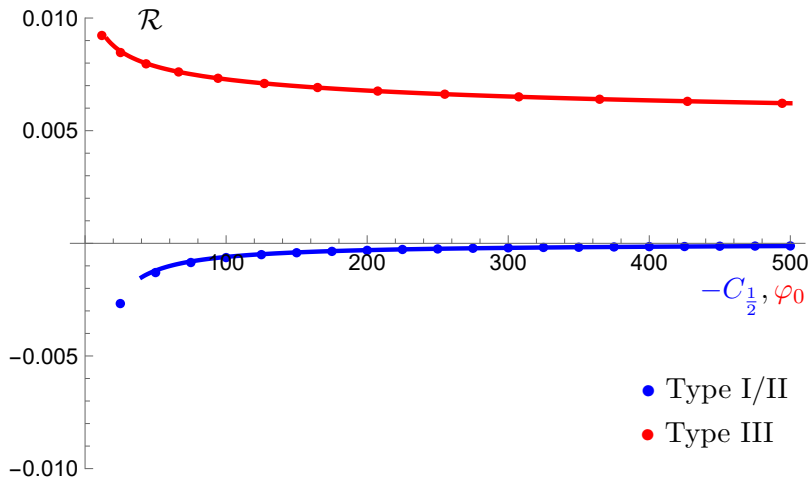

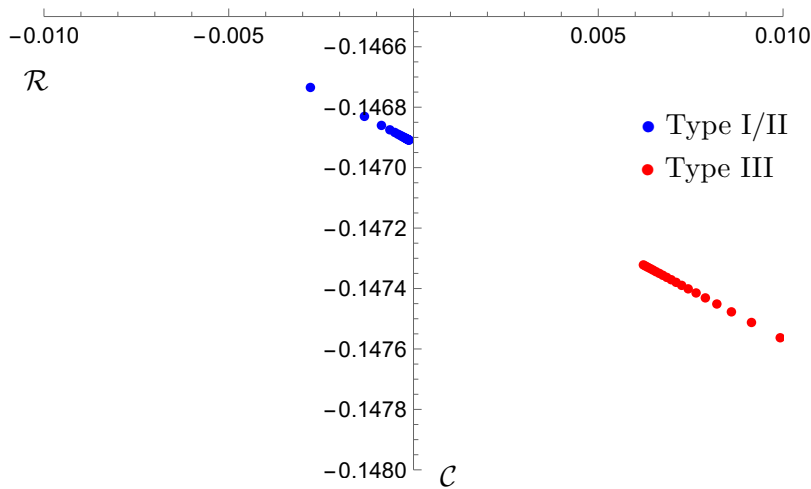
\begin{figure}[H]
    \centering
    \begin{subfigure}[t]{0.70\textwidth}
    \begin{tikzpicture}
            \node (img) {\includegraphics[width=\textwidth]{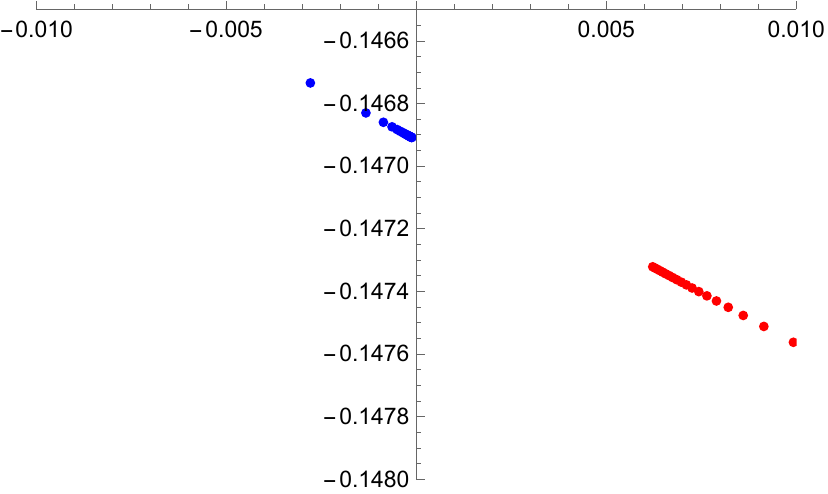}};
            \node[below=of img, node distance=0cm, xshift=0.5cm, yshift=1.5cm]{$\mathcal{C}$};
            \node[left=of img, node distance=0cm, rotate=0, anchor=center,xshift=1.5cm, yshift=2.2cm] {$\mathcal{R}$};
            \node(leg)[below=of img, node distance=0cm, xshift=4cm, yshift=6.2cm]{${\color{blue}\bullet}$ Type I/II};
            \node[below=of img, node distance=0cm, xshift=3.9cm, yshift=5.6cm]{${\color{red}\bullet}$ Type III};
    \end{tikzpicture}
    \end{subfigure}
    \caption{Dimensionless vev, $\mathcal{C}$, vs dimensionless curvature, $\mathcal{R}$ in the critical case $b=b_c$, $\alpha = \frac{1}{2}$}
    \label{fig:RCalpha12}
\end{figure}

In the critical case corresponding to IHQCD, shown in figures \ref{fig:Rconstalpha12} and \ref{fig:RCalpha12}, we have $T\leq 0$ for all type I/II  solutions, {as shown analytically in \ref{IHQCDsolT},} which implies $\mathcal{R}_{c\pm} \leqslant 0$. We find numerically that we have $\mathcal{R}_{c-} = 0$, {reached in the limit $C_\frac{1}{2} \to -\infty$.} These results suggests that type I/II solutions exist for IHQCD on any negatively curved space and in flat space.

The integration constant $C_{1\over 2}$ appears in the denominator with the log correction in \ref{IHQCDsolT}, making the numerical study of $\mathcal{R}$ for positive $C_{1\over 2}$ challenging due to a pole in the IR asymptotics as in (\ref{IHQCDdeltas}). {Because of this form, numerical results can only be trusted for values of $C_{1\over 2}$ such that $\frac{1}{\log(\f_{ini}) - C_{1\over 2}}$ is much smaller than $1$}.

The red solid curve in \ref{fig:Rconstalpha12} corresponds to a $\frac{1}{\log\varphi_0}$ fit. {Numerical results suggest that} Type III solutions {extend all the way to $\mathcal{R}_{c+} = 0$, however to obtain solutions with smaller positive curvature, an exponentially larger $\f_{ini}$ is required, which makes numerical computations exponentially slower.}
\

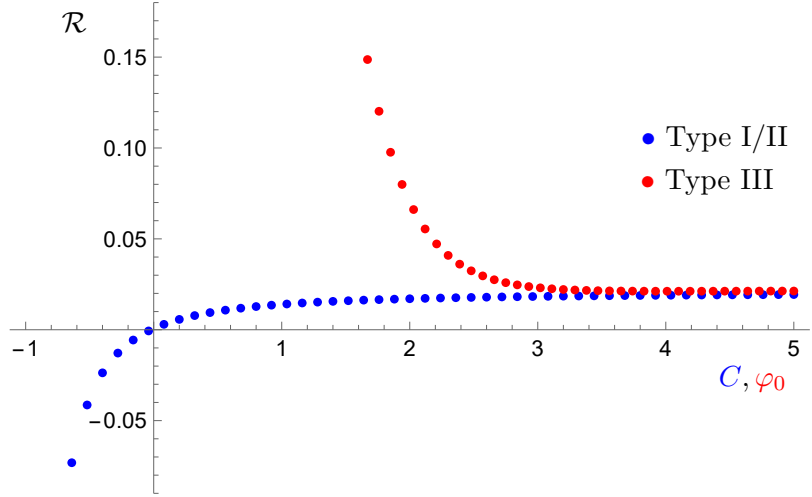
\begin{figure}[H]
    \centering
    \begin{subfigure}[t]{0.70\textwidth}
        \begin{tikzpicture}
            \node (img) {\includegraphics[width=\textwidth]{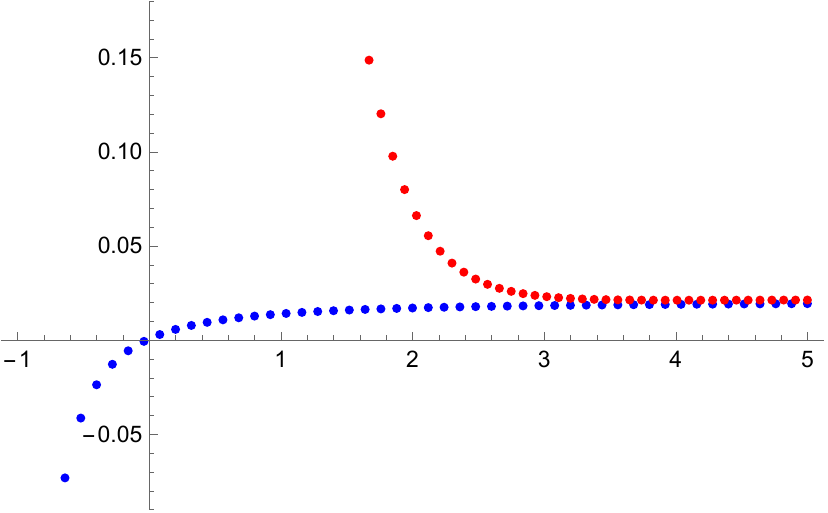}};
            \node[below=of img, node distance=0cm, xshift=4.5cm, yshift=3cm]{${\color{blue}C}, {\color{red}\varphi_0}$};
            \node[left=of img, node distance=0cm, rotate=0, anchor=center,xshift=2cm, yshift=3cm] {$\mathcal{R}$};
            \node(leg)[below=of img, node distance=0cm, xshift=4cm, yshift=6.2cm]{${\color{blue}\bullet}$ Type I/II};
            \node[below=of img, node distance=0cm, xshift=3.9cm, yshift=5.6cm]{${\color{red}\bullet}$ Type III};
            \end{tikzpicture}
    \end{subfigure}
    \caption{Dimensionless curvature $\mathcal{R}$ vs IR parameter ($C_{3\over 4}$ or $\f_0$) in the exponential-power law case, $b=b_c$, $\alpha = \frac{3}{4}$}
    \label{fig:Rconstalphasup12}
\end{figure}

\begin{figure}[H]
    \centering
    \begin{subfigure}[t]{0.70\textwidth}
        \begin{tikzpicture}
            \node (img) {\includegraphics[width=\textwidth]{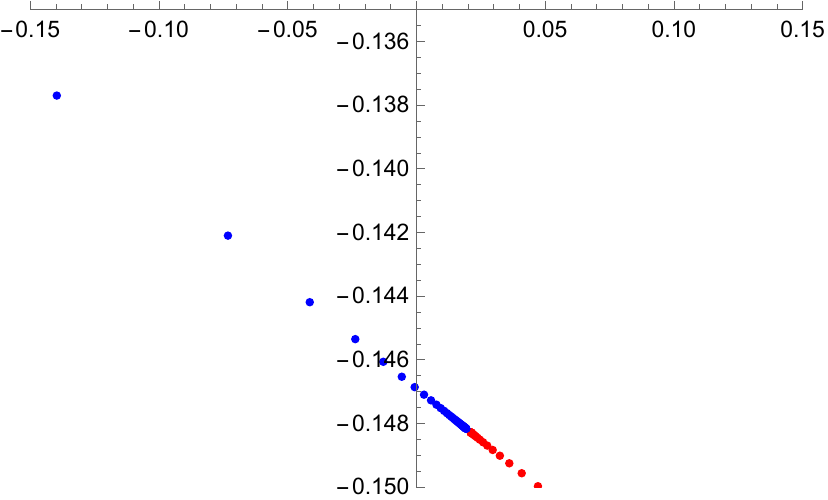}};
            \node[below=of img, node distance=0cm, xshift=0.5cm, yshift=1.5cm]{$\mathcal{C}$};
            \node[left=of img, node distance=0cm, rotate=0, anchor=center,xshift=1.5cm, yshift=2.2cm] {$\mathcal{R}$};
            \node(leg)[below=of img, node distance=0cm, xshift=4cm, yshift=4.2cm]{${\color{blue}\bullet}$ Type I/II};
\node[below=of img, node distance=0cm, xshift=3.9cm, yshift=3.6cm]{${\color{red}\bullet}$ Type III};
            \end{tikzpicture}
    \end{subfigure}
    \caption{Dimensionless vev, $\mathcal{C}$, vs dimensionless curvature, $\mathcal{R}$ in exponential-power law case, $b=b_c$, $\alpha = \frac{3}{4}$}
    \label{fig:RCalphasup12}
\end{figure}
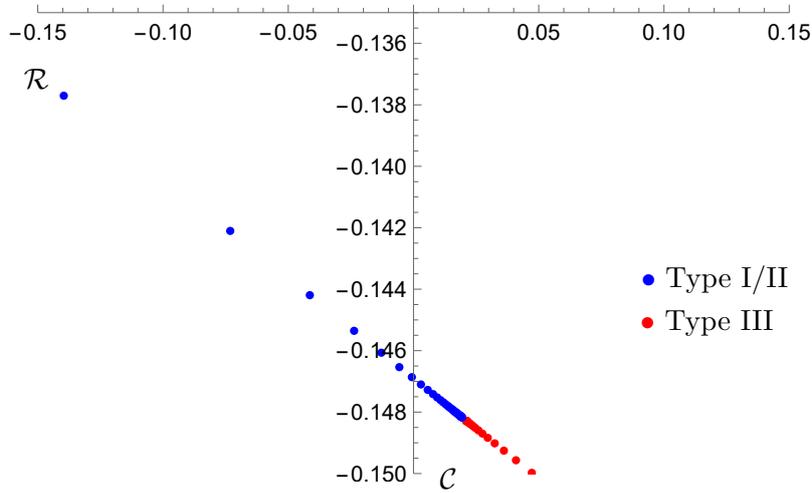

Figures \ref{fig:Rconstalphasup12} and \ref{fig:RCalphasup12} show the power law case for $\alpha = \frac{3}{4}$. Interestingly, the critical curvature $\mathcal{R}_{c-}$ is again found to be zero, while $\mathcal{R}_{c+}$ is positive. This is coherent with the asymptotics (\ref{TC}), where we found that the sign of the slice curvature was the sign of the integration constant. Notably, $\mathcal{R}_{c+}$ is now reached at $C_{3\over 4} \to +\infty$ instead of $C_{\a}=0$ for the case $\alpha < \frac{1}{2}$. {Moreover, the roles are also exchanged for $\mathcal{R}_{c-}$, which is obtained for $C_{3\over 4} \to 0$ instead of $C_{3\over 4} \to +\infty$ for $\alpha < \frac{1}{2}$}.

In the following, we investigate the behavior of the normalized UV parameters as a function of $\alpha$.

\subsubsection[Curvature as function of the parameter a]{Curvature $\mathcal{R}$ as function of the parameter $\alpha$}

Our numerics indicates  that for $\alpha \neq \frac{1}{2}$, the curvatures $\mathcal{R}_{c+}$ and $\mathcal{R}_{c-}$,
defined in subsection \ref{R+},
are obtained from the extremal particular type I/II solutions of the differential equation, ie. those with $C_{\a}=0,\pm\infty$.

\begin{figure}[H]
\centering
\begin{tikzpicture}
\node (img) {\includegraphics[width=0.8\textwidth]{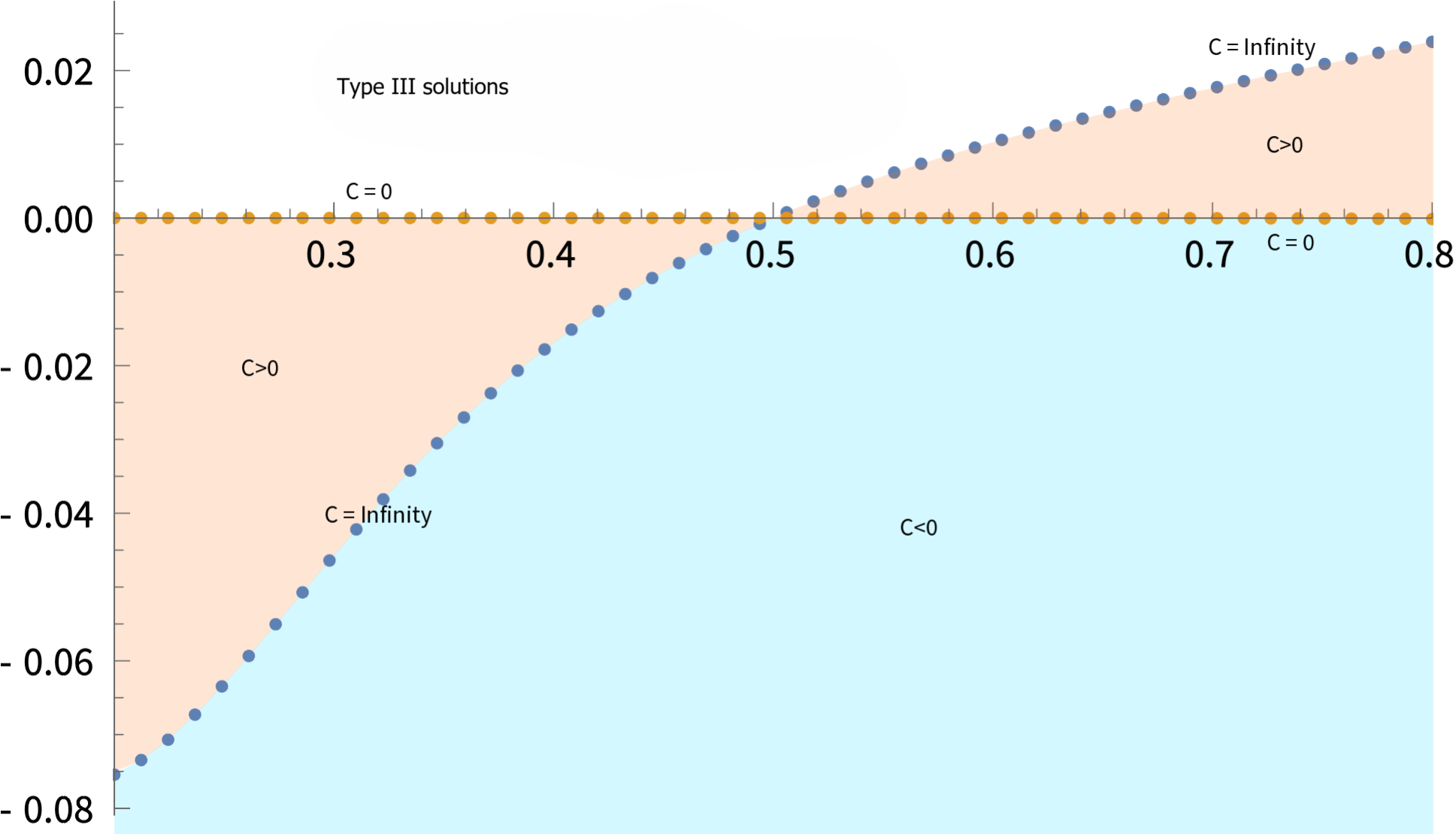}};
\node[below=of img, node distance=0cm, xshift=6.5cm, yshift=6.5cm]{$\alpha$};
\node[left=of img, node distance=0cm, rotate=0, anchor=center,xshift=1.5cm, yshift=3.5cm] {$\mathcal{R}$};
\end{tikzpicture}
\caption{A global summary between IR parameters and the UV dimensionless curvature $\mathcal{R}$ as a function of the exponent $\a$ of the asymptotic scalar potential.
The white area of the diagram is filled with type III solutions. The rest is filled by the type I/II solutions. There is no overlap between the two classes. The orange area corresponds to type I/II solutions with $C_{\a}>0$. The blue area corresponds to type I/II solutions with $C_{\a}<0$.}
\label{fig:Ralpha}
\end{figure}

Figure \ref{fig:Ralpha} gives a summary of the space of solutions as a function of the UV dimensionless curvature parameter $\mathcal{R}$ as a function of the type of potential asymptotics, parametrized by $\a$ in \ref{potentialasymp}. The vertical axis is $\mathcal{R}$ and includes both positive, zero and negative values. The horizontal axis is $\a\in [{0.2},0.8]$, all values being in the confining regime. The special value $\a={1\over 2}$ on this diagram corresponds to the IHQCD asymptotics and is special,  as it is the place of a double bifurcation on the space of Einstein-dilaton theories.

The white area of the diagram in figure \ref{fig:Ralpha} is filled with type III solutions. As discussed earlier, they only exist when $\mathcal{R}>0$. The rest is filled by the type I/II solutions. There is no overlap between the two classes, namely type III and type I/II. The orange area corresponds to type I/II solutions with $C_{\a}>0$. The blue area corresponds to type I/II solutions with $C_{\a}<0$.

These orange and the blue area, when $\a<{1\over 2}$,  meet  on the line
$C_{\a}\to \pm\infty$ which is common for the two signs.
On the other hand, for $\a>{1\over 2}$, these two areas meet each other on $C_{\a}=0$.

In all cases the limit $\f_0\to 0$ of type III solutions produces $\mathcal{R}\to +\infty$.
On the other hand the limit $\f_{0}\to \pm\infty$ of type III solutions depends on $\a$.
If $\a<{1\over 2}$ then the limit $\f_{0}\to \pm\infty$ of type III solutions produces $\mathcal{R}=0$.
If $\a>{1\over 2}$ then the limit $\f_{0}\to \pm\infty$ of type III solutions results in  the same limiting value $\mathcal{R}$ as the ones by type I/II solutions with $|C_{\a}|\to \infty$. {This limiting value is in fact  $\mathcal{R}_{c+}$, as we discus below.}

The values $\mathcal{R}_{c+}$ and $\mathcal{R}_{c-}$, defined in the beginning of section \ref{5.2}, can be read from figure \ref{fig:Ralpha}. They  are attained for a different value of $C_{\a}$ for $\alpha< \frac{1}{2}$ and $\alpha >\frac{1}{2}$ due to the bifurcation at $\alpha	= \frac{1}{2}$.
\begin{itemize}
\item For $\a<{1\over 2}$, $\mathcal{R}_{c+}=0$, and is attained with $C_{\a}=0$ corresponding to part of the orange-dotted line with $\a<{1\over 2}$.
  $\mathcal{R}_{c-}$ is negative and is given by the part of the blue-dotted line on the diagram with $\a<{1\over 2}$. It is attained with $|C_{\a}|\to \infty$.
\item For $\a>{1\over 2}$, $\mathcal{R}_{c+}$ is positive and it is given by the part of the blue-dotted line of the diagram with $\a>{1\over 2}$. It is attained with $|C_{\a}|\to \infty$.
On the other hand, $\mathcal{R}_{c-}=0$, and is attained with $C_{\a}=0$ corresponding to part of the orange-dotted line with $\a>{1\over 2}$.
\item At the bifurcation point $\a={1\over 2}$, the situation  is unusual.
As suggested by the diagram of figure \ref{fig:Ralpha}, all positive values of $\mathcal{R}$ are generated by the type III solutions with $0\leq \f_0<\infty$. All negative values of $\mathcal{R}$ are generated by type I/II solutions with $0>C_{1\over 2}>-\infty$. In this case, our numerics indicates that there are no type I/II solutions with $C_{1\over 2}>0$
 that end at the maximum of the potential.
 Instead, all such solutions go all the way from $\f\to +\infty$ to $\f\to -\infty$ and arrive there with generically bad singularities. We expect that at special values of $C_{1\over 2}>0$ one may arrive at $\f\to -\infty$ with acceptable singularities and such no-boundary solutions will have a similar interpretation as the ones found in the exponential case in \cite{AdS2}.
\end{itemize}

 Comparing with exponential asymptotics with $b_c<b<b_{G}$, studied recently in \cite{Jani}, we observe a similarity of the diagram found here for $\a>{1\over 2}$ and the general confining exponential case. There, as $b>b_c$ type I and type II asymptotics are distinct. The boundary between type III and type II asymptotics is formed by type I asymptotics, that sit at  a similar position as in figure  \ref{fig:Ralpha} with $\a<{1\over 2}$.

It is important to stress that for all $\a$ the {transition} between different types of solutions is continuous. Therefore we expect no first order quantum phase transitions by varying $\mathcal{R}$. However, there might be second or higher order transitions.

In the similar recent study for $b_c<b<b_G$, \cite{Jani}, two distinct cases were identified.
When $b_G>b>b_E$, with {$b_E$ defined in \ref{befimov}}
there is always a quantum first phase transition as we vary $\mathcal{R}$. This happens at the type I boundary between type III and type II solutions,  where $\mathcal{R}$ is positive. The regime in \cite{s2s2}  was coined  as the Efimov regime, because it portrays Efimov-like  phenomena, including an Efimov spiral for the scalar vev, \cite{Aharony2,s2s2,Jani}.
In the Efimov regime, a given $\mathcal{R}$ can be generated by a type III or a type II solution, unlike what we found in this paper.

In the monotonic regime,    $b_c<b<b_E$, it was found that to each value of $\mathcal{R}$ was associated a single solution, either type III or type II. Moreover, the transition was found to be continuous of order $1+[\delta]$ with $\delta$ given by the following formula
\be
\delta\equiv {(d-1)b+2\sqrt{1-{b^2\over b_E^2}}\over (d-1)b-2\sqrt{1-{b^2\over b_E^2}}}
\label{delta}
\ee
As $b\in [b_c,b_E]$, $\delta \in [\infty,1]$. This indicates that at $b=b_E$, the transition is second order, and as $b$ comes closer to $b_c$, its order becomes larger and larger.

It is plausible that in our case the transition is continuous higher order for $\a>{1\over 2}$ and it disappears or becomes a BKT transitions in the physically interesting case of $\a={1\over 2}$.
We shall reconsider this issue in the next section, after having computed the free energy.

\subsubsection{Free energy and Entropy}

In this subsection we shall  compute the free energy of holographic theories on spheres $S^d$, for different values of $\a$. General expressions for the free energy have already been worked out in \cite{C,F,dS}, which we shall briefly review. In this approach, the free energy $F$ is calculated using the standard dictionary,
\begin{equation}
    F(j, R) = -S_{on-shell}(\f_-, R),
    \label{defF}
\end{equation}
where $j$ is identified with $\f_-$, $R = R^{UV}$ (and we are suppressing the $UV$   superscript  in this section) and $S_{on-shell}$ is the action (including  a Gibbons-Hawking-York boundary term) calculated on a solution of the equations of motion. The free energy can be expressed as a functional of $A(u)$ as follows,
\begin{equation}
    F = 2(d-1)V_d \left[ e^{dA} \dot{A} \right]_{UV} - \frac{2R V_d}{d} \int_{UV}^{IR} du e^{(d-2)A}
    \label{eqF}
\end{equation}
where we have denoted by $V_d = \Tilde{\Omega}_d R^{-\frac{d}{2}}$ the volume of the $S^d$, with $\Tilde{\Omega}_d = 2 d^{\frac{d}{2}}(d-1)^{\frac{d}{2}} \pi^{\frac{d+1}{2}} / \Gamma(\frac{d+1}{2})$. Using the definitions (\ref{defw}),(\ref{defs}) and (\ref{defT}), the expression (\ref{eqF}) can be rewritten as,
\begin{equation}
    -\frac{F}{\Tilde{\Omega}_d} = [T^{-\frac{d}{2}} W]_{UV} + \frac{2}{d} \int_{UV}^{IR} d\f S^{-1} T^{-\frac{d}{2}+1}.
    \label{eqFWST}
\end{equation}
We next introduce a function $U(\f)$ satisfying the equation:
\begin{equation}
    S U' - \frac{d-2}{2(d-1)} W U = -\frac{2}{d}.
    \label{odeU}
\end{equation}
Using the definitions  (\ref{defw}-\ref{defT}), one can easily show that  the  integrand  in (\ref{eqFWST}) takes the form of a total derivative,
\be
{2\over d} S^{-1} T^{-\frac{d}{2}+1} = {d\over d\f} \left(- T^{-\frac{d}{2}+1} U \right).
\ee
and the second term in  (\ref{eqFWST}) can be written as the difference between an UV and an IR contribution.  The integration constant in $U$ introduced by the differential equation (\ref{odeU}) does not appear in the free energy, because it cancels between the two end-points of the integral. We  fix it in such a way that
$U\Big|_{IR}=0$, so that the IR contribution  vanishes  and the action can be written  in terms of UV data only \cite{F}:
\begin{equation}
    F = - \Tilde{\Omega}_d \left( [T^{-\frac{d}{2}} W]_{UV} + [T^{-\frac{d}{2} + 1} U]_{UV} \right).
    \label{Fdiverg}
\end{equation}

Importantly, the second term has been identified with the entanglement entropy $S_{E}$, calculated using the Ryu-Takayanagi (RT) prescription, in the sphere/de Sitter case, \cite{F}. On the one hand, when the slice manifold is de Sitter, written in global coordinates, one can compute the entanglement entropy between the two hemispheres of the spatial sphere at the minimum of the de Sitter scale factor using the RT prescription. This turns out to be given by:
\begin{equation}
    S_{E} = \Tilde{\Omega}_d \left(  T_{UV}^{-\frac{d}{2} + 1} U_{UV} \right)
    \label{defentropyreg1}
\end{equation}
When on  the other hand the slice manifold is the {\em static patch} of de Sitter, then there is a bulk horizon and a Bekenstein thermal entropy $S_{th}$. As shown in \cite{F}, this thermal entropy is exactly equal to the entanglement entropy (\ref{defentropyreg1}). Moreover, the various pieces of the action, $F$,  $S_{E}$ and $E$ defined by:
\be \label{defenergy}
E\equiv \Tilde{\Omega}_d ~ [T^{-\frac{d}{2}} W]_{UV}
\ee
were shown in \cite{F} to satisfy the standard thermodynamic identity,
\be \label{thermalid}
S_{E} =  \beta^2 \de_\beta (\beta^{-1} F)
\ee
where the inverse temperature $\beta$ is fixed  in terms the de Sitter radius  $L_{dS}$ by $\beta = 2\pi L_{dS}$.

When the slice metric is flat of with constant negative curvature we do not know a similar thermal interpretation for the three quantities $F,E,S_{E}$ as defined via (\ref{Fdiverg}), (\ref{defentropyreg1}) and (\ref{defenergy}).
However, it was shown numerically in \cite{AdS2} that even in the negative curvature case, for the leading branch of solutions,  thermodynamic identities like (\ref{thermalid}) hold.
For subleading branches (that we do not show in this paper), this relation (or more precisely,  equation (\ref{eqn:B})  below, which is derived from   (\ref{eqn:B})) is only   modified by  an  additive constant which changes as we move from one  branch to another .
It remains a challenge to interpret $E$ and $S_{E}$ thermodynamically in the negative curvature case.

We now come back to the calculation of the free energy (\ref{Fdiverg}) which, following \cite{F,dS}, we are going to rewrite in terms of appropriate functions of $\mathcal{R}$.  As usual, $S_{on-shell}$, as well as $F$ and $S_{E}$,  diverge  when the corresponding expressions are evaluated on the AdS boundary. For this we  introduce a UV cutoff $\Lambda$ (not to be confused with $\Lambda_{UV}$, which is the physical scale which breaks scale invariance). $S_{E}$ and $F$ then become:

\begin{equation}
    S_{E} = \Tilde{\Omega}_d \left(  T_\Lambda^{-\frac{d}{2} + 1} U_\Lambda \right)
    \label{defentropyreg}
\end{equation}
\begin{equation}
    F = - \Tilde{\Omega}_d \left( T_\Lambda^{-\frac{d}{2}} W_\Lambda +  T_\Lambda^{-\frac{d}{2} + 1} U_\Lambda \right)
    \label{defFreg}
\end{equation}
The renormalized expression for the entropy $S_{E}$ and the free energy in the case of $d=4$ can be reexpressed as \cite{dS}:
\begin{equation}
    \frac{S_E^{ren}}{\Tilde{\Omega}_4} = \frac{B(\mathcal{R}) - B_{ct}}{\mathcal{R}} - A_{ct}, \label{defrenSee},
\end{equation}
\begin{equation}
    \frac{F^{ren}}{\Tilde{\Omega}_4} = - \frac{\mathcal{C}(\mathcal{R}) - \mathcal{C}_{ct}}{\mathcal{R}^2}  - \frac{B(\mathcal{R}) - B_{ct}}{\mathcal{R}} + \frac{1}{192} + A_{ct}, \label{ren},
\end{equation}
where $A_{ct}$, $B_{ct}$, $C_{ct}$ are finite counterterms that have survived the subtraction of divergences and therefore control the scheme dependence of the renormalized free energy.
The function $\mathcal{C}$ was introduced {in equation (\ref{UVexp})}and {plotted in figures \ref{fig:RCexp}, \ref{fig:RCalphainf12}, \ref{fig:RCalpha12} and \ref{fig:RCalphasup12}},
and $B(\mathcal{R})$ is defined via the UV expansion of $U$, \cite{F},
\begin{equation}
    U(\varphi) = \frac{2}{d(d-2)} + B(\mathcal{R})\varphi^{\frac{d-2}{\Delta_-}} + \mathcal{O}(\varphi^2)
    \label{eqn:UB}
\end{equation}
The functions $B(\mathcal{R})$ and $ \mathcal{C}(\mathcal{R})$ are related by the following differential relation
\begin{equation}
    \mathcal{C}'(\mathcal{R}) = B(\mathcal{R}) - \mathcal{R}B'(\mathcal{R}) + \frac{\mathcal{R}}{96} \label{eqn:B}
\end{equation}
that is an avatar of the hidden thermodynamics of the problem, \cite{F,dS}.
The validity of the fit in (\ref{eqn:UB}) is checked by integrating the analytical relation (\ref{eqn:B}) between $ \mathcal{C}$ and $B$ and comparing the two values obtained for $B$.

Integrating equation (\ref{eqn:B}) generates an integration constant that can be absorbed in the value of the finite counterterms in the renormalized free energy. We use the value {$B_{ct} = -0.0662$} to make $S_{E}$ regular at $\mathcal{R} = 0$

\smallbreak

The functions $B$ and $ \mathcal{C}$ are fitted with the small curvature expansion in $d=4$ of $B(\mathcal{R})$ and $\mathcal{C}(\mathcal{R})$ worked out in \cite{dS}:

\begin{equation}
    B(\mathcal{R}) = B^{(0)} + B^{(1)} \mathcal{R} \left( 1 + \frac{\log(\mathcal{R})}{48}\right) + \mathcal{O}(\mathcal{R}^2)
    \label{lowRB}
\end{equation}
\begin{equation}
    \mathcal{C}(\mathcal{R}) = C^{(0)} +  C^{(1)}\mathcal{R} + { \mathcal{R}^2\over 192} + \mathcal{O}(\mathcal{R}^3)
    \label{lowRC}
\end{equation}
where we have imposed (\ref{eqn:B}). The free energy diverges as $\mathcal{R}\to 0^+$ as the volume of the manifold diverges.
$C^{(0)} -C_{ct}$ is proportional to the free energy density after we divide the free energy by the volume of the sphere.
The  expansions (\ref{lowRB}-\ref{lowRC})  were derived in \cite{dS}  for curved flows which reach a regular IR fixed point in the zero-curvature limit,  in other words all solutions were type III and   $\mathcal{R}\to 0^+$  when $\f_0$ approaches  a minimum of the potential away from $\varphi = 0$. In our case, there is no such IR fixed point for the Minkowski flows, and it is  not clear whether (\ref{lowRB}) and (\ref{lowRC}) are valid as $\f_0\to \infty$ in type III solutions for $\a\leq {1\over 2}$.  Although this seems plausible, obtaining an analytic small-curvature expansion in this case proved much more difficult than in \cite{dS}.

\begin{figure}[H]
\centering
\begin{subfigure}[t]{0.45\textwidth}
\begin{tikzpicture}
\node (img) {\includegraphics[width=\textwidth]{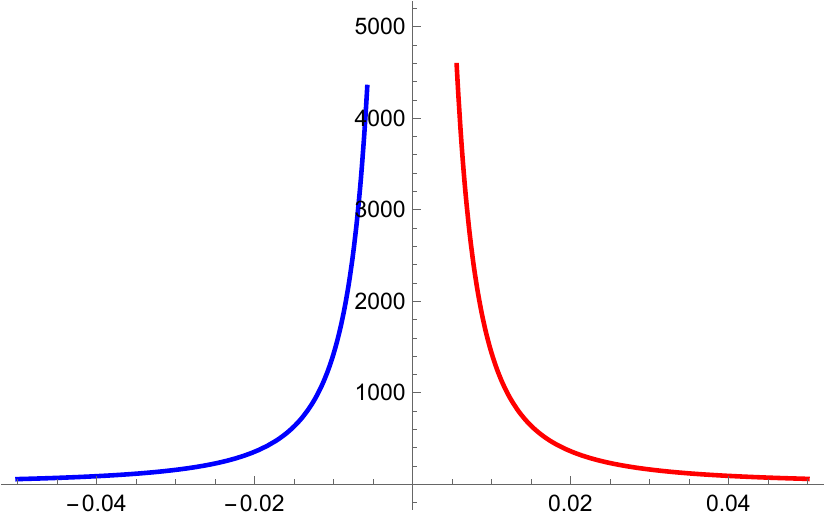}};
\node[below=of img, node distance=0cm, xshift=3cm, yshift=1.2cm]{$\mathcal{R}$};
\node[left=of img, node distance=0cm, rotate=0, anchor=center,xshift=4.5cm, yshift=2.5cm] {$F$};
\end{tikzpicture}
\caption{Exponential case $\alpha=0$}
\label{fig:F0full}
\end{subfigure}
\begin{subfigure}[t]{0.45\textwidth}
\begin{tikzpicture}
\node (img) {\includegraphics[width=\textwidth]{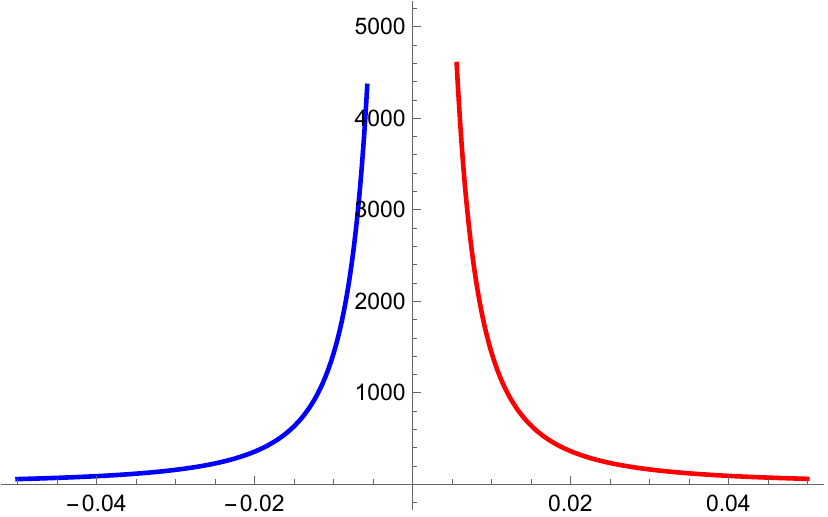}};
\node[below=of img, node distance=0cm, xshift=3cm, yshift=1.2cm]{$\mathcal{R}$};
\node[left=of img, node distance=0cm, rotate=0, anchor=center,xshift=4.5cm, yshift=2.5cm] {$F$};
\end{tikzpicture}
\caption{$\alpha = \frac{1}{4}$}
\label{fig:F14full}
\end{subfigure}

\begin{subfigure}[t]{0.45\textwidth}
\begin{tikzpicture}
\node (img) {\includegraphics[width=\textwidth]{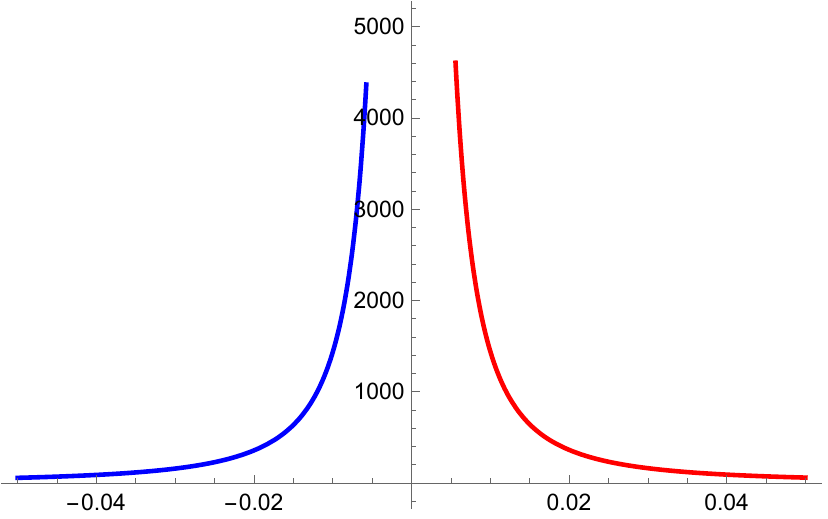}};
\node[below=of img, node distance=0cm, xshift=3cm, yshift=1.2cm]{$\mathcal{R}$};
\node[left=of img, node distance=0cm, rotate=0, anchor=center,xshift=4.5cm, yshift=2.5cm] {$F$};
\end{tikzpicture}
\caption{IHQCD : $\alpha = \frac{1}{2}$}
\label{fig:F12full}
\end{subfigure}
\begin{subfigure}[t]{0.45\textwidth}
\begin{tikzpicture}
\node (img) {\includegraphics[width=\textwidth]{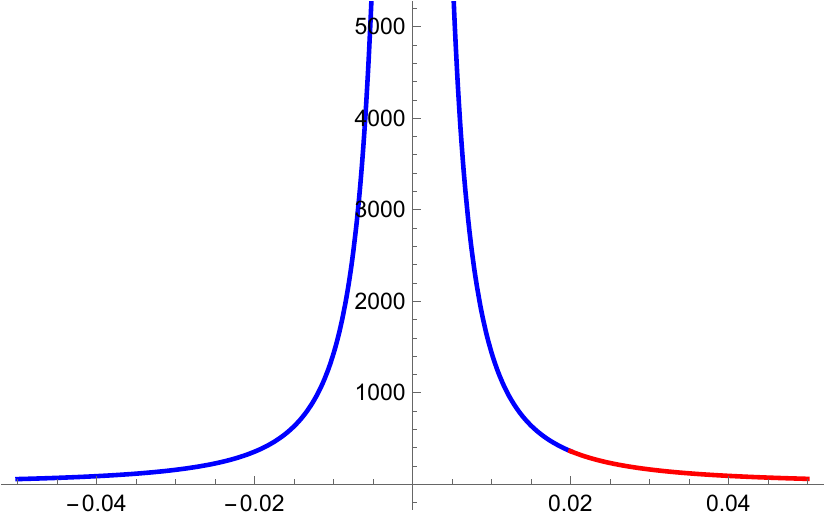}};
\node[below=of img, node distance=0cm, xshift=3cm, yshift=1.2cm]{$\mathcal{R}$};
\node[left=of img, node distance=0cm, rotate=0, anchor=center,xshift=4.5cm, yshift=2.5cm] {$F$};
\node(leg)[below=of img, node distance=0cm, xshift=3cm, yshift=5.2cm]{${\color{blue}\bullet}$ Type I/II};
\node[below=of img, node distance=0cm, xshift=2.9cm, yshift=4.6cm]{${\color{red}\bullet}$ Type III};
\end{tikzpicture}
\caption{$\alpha = \frac{3}{4}$}
\label{fig:F34full}
\end{subfigure}
\caption{Renormalized free energy at $b=b_c$ with $A_{ct} = -0.3$, $B_{ct} = -0.0662$ and $C_{ct}=0$}
\label{fig:freeenergy}
\end{figure}
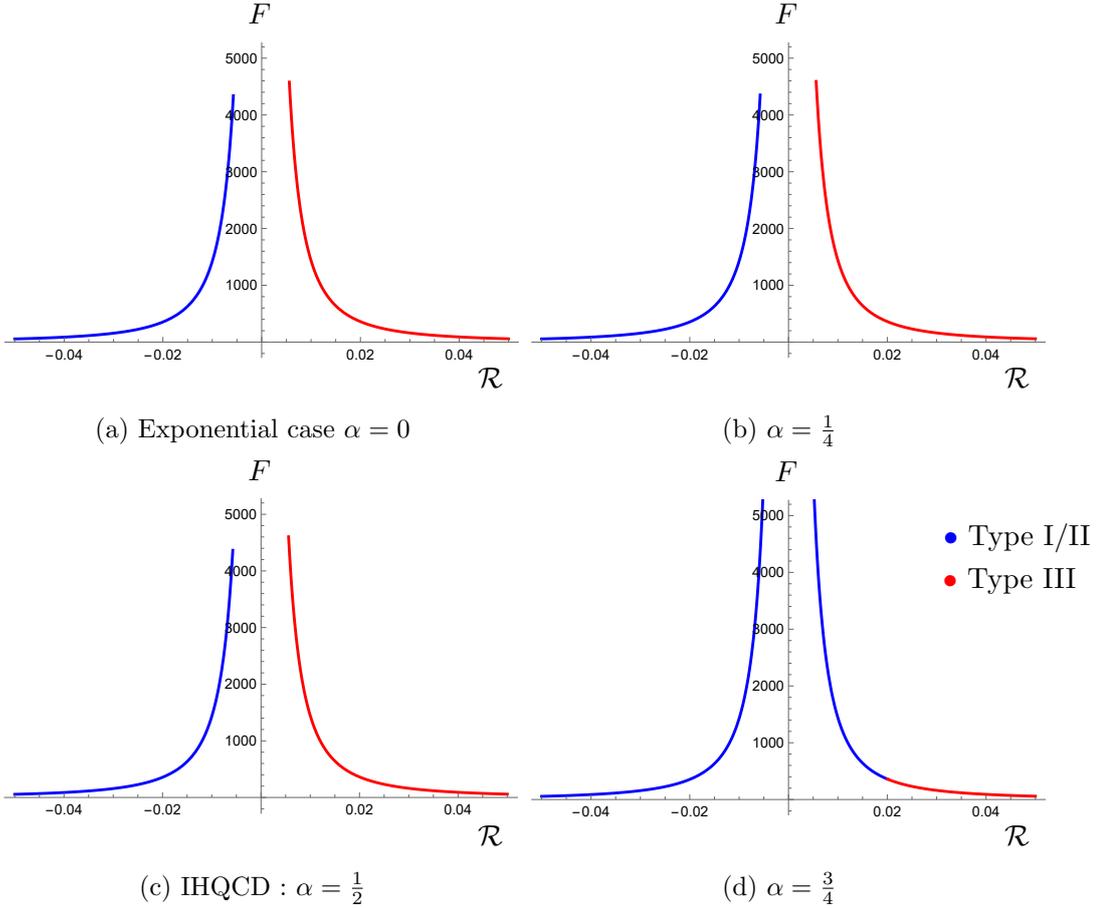

In figure \ref{fig:freeenergy} we plot the numerically obtained renormalized free energy for $d=4$, $b=b_c$ and for several values of $\a$,  as a function of $\mathcal{R}$, for a specific choice of the finite counterterms, $A_{ct} = -0.3$, $B_{ct} = -0.0662$ and $C_{ct}=0$. It should be noted that the free energy diverges as $\mathcal{R}\to 0$, both from positive and negative values. The divergences in the positive case was explained above in terms of the volume of the sphere becoming infinite in the limit. It is a large volume divergence.

In the negative curvature case, the manifold can have always infinite volume like AdS. It may also have finite volume like compact Schottky manifolds.
 This divergence was explained in \cite{AdS2}.

The curves {presented in figure \ref{fig:freeenergy}}
correspond to computing the renormalized free energy along a vertical slice of the space of solutions shown in figure \ref{fig:Ralpha}. $\mathcal{R}_{c+}$ therefore corresponds to the curvature of the change of color. Note that as expected from figure \ref{fig:Ralpha}, $\alpha = \frac{3}{4}$ is the only value for which $\mathcal{R}_{c+}$ occurs at non-zero value.

\begin{figure}[H]
\centering
\begin{subfigure}[t]{0.45\textwidth}
\begin{tikzpicture}
\node (img) {\includegraphics[width=\textwidth]{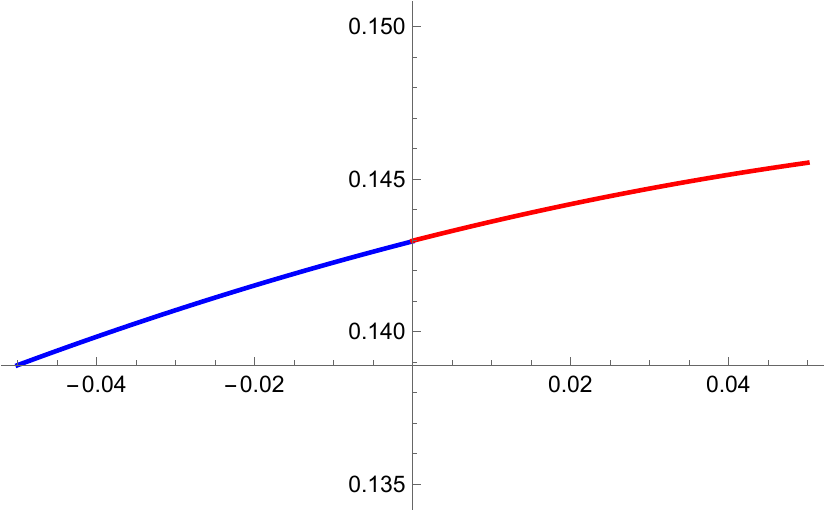}};
\node[below=of img, node distance=0cm, xshift=3cm, yshift=1.2cm]{$\mathcal{R}$};
\node[left=of img, node distance=0cm, rotate=0, anchor=center,xshift=4.5cm, yshift=2.5cm] {$f$};
\end{tikzpicture}
\caption{Exponential case $\alpha=0$}
\label{fig:F0vol}
\end{subfigure}
\begin{subfigure}[t]{0.45\textwidth}
\begin{tikzpicture}
\node (img) {\includegraphics[width=\textwidth]{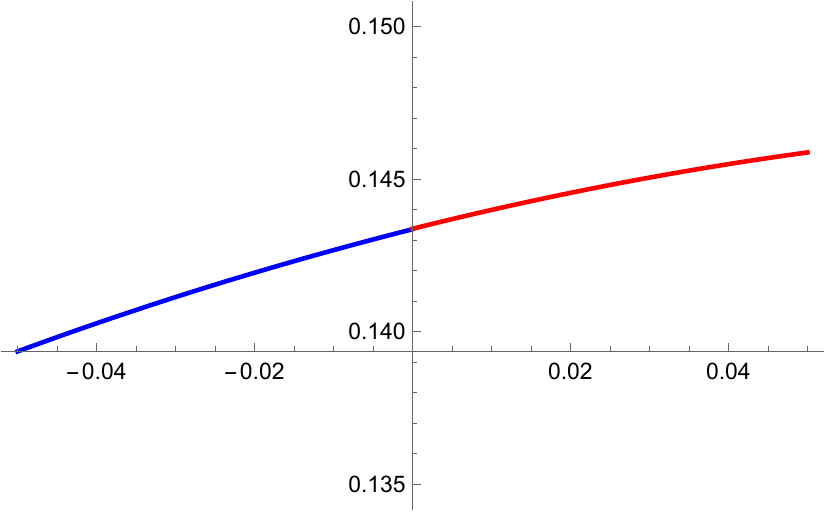}};
\node[below=of img, node distance=0cm, xshift=3cm, yshift=1.2cm]{$\mathcal{R}$};
\node[left=of img, node distance=0cm, rotate=0, anchor=center,xshift=4.5cm, yshift=2.5cm] {$f$};
\end{tikzpicture}
\caption{$\alpha = \frac{1}{4}$}
\label{fig:F14vol}
\end{subfigure}

\begin{subfigure}[t]{0.45\textwidth}
\begin{tikzpicture}
\node (img) {\includegraphics[width=\textwidth]{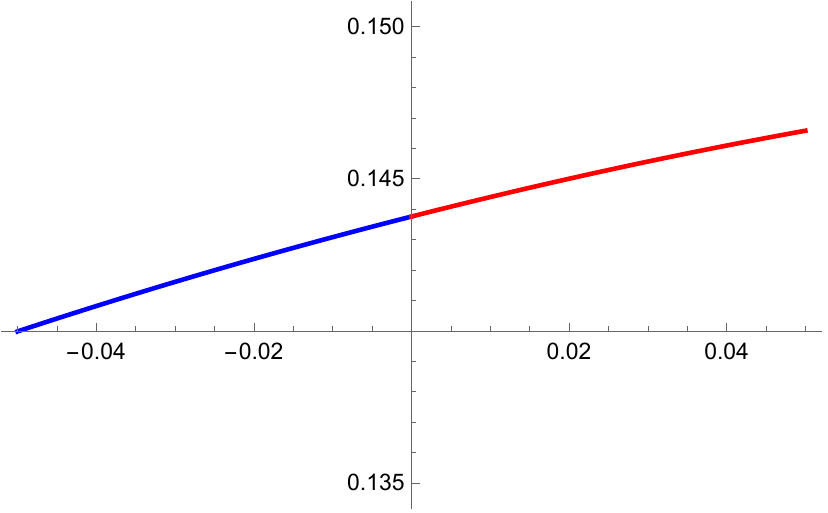}};
\node[below=of img, node distance=0cm, xshift=3cm, yshift=1.2cm]{$\mathcal{R}$};
\node[left=of img, node distance=0cm, rotate=0, anchor=center,xshift=4.5cm, yshift=2.5cm] {$f$};
\end{tikzpicture}
\caption{IHQCD : $\alpha = \frac{1}{2}$}
\label{fig:F12vol}
\end{subfigure}
\begin{subfigure}[t]{0.45\textwidth}
\begin{tikzpicture}
\node (img) {\includegraphics[width=\textwidth]{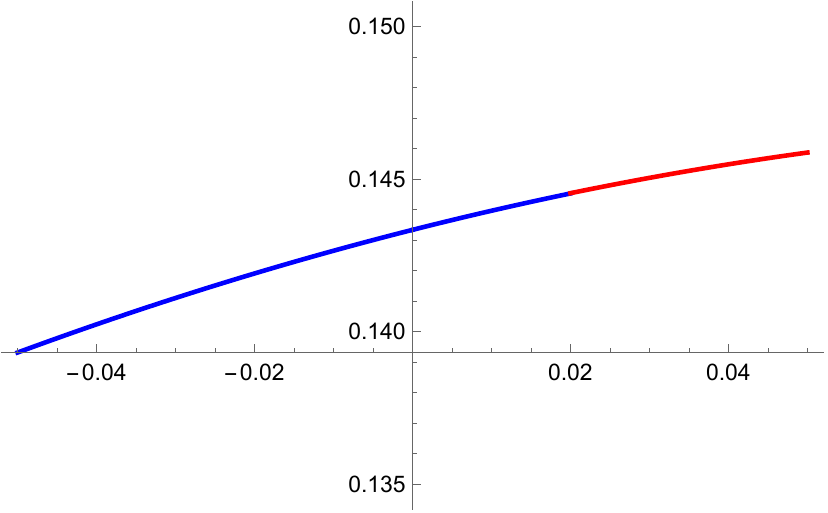}};
\node[below=of img, node distance=0cm, xshift=3cm, yshift=1.2cm]{$\mathcal{R}$};
\node[left=of img, node distance=0cm, rotate=0, anchor=center,xshift=4.5cm, yshift=2.5cm] {$f$};
\node(leg)[below=of img, node distance=0cm, xshift=2cm, yshift=5.2cm]{${\color{blue}\bullet}$ Type I/II};
\node[below=of img, node distance=0cm, xshift=1.9cm, yshift=4.6cm]{${\color{red}\bullet}$ Type III};
\end{tikzpicture}
\caption{$\alpha = \frac{3}{4}$}
\label{fig:F34vol}
\end{subfigure}
\caption{Renormalized free energy per unit volume at $b=b_c$ with $A_{ct} = -0.3$, $B_{ct} = -0.0662$ and $C_{ct}=0$}
\label{fig:freeenergyvol}
\end{figure}
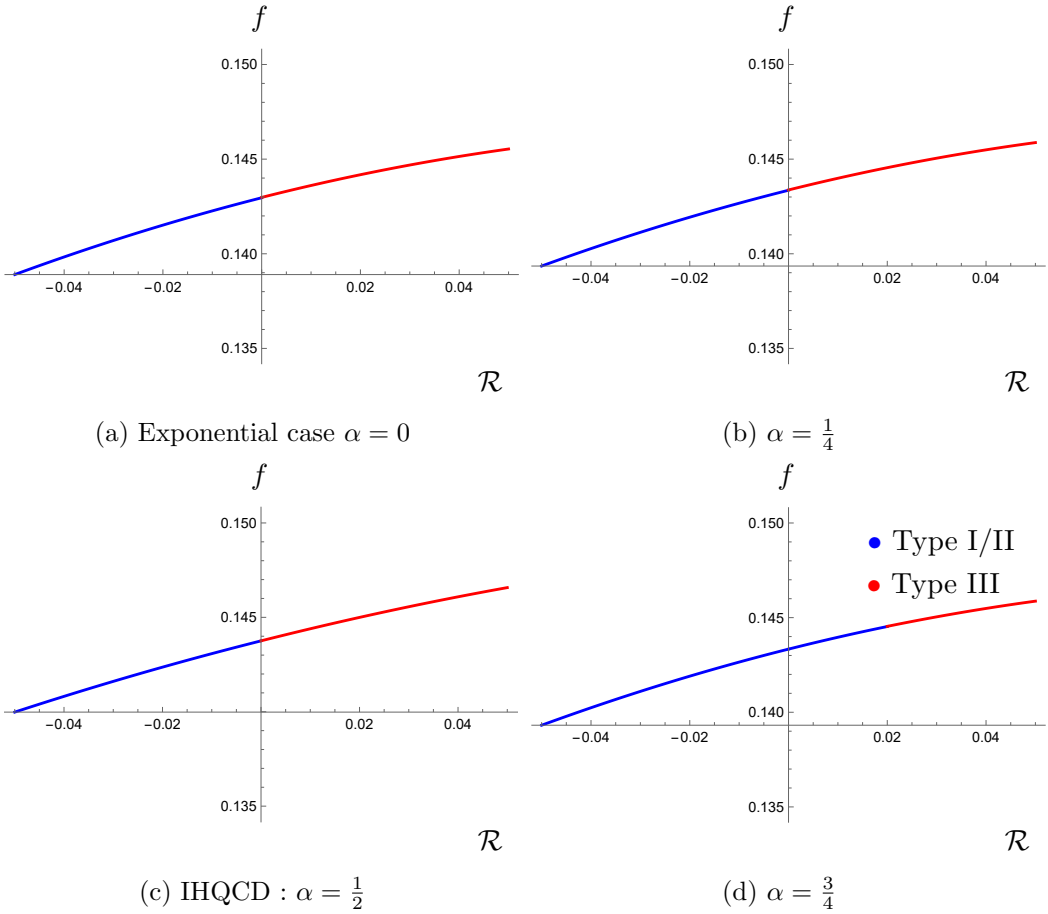

We obtain the free energy density $f$ and entropy density $s$  (per unit volume) by multiplying $F,S_{E}$ by $\mathcal{R}^2$.
The free energy per unit volume is plotted in figure \ref{fig:freeenergyvol} for $\a=0,{1\over 4},{1\over 2},{3\over 4}$. In the first three cases, the two distinct saddle points (namely type I/II and type III solutions) join  at $\mathcal{R}=0$. In the $\a>{1\over 2}$ case this happens at $\mathcal{R}_{c+}>0$.

All curves seem continuous at the transition point between blue and red, ie. type I/II and type III saddle points.

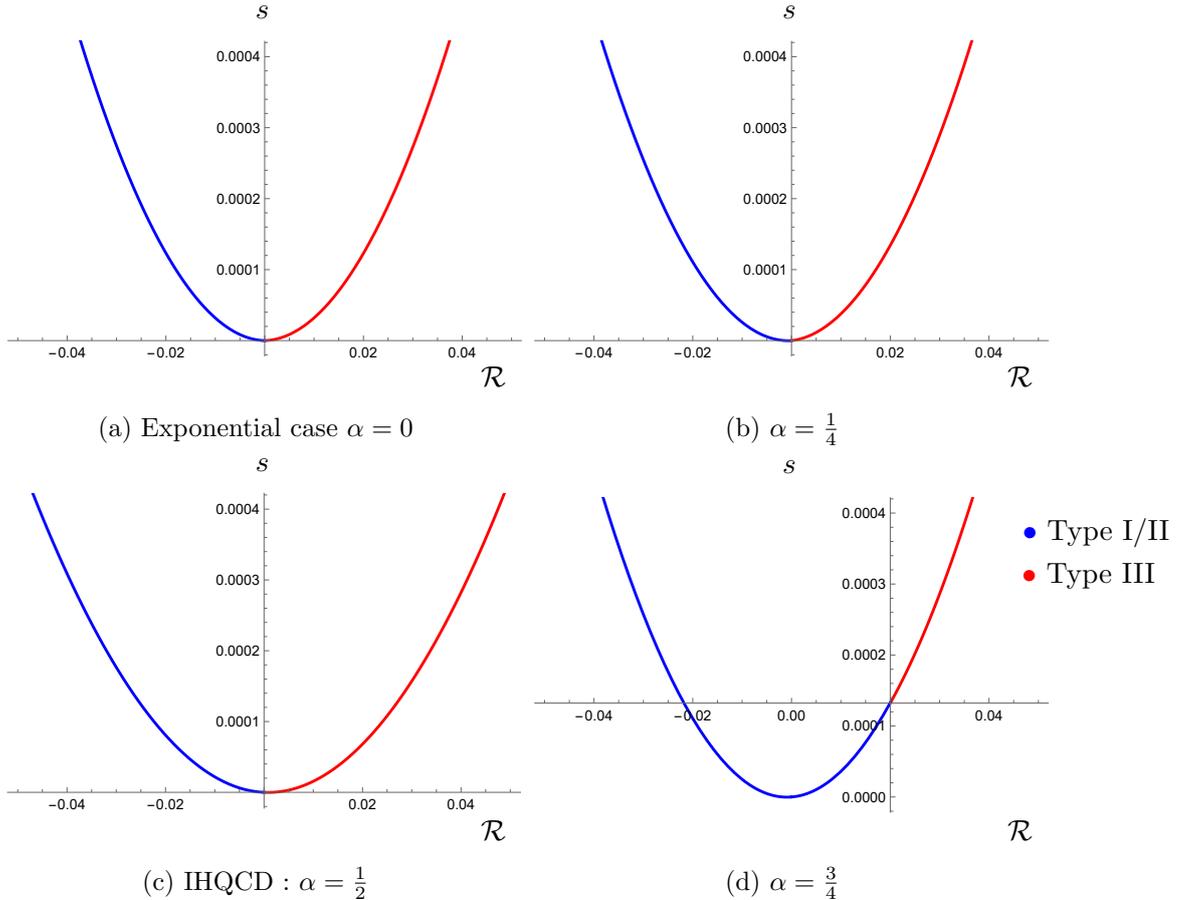
\begin{figure}[H]
\centering
\begin{subfigure}[t]{0.45\textwidth}
\begin{tikzpicture}
\node (img) {\includegraphics[width=\textwidth]{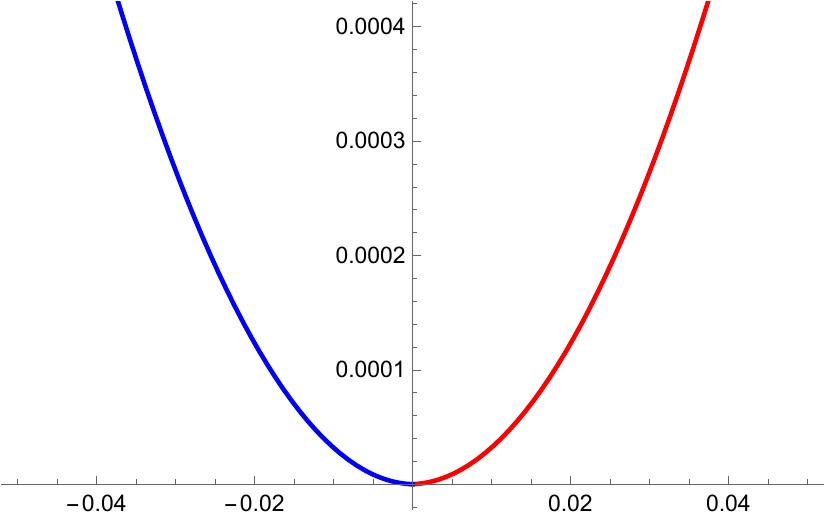}};
\node[below=of img, node distance=0cm, xshift=3cm, yshift=1.2cm]{$\mathcal{R}$};
\node[left=of img, node distance=0cm, rotate=0, anchor=center,xshift=4.5cm, yshift=2.5cm] {$s$};
\end{tikzpicture}
\caption{Exponential case $\alpha=0$}
\label{fig:s0vol}
\end{subfigure}
\begin{subfigure}[t]{0.45\textwidth}
\begin{tikzpicture}
\node (img) {\includegraphics[width=\textwidth]{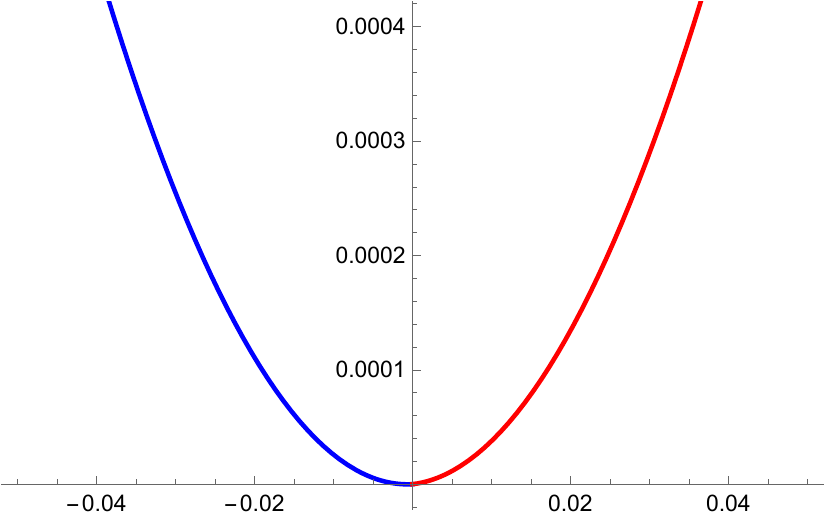}};
\node[below=of img, node distance=0cm, xshift=3cm, yshift=1.2cm]{$\mathcal{R}$};
\node[left=of img, node distance=0cm, rotate=0, anchor=center,xshift=4.5cm, yshift=2.5cm] {$s$};
\end{tikzpicture}
\caption{$\alpha = \frac{1}{4}$}
\label{fig:s14vol}
\end{subfigure}

\begin{subfigure}[t]{0.45\textwidth}
\begin{tikzpicture}
\node (img) {\includegraphics[width=\textwidth]{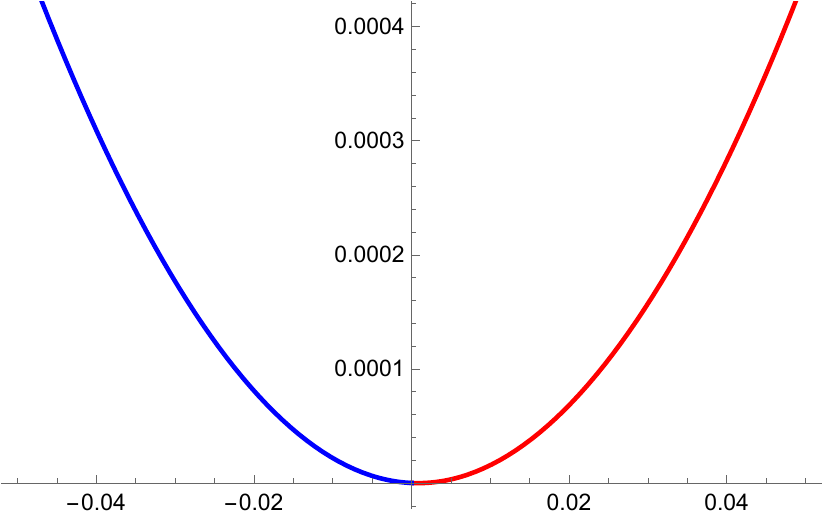}};
\node[below=of img, node distance=0cm, xshift=3cm, yshift=1.2cm]{$\mathcal{R}$};
\node[left=of img, node distance=0cm, rotate=0, anchor=center,xshift=4.5cm, yshift=2.5cm] {$s$};
\end{tikzpicture}
\caption{IHQCD : $\alpha = \frac{1}{2}$}
\label{fig:s12vol}
\end{subfigure}
\begin{subfigure}[t]{0.45\textwidth}
\begin{tikzpicture}
\node (img) {\includegraphics[width=\textwidth]{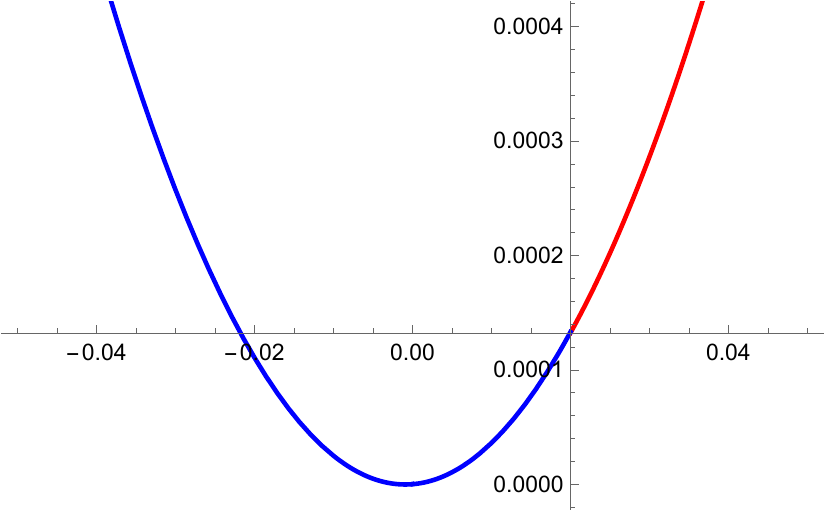}};
\node[below=of img, node distance=0cm, xshift=3cm, yshift=1.2cm]{$\mathcal{R}$};
\node[left=of img, node distance=0cm, rotate=0, anchor=center,xshift=4.5cm, yshift=2.5cm] {$s$};
\node(leg)[below=of img, node distance=0cm, xshift=4cm, yshift=5.2cm]{${\color{blue}\bullet}$ Type I/II};
\node[below=of img, node distance=0cm, xshift=3.9cm, yshift=4.6cm]{${\color{red}\bullet}$ Type III};
\end{tikzpicture}
\caption{$\alpha = \frac{3}{4}$}
\label{fig:s34vol}
\end{subfigure}
\caption{Renormalized entropy per unit volume at $b=b_c$ with $A_{ct} = -0.3$, $B_{ct} = -0.0662$ and $C_{ct}=0$}
\label{fig:entropyvol}
\end{figure}

The entropy  per unit volume is plotted in figure \ref{fig:entropyvol} for $\a=0,{1\over 4},{1\over 2},{3\over 4}$. { In all cases  it looks continuous at the transition point between the two different saddle point types.  Further numerical or analytical tools are required to check whether higher derivatives are discontinuous.
Already, as argued in the previous section, based on the results of \cite{Jani}, as $b\to {b_c}^+$ the order of the transition seems to move to $\infty$. It seems that either much better numerics are needed , or an analytical method that can provide the leading asymptotics of $F$ as $\mathcal{R}\to 0^{\pm}$.}

\section{Discussion and open questions}

In this work we have  obtained the asymptotic form of the solutions and constructed the phase diagram at finite boundary curvature for a particular class of confining holographic models which contains the phenomenologically relevant case of IHQCD. These models are five-dimensional Einstein-Dilaton theories in which the dilaton potential is characterized by a power-law-corrected exponential  asymptotics,
\be \label{conc1}
V \sim -\varphi^{\alpha} e^{2b_c \varphi} \qquad \varphi \to +\infty
\ee
and are at the ``boundary'' of a larger class of models which are dual to confining QFTs (for which $b>b_c$).

Because of the ``borderline'' nature of these models, the analysis of the continuous deformations of asymptotic solutions cannot be handled by a simple  linear analysis, as it was the case in previous work \cite{Jani}. To find the correct large-$\varphi$ expansion, we resorted to techniques from bifurcation theory, after reformulating the problem in terms of a (constrained)  dynamical system.

This is a novel (to our knowledge) application of dynamical system theory in holography. Although  the language of dynamical systems has appeared before in holography in relation to other problems  (see e.g. the recent works \cite{Marconnet:2022fmx,Andriot:2023wvg,DeClerck:2023fax}),  to our knowledge this is the first application of central manifold theory in this context. In our case, we used this method to analyze the large-$\varphi$ behavior of the solutions,  but there are no obstructions in using the same techniques for full RG-flows, as long as one introduces an auxiliary constrained variable  on top of the two ``physical'' superpotentials $W$ and $S$. This formulation could be useful to understand the structure of the solution space of more general holographic theories (e.g. multi-field theories, where the canonical formulation  can lead to interesting results \cite{multi-field}).

In the dynamical system language, the border between confining and non-confining regimes of exponential asymptotics appears as a bifurcation. Identifying the  central manifold at the non-linear level, we were able to systematically classify the structure of asymptotic solutions and identify their continuous deformation parameters, which map to the  ``dimensionless boundary curvature''  parameter of the dual QFT. This analytical control on the asymptotics of the solutions allowed us to numerically integrate the full holographic RG flow equations from the IR to the UV, to read-off the physical parameters of the  boundary QFT (boundary curvature and coupling) and to construct the phase diagram of the theory for $0< \alpha <1$ (which includes the IHQCD value $\alpha = 1/2$. Values of $\alpha$ outside of this interval require going to higher non-linear order, which is  eventually left for future work.

In the negative curvature regime,  we find similar results to those discussed in \cite{AdS2}: in this case there are no type-III solution (with a regular endpoint) but only solutions running to $\varphi \to \infty$. In principle, other classes of solutions may exist,  like the ones  discussed in \cite{AdS2} which connect different UV regions. However in this work  we only focused on solutions with a single boundary.

In the positive curvature regime, the  phase diagram (as a function of $\alpha$ and of the dimensionless boundary curvature) exhibits some expected features, but also some surprises. Like in the case of non-critical exponential $b>b_c$,   we find  solutions which end  regularly in the IR  (type III) and solutions which extend to the boundary of field space ($\varphi \to \infty$), and a phase transition between them at a critical curvature ${\cal R}_c$, which is finite as long as $\alpha >1/2$.  Similar results were obtained in \cite{Jani}  in the non-critical case for confining models ($b>b_c$).   When it exists, the phase transition is higher than first order, as in the case for non-critical $b$ close to $b_c$.

The surprise is that the phase transition ceases to exist at $\alpha = 1/2$,  still within the confining range (which extends down to $\alpha = 0$). Therefore,  in the range $0< \alpha \leq 1/2$ we have a class of  holographic models, which notably includes the phenomenologically interesting case of IHQCD, which are confining in flat space \cite{IHQCD}, display a finite-temperature phase transition  \cite{thermo}, but {\em not} a finite-curvature phase transition. This is unlike all confining holographic models that have been studied so far at finite curvature: all of them display a curvature-driven   transition at finite curvature between two different phases with distinct geometries, see e.g. \cite{Marolf:2010tg,Blackman:2011in,Jani}.

It is an interesting question, and one to be explored, whether this behavior of IHQCD may be interesting for phenomenology (e.g. in the early universe) or may be somehow checked using lattice field theory. One problem that needs to be answered to addressed the previous question is the following: it is not clear  how   the two geometrical phases (ending at finite $\varphi$ or extending to infinity) are told apart from the point of view of the dual QFT, as there is no obvious order parameter and both phases are gapped (by either the confinement scale or the curvature scale). It is therefore unclear how to characterize the phase transition  (or lack thereof) in terms of a physical QFT observable.

One piece of information which may shed light on the questions above (and may be interesting for phenomenology) is to compute the spectrum of excitations of the theory in the various phases, and how it changes as one varies the curvature. For this, one has to develop a general formalism of studying the perturbations about curved RG-flow geometry, which is the subject of  work in progress.

\section*{Acknowledgements}
\addcontentsline{toc}{section}{Acknowledgements}

We thank I. Basile, A. Ghodsi, C. Hoyos, D. Lust, D. Mateos, C. Montella, S. Morales-Tejera, E. Pr\'eau, J. Kastikainen, C. Rosen and J. Subils, for useful discussions.

 This
work is partially supported by the European MSCA grant HORIZON-MSCA-2022-PF-
01-01 ``BlackHoleChaos" No.101105116 and by the H.F.R.I call ``Basic research Financing
(Horizontal support of all Sciences)" under the National Recovery and Resilience Plan
``Greece 2.0" funded by the European Union  NextGenerationEU (H.F.R.I. Project Number:
15384.), by ANR grant ``XtremeHolo" (ANR project n.284452) and by the In2p3 grant
``Extreme Dynamics" .

\newpage

\appendix

\begin{appendix}
\renewcommand{\theequation}{\thesection.\arabic{equation}}
\addcontentsline{toc}{section}{Appendices}
\section*{APPENDIX}

\section{Type III asymptotic solutions}\label{III}

In this appendix we review the asymptotics associated with solutions ending at a finite value of $\f$, that we denote by $\f_0$. These solutions exist only for constant positive curvature (sphere or de Sitter). At this end-point, the constant curvature manifold's size  shrinks to zero, in a regular way, so that locally the geometry is smooth. Such asymptotics were described in detail in \cite{C}.

These are  asymptotic solutions of the equations \eqref{eqn:w}, (\ref{eqn:s}) where $\varphi$ asymptotes to a finite  $\varphi(u_0) = \varphi_0 < \infty$. In this case $ \varphi \rightarrow \varphi_{0}^{-} $ as $u\rightarrow u_0^{-}$ and the appropriate  ansatz is, \cite{C},
\begin{equation}
	S(\varphi) = \sqrt{\varphi_{0}-\varphi}\,\Bigl[S_0 + S_1\,\sqrt{\varphi_{0}-\varphi}+ S_2\,( \varphi_{0}-\varphi) + \mathcal{O}(\varphi_{0}-\varphi)^{3\slash 2}\Bigr]\,.
	\label{StypeA}
\end{equation}
\begin{equation}
	W(\varphi) = \frac{1}{\sqrt{\varphi_{0}-\varphi}}\Bigl[W_0 + W_1\,\sqrt{\varphi_{0}-\varphi}+ W_2\,( \varphi_{0}-\varphi)+ \mathcal{O}(\varphi_{0}-\varphi)^{3\slash 2}\Bigr]\,,
	\label{WexptypeA}
\end{equation}
Substituting this expansion to the equations, we obtain\footnote{There is also a second solution $ S_0 = \sqrt{-2V'(\varphi_{0})} $ which gives a finite size $ A(u_0)\neq 0 $ for the $S^{d}$ at the IR endpoint (see equation \eqref{eq:Au_type_A_lower_app} below).  We do not consider this solution, but it is relevant for hyperbolically sliced flows, where it would correspond to a bounce (see \cite{AdS1}).}
\begin{equation}
	S_0 = \sqrt{-\frac{2V'(\varphi_{0})}{d+1}}\,, \quad S_1 = 0\,, \quad S_2 = \frac{1}{2(d+3)}\biggl(3V''(\varphi_{0})+\frac{2V(\varphi_0)}{d-1}\biggr)\frac{1}{S_0}\,.
 \label{eq:ScoefficientsA}
\end{equation}
\begin{equation}
	W_0 = (d-1)\,S_0\,, \quad W_1 = 0\,, \quad W_2 = \frac{d-1}{2(d+3)}\biggl[\frac{2(d+4)}{d(d-1)}\,V(\varphi_{0})-V''(\varphi_{0})\biggr]\,.
	\label{W0expressionapp}
\end{equation}
Since $ S(\varphi) = \dot{\varphi} $, the expansion \eqref{StypeA} implies
\begin{equation}
	\varphi(u) = \varphi_{0}+\varphi_1\,(u_0-u)^{2} + \varphi_2\,(u_0-u)^{4}  + \mathcal{O}(u_0-u)^{5}\,, \quad u\rightarrow u_0^{-}\,,
 \label{eq:varphiu_typeA_expansion}
\end{equation}
with coefficients
\begin{equation}
	\varphi_1 = -\frac{S_0^{2}}{4}\,, \quad \varphi_{2} = \frac{S_0^{3}S_2}{40}\,.
\end{equation}
Substituting \eqref{eq:varphiu_typeA_expansion} to \eqref{WexptypeA} gives an expansion for $W$ in powers of $u_0-u$. Using the definition, $ W(\varphi) = -2(d-1)\,\dot{A} $, we obtain the expansion
\begin{equation}
	A(u) = \frac{W_0}{(d-1)\,S_0}\log{(u_0-u)} + A_0 + \frac{5S_0W_2+S_2W_0}{40(d-1)}\,(u_0-u)^{2} + \mathcal{O}(u_0-u)^{3}\,,
	\label{eq:Au_type_A_lower_app}
\end{equation}
where $ A_0 $ is an integration constant. Substituting the explicit values \eqref{W0expressionapp} for $W_0$ gives (focusing only on the leading term)
\begin{equation}
	A(u) = \log{\biggl(\frac{u_0-u}{\alpha}\biggr)} + \mathcal{O}(u_0-u)^{2}\,.
\end{equation}
where we have fixed $ A_0\equiv -\log{\alpha} $ by requiring the absence of a conical singularity in the metric at the IR endpoint.

For the $ T $ function  we obtain from\eqref{eqn:T}  the expansion
\begin{align}
	T(\varphi) = \frac{1}{\varphi_{0}-\varphi}\Bigl[T_0 + T_1\,\sqrt{\varphi_{0}-\varphi}+ \mathcal{O}( \varphi_{0}-\varphi)\Bigr]\,,\quad \varphi \rightarrow \varphi_{0}^{-}\,,
\end{align}
with the coefficients\footnote{The $ T_0 $ in \cite{C} has incorrectly an extra factor of $ (d+1)^{-1} $.}
\begin{equation}
	T_0 = \frac{d}{4}\, S_0 W_0 = \frac{d(d-1)}{4}\,S_0^{2}>0\,, \quad T_1 = 0\,.
	\label{T0U0expressionapp}
\end{equation}

\section{The flat-sliced solutions with critical asymptotics $b = b_c$}
\label{sec:A}

In this appendix, we give a geometrical interpretation of the flat-sliced case in terms of the dynamical system. We impose $T=0$, such that equation (\ref{eqn:solveT}) becomes:
\begin{equation}
        W' = S.
\end{equation}
We can use equation \ref{eqn:algT} and reexpress it in terms of the $\Tilde{S}$ and $\Tilde{W}$. We place ourselves in the critical exponential case $b=b_c$. We choose the solution with $\Tilde{W}>0$, and we obtain:
\begin{equation}
    \Tilde{W} = \sqrt{\Tilde{S}^2+2} \sqrt{\frac{2(d-1)}{d}} \label{Fixcurv}
\end{equation}
This equation can be interpreted as defining a flat submanifold of the phase space $(\Tilde{W}, \Tilde{S})$. We can then pullback the dynamical system on this submanifold, which gives:
\begin{equation}
    \Tilde{S}' = -b_c \Tilde{S} + b_G \sqrt{\Tilde{S}^2+2} - \frac{2b_c}{\Tilde{S}}
\end{equation}
Though this is separable and could be integrated, we choose to expand it. After repeating the change of variables (\ref{defdelta}) from $\Tilde{S}$ to $\delta \Tilde{S}$ and expanding $\delta \Tilde{S}'$, we obtain:
\begin{equation}
    \delta \Tilde{S}' = \sqrt{\frac{d-1}{2}} \delta \Tilde{S} + \mathcal{O}(\delta \Tilde{S}^2).
\end{equation}
This equation corresponds to expanding $\delta \Tilde{S}'$ along an unstable direction. Interestingly, the coefficient is exactly $\lambda_y$, so equation (\ref{Fixcurv}) projects the differential equation on the unstable manifold of the critical point, so that the flat submanifold and the unstable manifold coincide to first order. This is expected, as this solution exists for only one curvature of the boundary, $\mathcal{R} = 0$, therefore the solution should have no deformation parameter. This result was indeed already obtained in the study of the flat case in \cite{multirg}.

Though the calculations of this appendix were performed in the exponential case, the conclusion is more general and a similar situation occurs for more complicated potential asymptotics.

\section{Explicit solution of the third order differential equation in the exponential case}
\label{sec:B}

In this appendix we solve analytically the differential equation expanded to  third order in the IR in the exponential case, in order to derive the third order expansion of the solution obtained in equation (\ref{asympexpS}).
Starting from equation (\ref{eqn:deltasp}):
\begin{equation}
    \delta \Tilde{S}' = -\frac{2(d-1)}{d} \delta \Tilde{S}^2 + \frac{4(d-1) \sqrt{2(d-1)}}{d^2} \delta \Tilde{S}^3 + \mathcal{O}(\delta \Tilde{S}^4).
\end{equation}

We first solve the second order expansion of this equation. Neglecting the $\delta \Tilde{S}^3$ term, we obtain:
\begin{equation}
    \frac{d}{2(d-1) \delta \Tilde{S}(\varphi)} = \varphi - \f_\infty - C_0,
    \label{asympwrongeq}
\end{equation}
And the expansion we obtain for $\delta \Tilde{S}$ is:
\begin{equation}
    \delta \Tilde{S}(\varphi) = \frac{d}{2(d-1)} \frac{1}{\varphi} + \left( \frac{d}{2(d-1)}\right)^2\frac{\f_\infty + C_0}{\varphi^2} + \mathcal{O}(\varphi^{-3}) \label{eqn:asympswrong}
\end{equation}
The first subleading term we have included in this solution cannot be trusted: $\delta \Tilde{S}'$ obtains a contribution $\frac{\f_\infty + C_0}{\varphi^3}$ for this term, which is of the same order as the leading contribution from the $\mathcal{O}(\delta \Tilde{S}^3)$ in equation (\ref{eqn:deltasp}), which we have neglected to obtain (\ref{eqn:asympswrong}). Therefore, the asymptotics (\ref{eqn:asympswrong}) is {\em incorrect}, and the $\delta \Tilde{S}^3$ terms in equation (\ref{eqn:deltasp}) need to be taken into account, in order to obtain the correct asymptotics to order $\frac{1}{\varphi^2}$. We obtain instead:
\begin{equation}
        \sqrt{\frac{2}{d-1}} \log\left(\left| -2\frac{d-1}{d \delta \Tilde{S}(\varphi)} + \frac{4 (d-1) \sqrt{2(d-1)}}{d^2}\right|\right) + \frac{d}{2(d-1) \delta \Tilde{S}(\varphi)} = \varphi - \f_\infty - C_0
\end{equation}
we perform the following change of variables:
\begin{equation}
    X = 2\frac{d-1}{d \delta \Tilde{S}(\varphi)} - \frac{4 (d-1) \sqrt{2(d-1)}}{d^2}
\end{equation}
Which leads to:
\begin{equation}
    \sqrt{\frac{2}{d-1}} \log\left(\left| X \right|\right) + \frac{d^2}{4(d-1)^2} X + \left(\sqrt{\frac{2}{d-1}} - \varphi + \f_\infty + C_0\right) = 0
\end{equation}
{In this appendix we solve only the case $X>0$. Sufficiently close to the critical point, this corresponds to $\delta \Tilde{S}>0$.} Redefining $e^z = X$, we obtain
\begin{equation}
     \frac{d^2}{4(d-1)^2} e^z  +  \sqrt{\frac{2}{d-1}} z + \left(\sqrt{\frac{2}{d-1}} - \varphi +  \f_\infty + C_0 \right) = 0
     \label{eqBlamb}
\end{equation}
This is  an equation of the type $$A e^z + B z + C = 0\;,$$ with
\begin{align}
    A &= \frac{d^2}{4(d-1)^2} & B &= \sqrt{\frac{2}{d-1}} & C&=\left(\sqrt{\frac{2}{d-1}} - \varphi + \f_\infty + C_0\right)
\end{align}
We define the Lambert function, $W$, as the solution to the equation:
\begin{align}
    z &= we^w  & W(z) &= w
\end{align}
This function is multivalued on $]\frac{-1}{e}, 0[$. It is uniquely defined on $\left\{\frac{-1}{e}\right\} \cup [0, +\infty[$, and it is not defined on $]-\infty, \frac{-1}{e}[$. To determine the number of solutions of this equation, we use a discriminant:
\begin{equation}
    \Delta = \frac{A}{B}e^{-\frac{C}{B}} = \frac{d^2}{4(d-1)\sqrt{2(d-1)} } e^{\sqrt{\frac{d-1}{2}}(\varphi - \f_\infty - C_0) - 1}
\end{equation}
In our case, it is always positive, which implies that the equation has exactly one solution. It is given by:
\begin{equation}
    z = -W(\Delta) - \frac{C}{B} = -W\left(\frac{d^2}{4(d-1)\sqrt{2(d-1)} } e^{\sqrt{\frac{d-1}{2}}(\varphi - \f_\infty - C_0) - 1}\right) + \sqrt{\frac{d-1}{2}}(\varphi - \f_\infty - C_0) - 1
\end{equation}
Solving for $\delta \Tilde{S}$ in $z$, we obtain:
\begin{align}
   {2(d-1)\over d \delta \Tilde{S}} &=  \exp\left[-W\left(\frac{d^2}{4(d-1)\sqrt{2(d-1)} } e^{\sqrt{\frac{d-1}{2}}(\varphi - \f_\infty - C_0) - 1}\right) \right.   \nonumber \\
    &\left.+ \sqrt{\frac{d-1}{2}}(\varphi - \f_\infty - C_0) - 1\right]  + \frac{4(d-1)\sqrt{2(d-1)}}{d^2}
\end{align}
{Inverting this equation and expanding $\delta \Tilde{S}$ as $\varphi \to +\infty$ we obtain (\ref{asympexpS}). The case considered in this appendix, $X>0$, provides an unique solution, which has $\delta \Tilde{S} \to 0$ in the IR. The other case, $X<0$, corresponds to values $\delta \Tilde{S} <0$ and gives an equation analogous to(\ref{eqBlamb}) but with a negative discriminant, for which the Lambert $W$ function is multivalued. In this case, there is a branch of the function for which $\delta \Tilde{S}$ does not go to zero. This dependence of the solution on the sign of integration constant illustrates the semistability phenomenon discussed in \ref{sec343}}

\section{Analytical solution of the exponential critical case in $d=4$}
\label{sec:C}

In this appendix we derive an exact solution to the case $V = -V_{\infty} e^{2 b_c \varphi}$ in $d=4$.

With $d=4$, equation (\ref{eqn:derWS}) shows that the second and third derivatives of $W$ with respect to $S$ vanish. In fact, the first order expansion of $W$ gives an exact solution to (\ref{eqn:traj}). This solution corresponds to replacing $d=4$ in the expressions of Section 3:
\begin{equation}
    \Tilde{W} = 2 - \sqrt{\frac{3}{2}} \delta \Tilde{S}
\end{equation}
It is possible to reinject this expression directly in (\ref{eqn:Stilde}), to obtain the differential equation
\begin{equation}
    \delta \Tilde{S}' = \frac{-1}{1 + \sqrt{\frac{3}{2}}\delta \Tilde{S}} + 1 - \sqrt{\frac{3}{2}}\delta \Tilde{S}
\end{equation}
This differential equation is separable. It can be integrated into the algebraic equation depending on an integration constant $C_0$
\begin{equation}
    \sqrt{\frac{2}{3}} \log(\delta \Tilde{S}) - \frac{2}{3\delta \Tilde{S}} = -\varphi + \f_\infty + C_0
\end{equation}
This algebraic equation can be solved using a method similar to appendix \ref{sec:B}. Indeed, defining $Y = - \log(\delta \Tilde{S})$ we obtain
\begin{equation}
    \frac{2}{3} e^Y = \varphi -\f_\infty - C_0 - \sqrt{\frac{2}{3}} Y
\end{equation}
To solve it, we use the Lambert $W$ function introduced in the previous appendix
\begin{equation}
    \sqrt{\frac{2}{3}} e^{\sqrt{\frac{3}{2}}(\varphi - \f_\infty - C_0)} = \left( \sqrt{\frac{3}{2}}(\varphi - \f_\infty - C_0) - Y\right) e^{\left( \sqrt{\frac{3}{2}}(\varphi - \f_\infty - C_0) - Y\right)}
\end{equation}
\begin{equation}
    W\left(\sqrt{\frac{2}{3}} e^{\sqrt{\frac{3}{2}}(\varphi - \f_\infty - C_0)}\right) = \left( \sqrt{\frac{3}{2}}(\varphi - \f_\infty - C_0) - Y\right)
\end{equation}
Using an identity of the $W$ function, $\exp(W(x)) = \frac{x}{W(x)}$, we obtain:
\begin{equation}
    \delta \Tilde{S}(\varphi) = \frac{\sqrt{2}}{\sqrt{3} W\left(\sqrt{\frac{2}{3}} e^{\sqrt{\frac{3}{2}}(\varphi - \f_\infty - C_0)}\right)}
\end{equation}
The potential $V = -V_{\infty} e^{\sqrt{\frac{2}{3}}\varphi}$ therefore admits the exact family of solutions in $d=4$:
\begin{equation}
    S_c(\varphi) = \sqrt{\frac{2 V_{\infty}}{3}} e^{\frac{\varphi}{\sqrt{6}}} \left( 1 + \frac{1}{W\left(\sqrt{\frac{2}{3}} e^{\sqrt{\frac{3}{2}}(\varphi - \f_\infty - C_0)}\right)} \right)
\end{equation}
\begin{equation}
    W_c(\varphi) = \sqrt{V_{\infty}} e^{\frac{\varphi}{\sqrt{6}}} \left( 2- \frac{1}{W\left(\sqrt{\frac{2}{3}} e^{\sqrt{\frac{3}{2}}(\varphi - \f_\infty - C_0)}\right)} \right)
\end{equation}
This solution has of course the same second order asymptotics as the ones derived for general $d$ in the text. Using $T=\frac{d}{2} S(W'-S)$ we obtain:
\begin{equation}
    T_c(\varphi) = - 2V_{\infty} e^{\sqrt{\frac{2}{3}} \varphi} \left( \frac{1+ W\left(\sqrt{\frac{2}{3}} e^{\sqrt{\frac{3}{2}}(\varphi - \f_\infty - C_0)}\right)}{W\left(\sqrt{\frac{2}{3}} e^{\sqrt{\frac{3}{2}}(\varphi - \f_\infty - C_0)}\right)^2}\right)
\end{equation}
{This is the slice curvature for a continuous one-parameter family of solutions. In this expression, the flat solution is recovered in the limit $C_0\to -\infty$}

\section{Dynamical system results}
\label{sec:D}

{In this appendix we collect some mathematical results we use in the main text for our purposes.} We use the definition of a dynamical system from a vector field $f$:
\be
 X' = f(\varphi, X),
\label{fX}
\ee
where $X'$ designates the derivative of $X \in \mathbb{R}^n$ with respect to $\varphi$. A particular class of systems for which $f$ does not depend explicitly on $\varphi$ is called autonomous, and have several additional properties that we used in this article. Their critical points $X_*$ are simply solutions of $f(X_*) = 0$.

For every dynamical system, one can define a semigroup structure and the associated orbits. As a reminder, a semigroup is an algebraic structure with an associative operation, but without an inverse for every element.
Given a solution $X(\f)$ of a differential equation, one can define a one parameter family $T(\tau)$ of ``forward translation" operators, that act on the solution and return the solution $X(\f+\tau)$. This set of translations following the flow of the vector field generate a semigroup.

For autonomous linear dynamical systems, i.e. dynamical systems for which $f(\f, X) = A\cdot X$ with $A$ an $n\times n$ matrix, the trajectories can be obtained analytically by using $T(\f) = \exp(\f A)$. The solutions to these dynamical systems are entirely classified, and everything is known about their critical points\footnote{These are points where the dynamical vector field vanishes and the associated flow is stationary.}, stability and nature of solutions, as a function of the eigenvalues at critical points \cite{strogatz, dyna}.

The study of nonlinear autonomous dynamical systems is however more challenging, and complex behaviors can arise \cite{dyna}. An important result in the nonlinear case is the Hartman-Grobman theorem, which gives a condition for these complex phenomena to happen. This theorem states that if the eigenvalues have non-zero real part, the orbits of the nonlinear system can be continuously deformed into orbits of the linearized system, and the nonlinear orbits asymptote to the linearized ones at the critical point. In this case, the critical point is named hyperbolic, and the system can be understood qualitatively by linearizing $f$ at the critical point.

However, linearization fails when the real part of an eigenvalue vanishes, which is known as a {\em local bifurcation}. The behavior of solutions next to critical points at a bifurcation is harder to determine using only analysis tools. However, one can use geometrical tools to obtain more information about the solution. One of these tools is the {\em center manifold}.

We shall consider a dynamical system defined by a smooth vector field $f$ with a critical point, and let $A$ be the gradient of $f$ at the critical point. Let $\sigma_{u/s/c}$ be the subsets of the spectrum of $A$ with respectively positive, negative and zero real part eigenvalues, and let $E_{u/s/c}$ be the associated direct sum of eigenspaces. These linear tools can be used to define nonlinear counterparts known as unstable, stable and center manifolds, $W_u$, $W_s$ and $W_c$:

\begin{itemize}
    \item There exists a unique smooth manifold $W_u$, called {\em unstable manifold}, that is invariant under the flow of $f$ and is tangent to $E_u$ at the critical point.

    \item There exists a unique smooth manifold $W_s$, called {\em stable manifold}, that is invariant under the flow of $f$ and is tangent to $E_s$ at the critical point.

    \item There exists a smooth manifold, $W_c$, called {\em center manifold}, that is invariant under the flow of $f$ and is tangent to $E_c$ at the critical point.
\end{itemize}

$W_c$ may in general fail to be unique. In the cases that arise in this article it is unique and it is possible to find coordinates that span $W_u$,$W_s$ and $W_c$ (Respectively $X_1$, $X_2$ and $Y$ in section \ref{sec:4.3}), and graph the center manifold as function of these coordinates (i.e. find $f$ such that $Y = h(X_1, X_2)$). In these coordinates, one can write a partial differential equation for the function $h$. The pullback of $f$ on the center manifold defines a new dynamical system which has the dimension of the center manifold. This new system has lower dimension, thus it is easier to solve and one may resort again to analytical techniques.
\smallbreak

The center manifold can also be sensitive to some non-local information. When it crosses another critical point, another phenomenon known as a global bifurcation can occur. These bifurcations are harder to see analytically and generally require some topological argument to be detected. In our case, one can use index arguments to obtain information about the topological structure of the phase portrait.

{A tool to study this  is Conley index theory, which is reviewed in \cite{Lappicy}.} It can be used to prove that the critical solutions I and II are connected by {an} orbit for $b \neq b_c$ with $b<b_G$. Indeed, the type-0 generic solutions and type II solutions  exhibit a bifurcation behavior at infinity at $b=b_G$, which destroys the orbit connecting type II and type I. The article of Gukov, \cite{Gukov}, reviews applications of this index to renormalization group flows.

\section{Numerical solutions}
\label{sec:E}

\begin{figure}[H]
\centering
\begin{subfigure}[t]{0.45\textwidth}
\begin{tikzpicture}
\node (img) {\includegraphics[width=\textwidth]{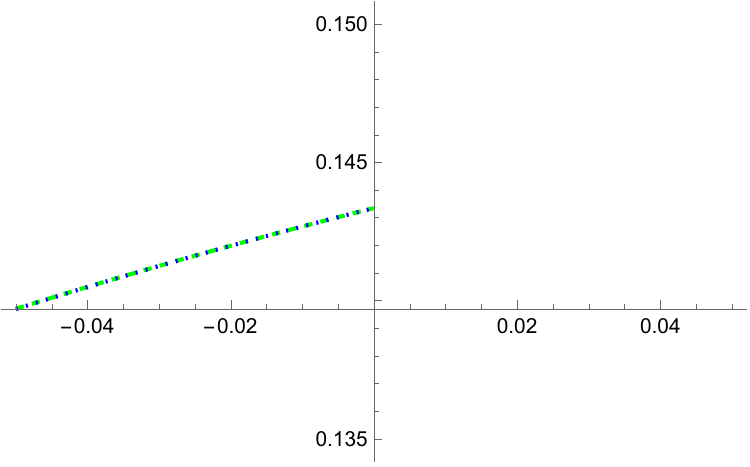}};
\node[below=of img, node distance=0cm, xshift=3cm, yshift=1.2cm]{$\mathcal{R}$};
\node[left=of img, node distance=0cm, rotate=0, anchor=center,xshift=4.5cm, yshift=2.5cm] {$f$};
\end{tikzpicture}
\caption{ }
\end{subfigure}
\begin{subfigure}[t]{0.45\textwidth}
\begin{tikzpicture}
\node (img) {\includegraphics[width=\textwidth]{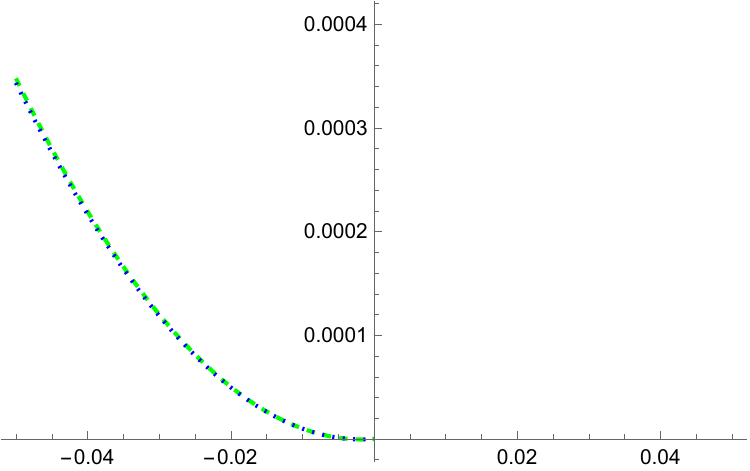}};
\node[below=of img, node distance=0cm, xshift=3cm, yshift=1.2cm]{$\mathcal{R}$};
\node[left=of img, node distance=0cm, rotate=0, anchor=center,xshift=4.5cm, yshift=2.5cm] {$s$};
\end{tikzpicture}
\caption{ }
\end{subfigure}
\caption{Free energy (a) and Entropy (b) per unit volume in the IHQCD case, for full asymptotics (blue dots) and lower order asymptotics (green dashes).}
\label{fig:checkihqcdorder}
\end{figure}
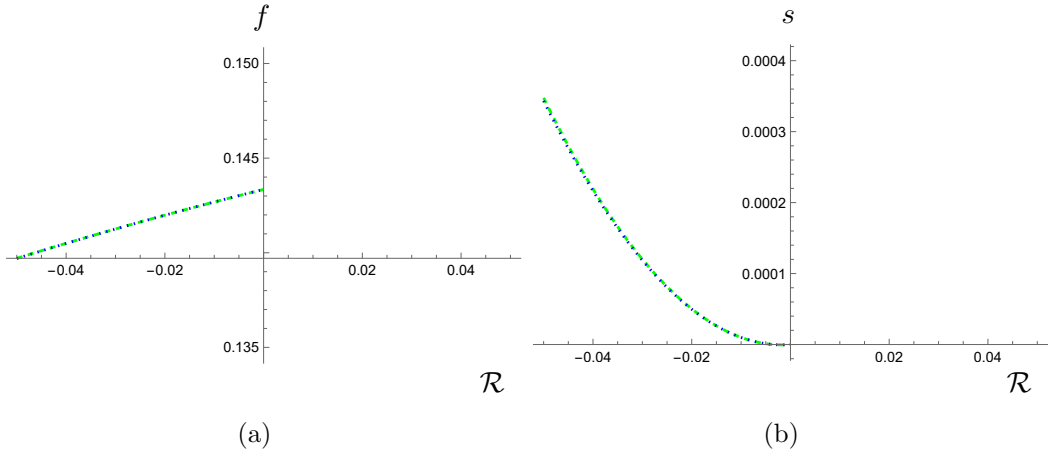

In this appendix, we present how the numerical RG flows were obtained, and we study the sensitivity of the results obtained to the numerical parameters. We use the mathematica function "ParametricNDSolve", with a precision objective set to $10^{-25}$ in the intermediate numerical calculations. We solve the ODE (\ref{odesecorderS}) for chosen values of the IR parameter $c$, and calculate $W$ from (\ref{eqn:WfunS}). For solutions that have $\f \to \infty$, we use a value $\varphi_{ini} = 20000$ for IHQCD, where a logarithmic behavior is expected, and $10000$ for every other case to simulate an infrared at $\f\to+\infty$. For type III solutions the initialisation depends on a parameter $$\varepsilon \equiv \varphi_0 - \varphi. \label{defepsilon}$$ This parameter is initialized at $\varepsilon=10^{-6}$. Once $S$ and $W$ are obtained, we sample them with a step $\delta \varphi_{fit}$ in a region $\varphi_{fit, 1}<\varphi_{fit, 2}$ close to the UV, and fit them with the asymptotics derived theoretically in the UV to obtain $\mathcal{C}$ from $S$ and $\mathcal{R}$ from $W$. The fit parameters used are adjusted to avoid numerical inaccuracy when calculating derivatives:
\be
    \varphi_{fit, 1}= 5\times 10^{-6}\sp  \varphi_{fit, 2}=  5\times 10^{-5}\sp   \delta\varphi_{fit} = 5\times 10^{-7} \label{numparam}
\ee
In the following, we work with renormalized values, for counter-terms $A_{ct} = -0.3$, $B_{ct} = -0.0662$ and $C_{ct}=0$ defined in (\ref{defrenSee}-\ref{ren}).
\smallbreak
Firstly, we check the sensitivity of the results to the precision of the IR initial conditions. To do so, we derive an approximated form to the IHQCD potential (\ref{IHQCDdeltas}) in the region $c>>\log(\varphi_{ini})$:
\begin{equation}
    \delta \Tilde{S}(\varphi) = \frac{d}{2(d-1) \varphi} \left(  1 - \frac{1}{c} - \frac{\log(\varphi)}{c^2} + \mathcal{O}\left(\frac{\log(\varphi)^2}{c^3} \right)\right),
    \label{IHQCDexpand}
\end{equation}
This amounts to neglecting an infinite number of terms in $ \mathcal{O}(\log(\varphi)^n), n \geqslant2$, which are leading compared to the $\mathcal{O}(\frac{1}{\varphi})$ expansion obtained in (\ref{IHQCDdeltas}). We show that the expansion (\ref{IHQCDdeltas}) is sufficient to obtain both the qualitative and quantitative behavior of the free energy and entropy by comparing these results to the ones obtained with the expanded asymptotics (\ref{IHQCDexpand}).
The results are very similar after this modification, as shown in figure \ref{fig:checkihqcdorder}. Asymptotics from (\ref{IHQCDdeltas}) are shown in blue dots, and the lower order expansion (\ref{IHQCDexpand}) is shown in green dashes.

{This indicates that the numerics is not very sensitive to the corrections in the initial conditions specified in the IR, indicating that the next orders are not necessary to grasp the qualitative behavior or the quantitative results of the free energy.}

The precision of the numerical integration might lead to significant errors for the physical quantities. The values obtained with $10^{-25}$ precision are shown in blue and red dots respectively for critical and type III solutions in figure \ref{fig:checkihqcdprecision}. We increase the precision objective to $10^{-35}$ and show it in green and cyan dashes respectively on the same figure. The results obtained for the free energy and entropy are again quantitatively similar:

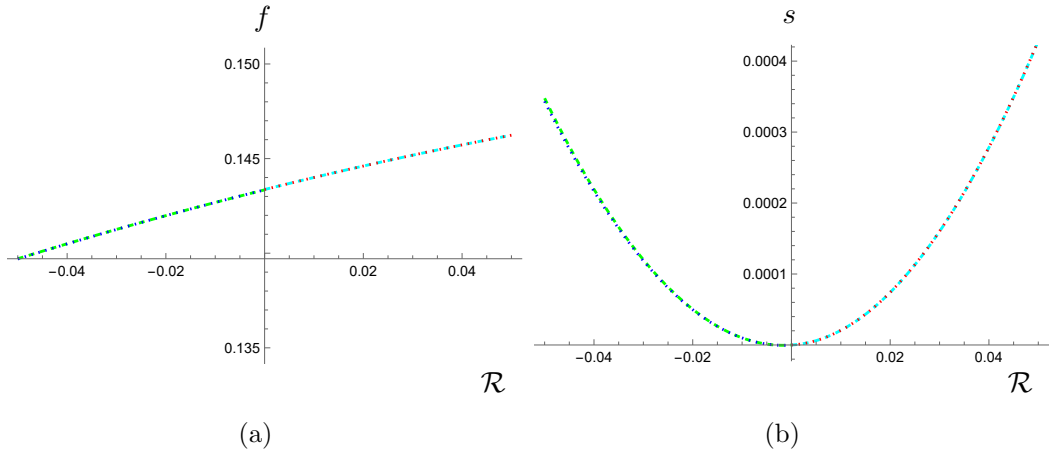
\begin{figure}[H]
\centering
\begin{subfigure}[t]{0.45\textwidth}
\begin{tikzpicture}
\node (img) {\includegraphics[width=\textwidth]{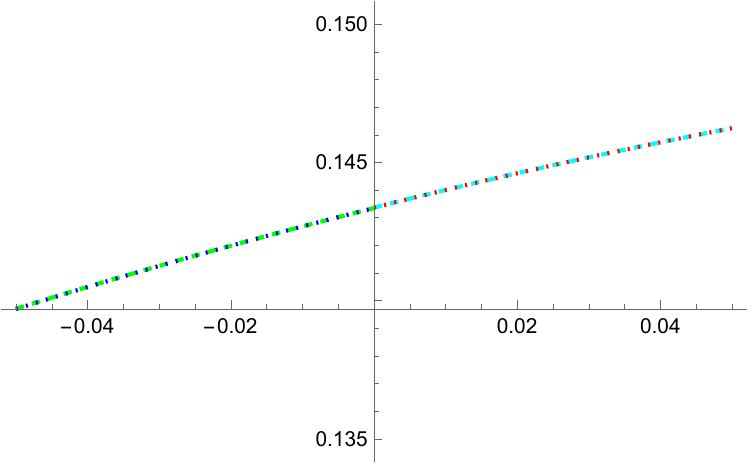}};
\node[below=of img, node distance=0cm, xshift=3cm, yshift=1.2cm]{$\mathcal{R}$};
\node[left=of img, node distance=0cm, rotate=0, anchor=center,xshift=4.5cm, yshift=2.5cm] {$f$};
\end{tikzpicture}
\caption{ }
\end{subfigure}
\begin{subfigure}[t]{0.45\textwidth}
\begin{tikzpicture}
\node (img) {\includegraphics[width=\textwidth]{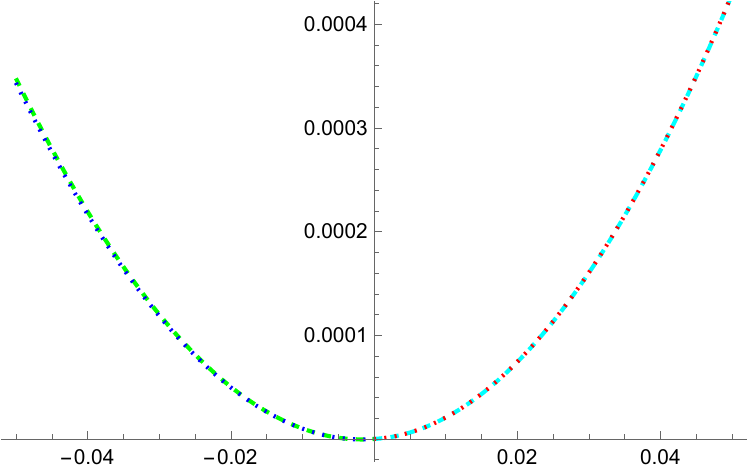}};
\node[below=of img, node distance=0cm, xshift=3cm, yshift=1.2cm]{$\mathcal{R}$};
\node[left=of img, node distance=0cm, rotate=0, anchor=center,xshift=4.5cm, yshift=2.5cm] {$s$};
\end{tikzpicture}
\caption{ }
\end{subfigure}
\caption{Free energy (a) and Entropy (b) per unit volume in the IHQCD case, comparing the precision used in the text(blue/red dots) and increased precision(green/cyan dashes)}
\label{fig:checkihqcdprecision}
\end{figure}

\begin{figure}[H]
    \centering
    \begin{subfigure}[t]{0.45\textwidth}
    \begin{tikzpicture}
    \node (img) {\includegraphics[width=\textwidth]{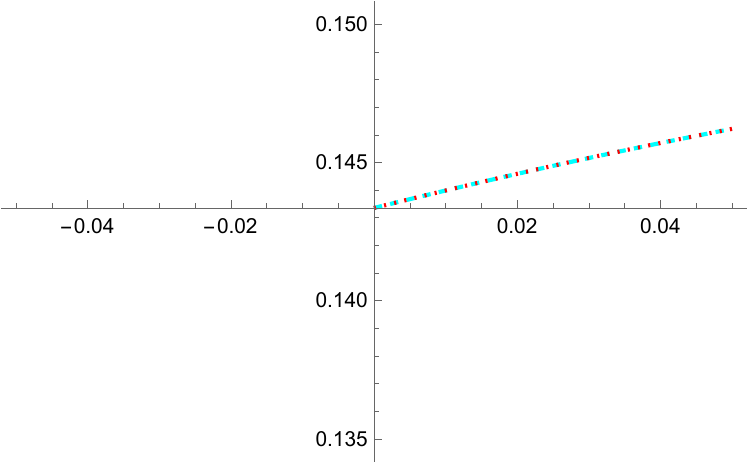}};
    \node[below=of img, node distance=0cm, xshift=3cm, yshift=1.2cm]{$\mathcal{R}$};
    \node[left=of img, node distance=0cm, rotate=0, anchor=center,xshift=4.5cm, yshift=2.5cm] {$f$};
    \end{tikzpicture}
    \caption{ }
    \end{subfigure}
    \begin{subfigure}[t]{0.45\textwidth}
    \begin{tikzpicture}
    \node (img) {\includegraphics[width=\textwidth]{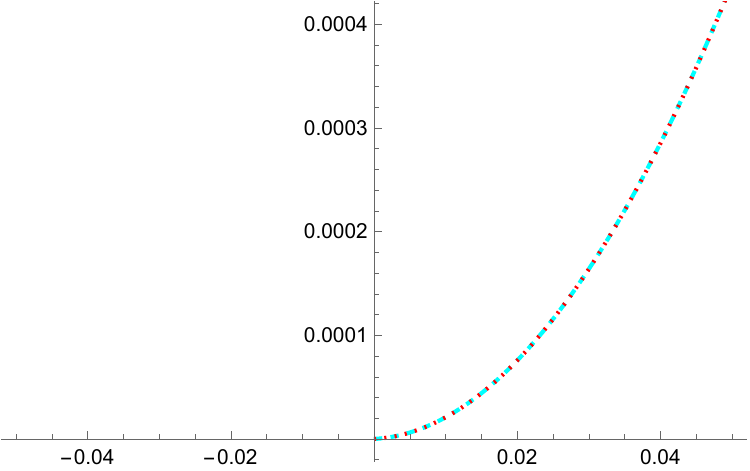}};
    \node[below=of img, node distance=0cm, xshift=3cm, yshift=1.2cm]{$\mathcal{R}$};
    \node[left=of img, node distance=0cm, rotate=0, anchor=center,xshift=4.5cm, yshift=2.5cm] {$s$};
    \end{tikzpicture}
    \caption{ }
    \end{subfigure}
    \caption{Free energy (a) and Entropy (b) per unit volume in the IHQCD case, comparing results with $\varepsilon = 10^{-6}$ (red dots) used in the text and $\varepsilon = 10^{-8}$ in (cyan dashes).}
    \label{fig:checkihqcdinitialcondition}
    \end{figure}
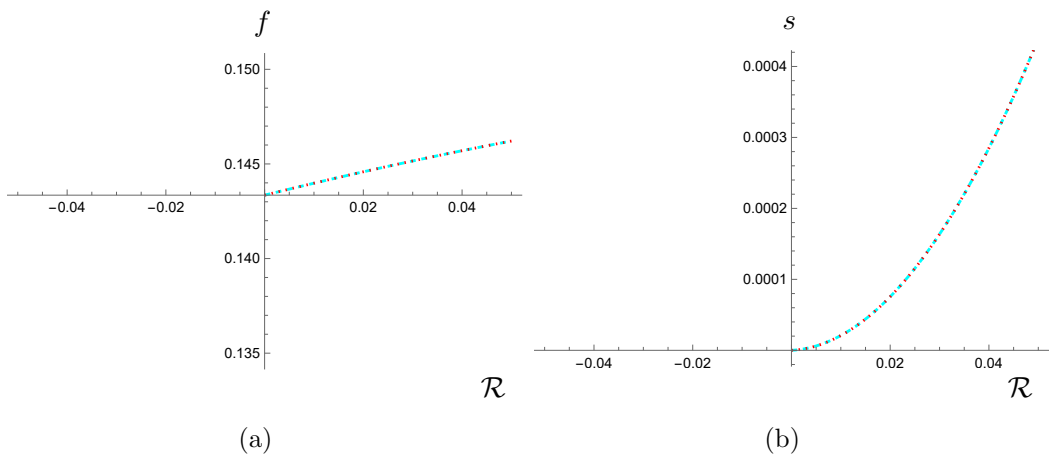

We perform a third check for the type III solutions. We show the difference between $\varepsilon = 10^{-6}$ (red dots) to $\varepsilon=10^{-8}$ (cyan dashes). The results are shown in \ref{fig:checkihqcdinitialcondition}.

\section{Solution to the IHQCD differential equation}
\label{sec:F}

In this appendix we solve analytically equation (\ref{mainU}) and derive $\delta \Tilde{S}$ and $T$ for both the generic power-law case and IHQCD case.
\subsection{Generic power-law case}
We first cover the generic asymptotic power-law correction with exponent $\alpha$ as in equation (\ref{potentialasymp}). In this case the roots $U_1$ and $U_2$ are distinct, we perform a partial fraction expansion:
\begin{equation}
    \left( \frac{ U' }{U - 2\alpha} - \frac{ U' }{U - 1}\right) = \frac{(1-2\alpha)}{\varphi}  + \mathcal{O}(\varphi^{-2})
\end{equation}
Integrating this expansion, we obtain:
\begin{equation}
    \log \left(\left|\frac{U - 2\alpha}{U - 1}\right|\right) = (1-2\alpha)\log(\varphi) + c
\end{equation}
where $c$ is a (real) integration constant which we shall connect shortly to $C_\a$. Here we have two branches due to the absolute values inside the log. Exponentiating this equation, we obtain:
\begin{equation}
    U(\varphi) = \frac{2\alpha \pm e^c \varphi^{1-2\alpha}}{1 \pm e^c \varphi^{1-2\alpha}} \times (1+ \mathcal{O}(\varphi^{-1}))
\end{equation}
Here, the $\pm$ sign distinguishes to two branches: The $+$ sign corresponds to values of $U$ in the interval between the roots $(U_1, U_2)$; the other branch, correponding to $U$ outside this interval, gives the $-$ sign. We can patch up these branches together introducing a new constant $C_\a$ having an arbitrary sign:
\begin{equation}
    U(\varphi) = \frac{2\alpha +  C_\a \varphi^{1-2\alpha}}{1 +  C_\a \varphi^{1-2\alpha}} \times (1+ \mathcal{O}(\varphi^{-1})).
\end{equation}
We obtain the corresponding solutions for the original variables $S, T, W$ by using $\delta \Tilde{S} = X_1 + Y$ and that $Y = h(X)$ is of order $\frac{1}{\varphi^2}$. Hence, until we reach order $\frac{1}{\varphi^2}$, $\delta \Tilde{S}$ and $X_1$ have same asymptotics. Using the definition of $X_1$, we obtain:
\begin{equation}
    \delta \Tilde{S}(\varphi) = \frac{d}{2(d-1) \varphi} \frac{2\alpha + C_\a \varphi^{1-2\alpha}}{1 +  C_\a \varphi^{1-2\alpha}}\times (1+ \mathcal{O}(\varphi^{-1})).
\end{equation}
We can then calculate $T$:
\begin{equation}
    T = \frac{d}{\sqrt{2(d-1)}} e^{\sqrt{\frac{2}{d-1}} \varphi} \varphi^{\alpha - 1}\left( 2\alpha - \frac{2\alpha + C_\a  \varphi^{1-2\alpha}}{1+  C_\a \varphi^{1-2\alpha}} + \mathcal{O}(\varphi^{-1}) \right)
\end{equation}

\subsection{IHQCD case}

We cover here the IHQCD case, corresponding to the critical value $\a={1\over 2}$. In this case, the polynomial in the denominator of the left hand side of equation (\ref{mainU}) has only one root with multiplicity two. The equation becomes:
\begin{equation}
    \frac{U'}{(U - 1)^2} = -\frac{1}{\varphi} + \mathcal{O}(\varphi^{-2})
\end{equation}
Integrating and simplifying the result now gives:
\begin{equation}
    U(\varphi) = 1 + \frac{1}{\log(\varphi)- C_{\frac{1}{2}}} +\mathcal{O}(\varphi^{-1}),
\end{equation}
where $C_{\frac{1}{2}}$ is a real integration constant. We obtain the following asymptotics for $\delta \Tilde{S}$:
\begin{equation}
    \delta \Tilde{S}(\varphi) = \frac{d}{2(d-1) \varphi} \left(  1 + \frac{1}{\log(\varphi)- C_{\frac{1}{2}}} +\mathcal{O}(\varphi^{-1})\right),
\end{equation}
where there is again no need to expand to third order, as the second order solution does not reach order $\mathcal{O}(\varphi^{-2})$. We arrive at the following expression for $T$:
\begin{equation}
    T = \frac{d}{\sqrt{2(d-1)}} e^{\sqrt{\frac{2}{d-1}}\varphi} \left( -\frac{1}{\log(\varphi) - C_{\frac{1}{2}}}  + \mathcal{O}(\varphi^{-1}) \right).
\end{equation}

\end{appendix}


\addcontentsline{toc}{section}{References}

\begin{thebibliography}{110}



\bibitem{magoo}
O.~Aharony, S.~S.~Gubser, J.~M.~Maldacena, H.~Ooguri and Y.~Oz,
{\em ``Large N field theories, string theory and gravity,''}
Phys. Rept. \textbf{323} (2000), 183-386
doi:10.1016/S0370-1573(99)00083-6
\hre{hep-th}{9905111}

\bibitem{witten}
E.~Witten,
{\em ``Anti-de Sitter space, thermal phase transition, and confinement in gauge theories,''}
Adv. Theor. Math. Phys. \textbf{2} (1998), 505-532
doi:10.4310/ATMP.1998.v2.n3.a3
\hre{hep-th}{9803131}

\bibitem{Malda-nunez}
J.~M.~Maldacena and C.~Nunez,
{\em ``Towards the large N limit of pure N=1 superYang-Mills,''}
Phys. Rev. Lett. \textbf{86} (2001), 588-591
doi:10.1103/PhysRevLett.86.588
\hre{hep-th}{0008001}

\bibitem{GPPZ}
L.~Girardello, M.~Petrini, M.~Porrati and A.~Zaffaroni,
{\em ``Confinement and condensates without fine tuning in supergravity duals of gauge theories,''}
JHEP \textbf{05} (1999), 026
doi:10.1088/1126-6708/1999/05/026
\hre{hep-th}{9903026}\\
L.~Girardello, M.~Petrini, M.~Porrati and A.~Zaffaroni,
{\em ``The Supergravity dual of N=1 superYang-Mills theory,''}
Nucl. Phys. B \textbf{569} (2000), 451-469
doi:10.1016/S0550-3213(99)00764-6
\hre{hep-th}{9909047}

\bibitem{IHQCD}
 U.~Gursoy    and E.~Kiritsis,
  {\em ``Exploring improved holographic theories for QCD: Part I,''}
  JHEP {\bf 0802} (2008) 032
  doi:10.1088/1126-6708/2008/02/032
\hri{0707.1324}{[hep-th]};\\
U.~Gursoy, E.~Kiritsis and F.~Nitti,
{\em ``Exploring improved holographic theories for QCD: Part II,''}
JHEP {\bf 0802} (2008) 019
doi:10.1088/1126-6708/2008/02/019
\hri{0707.1349}{[hep-th]};\\

\bibitem{VQCD}
M.~Jarvinen and E.~Kiritsis,
{\em ``Holographic Models for QCD in the Veneziano Limit,''},
\hrj{10.1007/JHEP03(2012)002}{JHEP \textbf{03} (2012), 002};
\hri{1112.1261}{ [hep-ph]};\\
T.~Alho, M.~J\"arvinen, K.~Kajantie, E.~Kiritsis and K.~Tuominen,
{\em ``On finite-temperature holographic QCD in the Veneziano limit,''}
\hrj{10.1007/JHEP01(2013)093}{JHEP \textbf{01} (2013), 093};
\hri{1210.4516}{ [hep-ph]};\\
T.~Alho, M.~J\"arvinen, K.~Kajantie, E.~Kiritsis, C.~Rosen and K.~Tuominen,
{\em ``A holographic model for QCD in the Veneziano limit at finite temperature and density,''}
\hrj{10.1007/JHEP04(2014)124}{JHEP \textbf{04} (2014), 124};
[erratum: JHEP \textbf{02} (2015), 033];
\hri{1312.5199}{ [hep-ph]}.

\bibitem{Jani}
J.~Kastikainen, E.~Kiritsis and F.~Nitti,
{\em ``Holographic confining theories on space-times with constant positive curvature,''}
\hri{2502.04036}{ [hep-th]}.


\bibitem{data}
U.~Gursoy, E.~Kiritsis, L.~Mazzanti and F.~Nitti,
{\em ``Improved Holographic Yang-Mills at Finite Temperature: Comparison with Data,''}
\hrj{10.1016/j.nuclphysb.2009.05.017}{Nucl. Phys. B \textbf{820} (2009), 148-177};
\hri{0903.2859}{ [hep-th]};\\
U.~Gursoy, E.~Kiritsis, G.~Michalogiorgakis and F.~Nitti,
{\em ``Thermal Transport and Drag Force in Improved Holographic QCD,''}
\hrj{10.1088/1126-6708/2009/12/056}{JHEP \textbf{12} (2009), 056}
\hri{0906.1890}{ [hep-ph]}.

\bibitem{Panero}
M.~Panero,
{\em ``Thermodynamics of the QCD plasma and the large-N limit,''}
\hrj{10.1103/PhysRevLett.103.232001}{Phys. Rev. Lett. \textbf{103} (2009), 232001};
\hri{0907.3719}{ [hep-lat]};\\
B.~Lucini and M.~Panero,
{\em ``SU(N) gauge theories at large N,''}
\hrj{10.1016/j.physrep.2013.01.001}{Phys. Rept. \textbf{526} (2013), 93-163};
\hri{1210.4997}{ [hep-th]}.

\bibitem{review}
  U.~Gursoy, E.~Kiritsis, L.~Mazzanti, G.~Michalogiorgakis and F.~Nitti,
  {\em ``Improved Holographic QCD,''}
  Lect.\ Notes Phys.\  {\bf 828} (2011) 79
  doi:10.1007/978-3-642-04864-7
\hri{1006.5461}{[hep-th]}.

\bibitem{disect}
E.~Kiritsis,
{\em ``Dissecting the string theory dual of QCD,''}
\hrj{10.1002/prop.200900011}{Fortsch. Phys. \textbf{57} (2009), 396-417};
\hri{0901.1772}{ [hep-th]}.

\bibitem{GK}
B.~Gouteraux and E.~Kiritsis,
{\em ``Generalized Holographic Quantum Criticality at Finite Density,}
JHEP \textbf{12}, 036 (2011)
doi:10.1007/JHEP12(2011)036
\hri{1107.2116}{[hep-th]}.

\bibitem{GK1}
B.~Gouteraux, J.~Smolic, M.~Smolic, K.~Skenderis and M.~Taylor,
{\em ``Holography for Einstein-Maxwell-dilaton theories from generalized dimensional reduction,''}
\hrj{10.1007/JHEP01(2012)089}{JHEP \textbf{01}, 089 (2012)}
\hri{arXiv:1110.2320}{[hep-th]}.

\bibitem{hy2}
L.~Huijse, S.~Sachdev and B.~Swingle,
{\em ``Hidden Fermi surfaces in compressible states of gauge-gravity duality,''}
\hrj{10.1103/PhysRevB.85.035121}{Phys. Rev. B \textbf{85} (2012), 035121};
\hri{1112.0573}{ [cond-mat.str-el]}.


\bibitem{Basile}
I.~Basile, D.~L\"ust and C.~Montella,
{\em ``Shedding black hole light on the emergent string conjecture,''}
\hrj{10.1007/JHEP07(2024)208}{JHEP \textbf{07} (2024), 208};
\hri{2311.12113}{ [hep-th]}.

\bibitem{AdS2}
A.~Ghodsi, E.~Kiritsis and F.~Nitti,
{\em ``On holographic confining QFTs on AdS,''}
\hri{2409.02879}{ [hep-th]}.

\bibitem{C}
J.~K.~Ghosh, E.~Kiritsis, F.~Nitti and L.~T.~Witkowski,
{\em``Holographic RG flows on curved manifolds and quantum phase transitions,''}
\hrj{10.1007/JHEP05(2018)034}{JHEP \textbf{05} (2018), 034};
\hri{1711.08462}{[hep-th]}.


\bibitem{F}
J.~K.~Ghosh, E.~Kiritsis, F.~Nitti and L.~T.~Witkowski,
{\em ``Holographic RG flows on curved manifolds and the $F$-theorem,''}
\hrj{10.1007/JHEP02(2019)055}{JHEP \textbf{02} (2019), 055};
\hri{1810.12318}{[hep-th]}

\bibitem{dS}
J.~K.~Ghosh, E.~Kiritsis, F.~Nitti and L.~T.~Witkowski,
{\em ``Back-reaction in massless de Sitter QFTs: holography, gravitational DBI action and f(R) gravity,''}
JCAP \textbf{07} (2020), 040
doi:10.1088/1475-7516/2020/07/040
\hri{2003.09435}{[hep-th]}.

\bibitem{s2s2}
E.~Kiritsis, F.~Nitti and E.~Pr\'eau,
{\em ``Holographic QFTs on $S^{2}\times S^{2}$, spontaneous symmetry breaking and Efimov saddle points,''}
\hrj{10.1007/JHEP08(2020)138}{JHEP \textbf{08} (2020), 138};
\hri{2005.09054}{ [hep-th]}.

\bibitem{Jani2}
N.~Jokela, J.~Kastikainen, E.~Kiritsis and F.~Nitti,
{\em ``Flavored ABJM theory on the sphere and holographic F-functions,''}
\hrj{10.1007/JHEP03(2022)091}{JHEP \textbf{03} (2022), 091};
\hri{2112.08715}{ [hep-th]}.

\bibitem{Gubser}
S.~S.~Gubser,
{\em ``Curvature singularities: The good, the bad, and the naked,''}
Adv.\ Theor.\ Math.\ Phys.\  {\bf 4} (2000) 679
\hre{hep-th}{0002160}.

\bibitem{dyna}
J.~Carr,
``\href{https://link.springer.com/book/10.1007/978-1-4612-5929-9}{Applications of center Manifold Theory,''}
Applied Mathematical Sciences, volume 35, ISBN:9781461259305 (1981).

\bibitem{dynaproof}
J.~Guckenheimer and P.~Holmes,
"\href{https://link.springer.com/book/10.1007/978-1-4612-1140-2}{Nonlinear oscillations, dynamical systems, and bifurcations of
vector fields}",
Applied Mathematical Sciences, volume 42, ISBN : 0-387-90819-6, (1983).
%

\bibitem{Gukov}
S.~Gukov,
  {\em ``RG Flows and Bifurcations''}
  Nuclear Physics B (2017) 919
  \hri{arXiv:1608.06638}{[hep-th]}.


\bibitem{strogatz}
S.H.~Strogatz,
"\href{https://doi.org/10.1201/9780429492563 }{Nonlinear Dynamics and Chaos}: With Applications to Physics, Biology, Chemistry, and Engineering (2nd ed.),"
CRC Press (2015)

\bibitem{GN}
S.~S.~Gubser, A.~Nellore, S.~S.~Pufu and F.~D.~Rocha,
{\em ``Thermodynamics and bulk viscosity of approximate black hole duals to finite temperature quantum chromodynamics,''}
\hrj{10.1103/PhysRevLett.101.131601}{Phys. Rev. Lett. \textbf{101} (2008), 131601};
\hri{0804.1950}{ [hep-th]};
S.~S.~Gubser and A.~Nellore,
{\em ``Mimicking the QCD equation of state with a dual black hole,''}
\hrj{10.1103/PhysRevD.78.086007}{Phys. Rev. D \textbf{78} (2008), 086007};
\hri{0804.0434}{ [hep-th]}.

\bibitem{thermo}
U.~Gursoy, E.~Kiritsis, L.~Mazzanti and F.~Nitti,
{\em ``Deconfinement and Gluon Plasma Dynamics in Improved Holographic QCD,''}
\hrj{10.1103/PhysRevLett.101.181601}{Phys. Rev. Lett. \textbf{101} (2008), 181601};
\hri{0804.0899}{ [hep-th]};\\
U.~Gursoy, E.~Kiritsis, L.~Mazzanti and F.~Nitti,
  {\em ``Holography and Thermodynamics of 5D Dilaton-gravity,''}
  JHEP {\bf 0905}, 033 (2009)
\hri{0812.0792}{[hep-th]}.

\bibitem{multirg}
  E.~Kiritsis, F.~Nitti and L.~S.~Pimenta,
  {\em ``Exotic RG Flows from Holography,''}
\hrj{10.1002/prop.201600120}{Fortsch. Phys. \textbf{65} (2017) no.2, 1600120};
\hri{1611.05493}{[hep-th]};\\
F.~Nitti, L.~Silva Pimenta and D.~A.~Steer,
{\em ``On multi-field flows in gravity and holography,''}
\hrj{10.1007/JHEP07(2018)022}{JHEP \textbf{07} (2018), 022};
\hri{1711.10969}{ [hep-th]}.




\bibitem{AdS1}
A.~Ghodsi, J.~K.~Ghosh, E.~Kiritsis, F.~Nitti and V.~Nourry,
{\em ``Holographic QFTs on AdS$_{d}$, wormholes and holographic interfaces,''}
\hrj{10.1007/JHEP01(2023)121}{JHEP \textbf{01} (2023), 121};
\hri{2209.12094}{ [hep-th]}.

\bibitem{BKL}
  V.A. Belinsky, I.M. Khalatnikov, E.M. Lifshitz,
{\em ``Oscillatory approach to a singular point in the relativistic cosmology"},
\href{https://www.tandfonline.com/doi/abs/10.1080/00018737000101171}{Adv.Phys. 19 (1970) 525-573}.


\bibitem{BKL2}
S.~Foster,
{\em ``Scalar field cosmological models with hard potential walls,''}
\hre{gr-qc}{9806113}.


\bibitem{Umut}
U.~Gursoy,
{\em ``Gravity/Spin-model correspondence and holographic superfluids,''}
\hrj{10.1007/JHEP12(2010)062}{JHEP \textbf{12} (2010), 062};
\hri{1007.4854}{ [hep-th]};\\
{\em ``Continuous Hawking-Page transitions in Einstein-scalar gravity,''}
\hrj{10.1007/JHEP01(2011)086}{JHEP \textbf{01} (2011), 086};
\hri{1007.0500}{ [hep-th]}.

\bibitem{Aharony2}
O.~Aharony, E.~Y.~Urbach and M.~Weiss,
{\em ``Generalized Hawking-Page transitions,''}
\hrj{10.1007/JHEP08(2019)018}{JHEP \textbf{08} (2019), 019}
\hri{1904.07502}{[hep-th]}


\bibitem{Marconnet:2022fmx}
P.~Marconnet and D.~Tsimpis,
{\em ``Universal accelerating cosmologies from 10d supergravity,''}
JHEP \textbf{01}, 033 (2023)
doi:10.1007/JHEP01(2023)033
\hri{2210.10813}{[hep-th]}.

\bibitem{Andriot:2023wvg}
D.~Andriot, D.~Tsimpis and T.~Wrase,
{\em ``Accelerated expansion of an open universe and string theory realizations,''}
Phys. Rev. D \textbf{108}, no.12, 123515 (2023)
doi:10.1103/PhysRevD.108.123515
\hri{2309.03938}{[hep-th]}.

\bibitem{DeClerck:2023fax}
M.~De Clerck, S.~A.~Hartnoll and J.~E.~Santos,
JHEP \textbf{07}, 202 (2024)
doi:10.1007/JHEP07(2024)202
\hri{2312.11622}{[hep-th]}.

\bibitem{multi-field}
F.~Nitti, L.~Silva Pimenta and D.~A.~Steer,
{\em ``On multi-field flows in gravity and holography,''}
JHEP \textbf{07}, 022 (2018)
doi:10.1007/JHEP07(2018)022
\hri{1711.10969}{[hep-th]}.

\bibitem{Marolf:2010tg}
D.~Marolf, M.~Rangamani and M.~Van Raamsdonk,
{\em ``Holographic models of de Sitter QFTs,''}
Class. Quant. Grav. \textbf{28}, 105015 (2011)
doi:10.1088/0264-9381/28/10/105015
\hri{1007.3996}{[hep-th]}.

\bibitem{Blackman:2011in}
J.~Blackman, M.~B.~McDermott and M.~Van Raamsdonk,
{\em ``Acceleration-Induced Deconfinement Transitions in de Sitter Spacetime,''}
JHEP \textbf{08}, 064 (2011)
doi:10.1007/JHEP08(2011)064
\hri{1105.0440}{[hep-th]}.


\bibitem{Lappicy}
P.~Lappicy,
"Conley's index and connection matrices for non-experts,"
\hri{1901.05565}{[math.DS]}



\end{thebibliography}

\end{document}